\documentclass[11pt]{article}

\usepackage[a4paper,margin=20mm]{geometry}
\usepackage{seqsplit}
\usepackage{setspace}

\usepackage{amsmath,amssymb,bm}
\usepackage{graphicx}
\usepackage{booktabs}
\usepackage{array}
\usepackage{multirow}
\usepackage{lmodern}

\usepackage{xcolor}
\usepackage{caption}
\usepackage{subcaption}
\usepackage{enumitem}
\usepackage{url}

\usepackage[hidelinks]{hyperref}
\usepackage[T1]{fontenc}
\usepackage[utf8]{inputenc}
\usepackage[super,comma,sort&compress]{natbib}

\DeclareCaptionFont{figurecompact}{\fontsize{8.5}{10}\selectfont}
\setlist{nosep}

\newcommand{\R}{\mathcal{R}}
\newcommand{\TV}{d_{\mathrm{TV}}}

\newcommand{\modelid}[1]{\texttt{\seqsplit{#1}}}

\graphicspath{{figures/}}

\newcommand{\draftfigure}[2]{%
  \includegraphics[width=\textwidth]{#1}%
}

\newcommand{\FigOnePath}{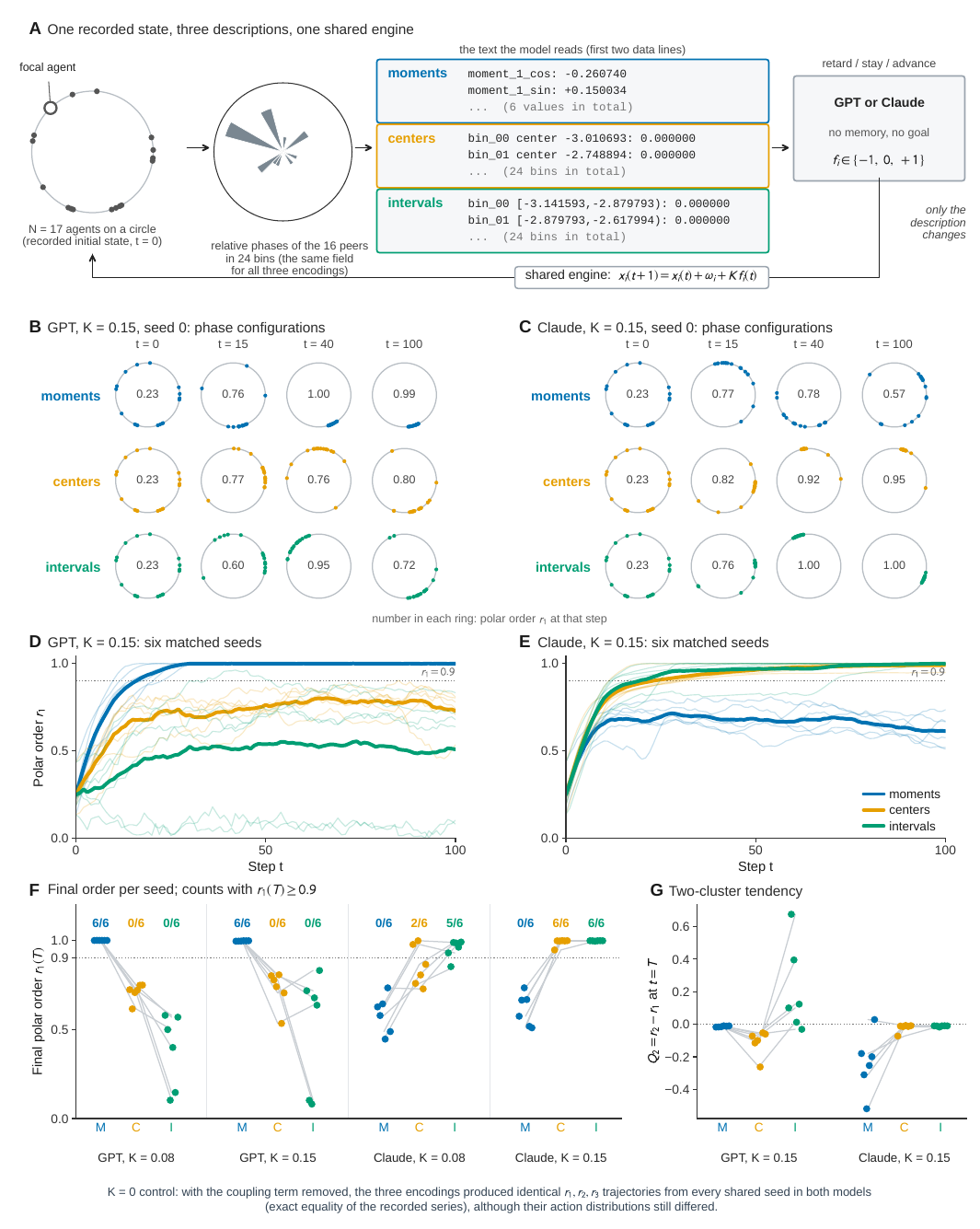}
\newcommand{\FigTwoPath}{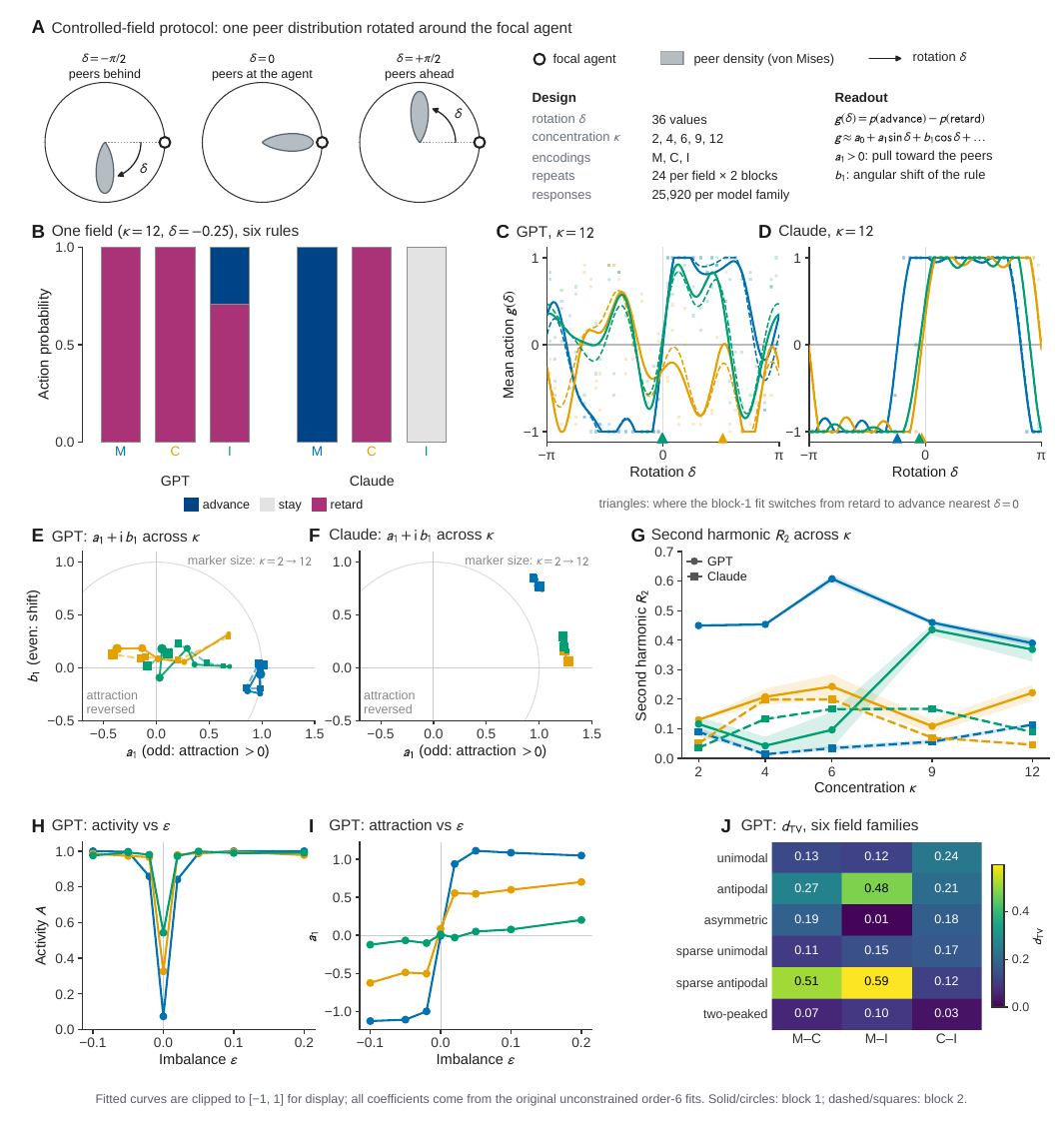}
\newcommand{\FigThreePath}{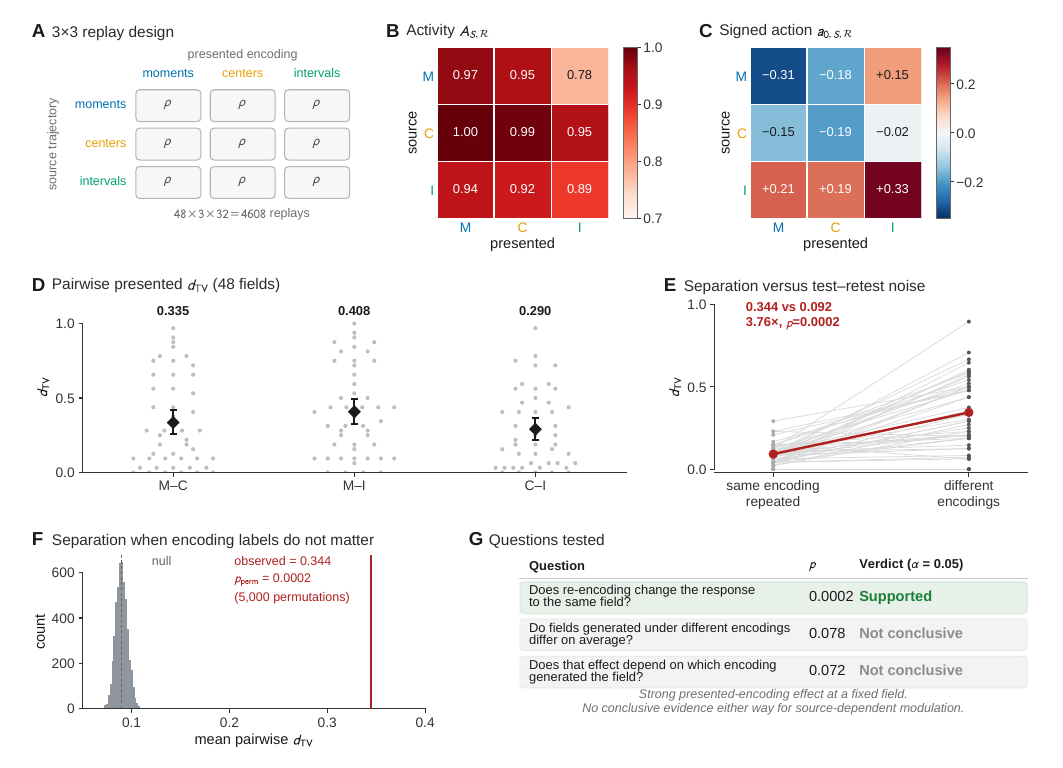}
\newcommand{\FigFourPath}{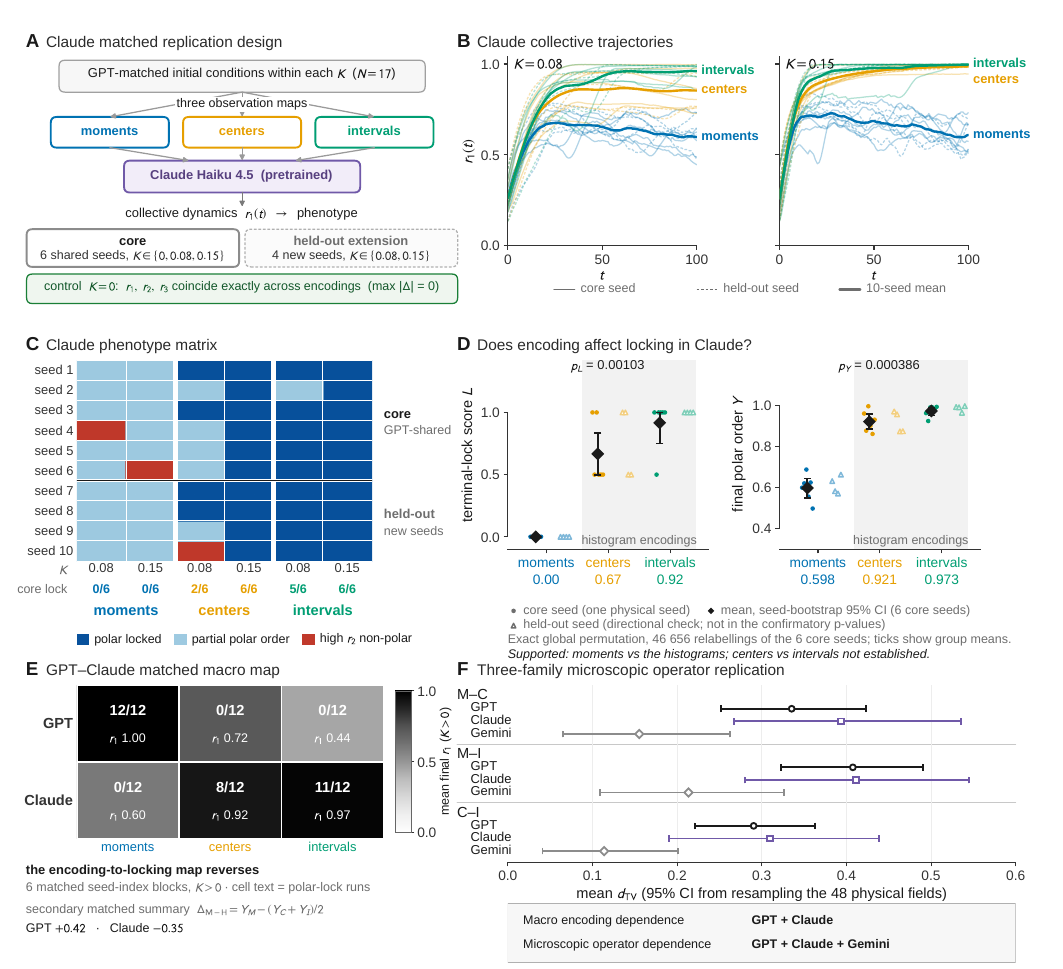}
\newcommand{\FigFivePath}{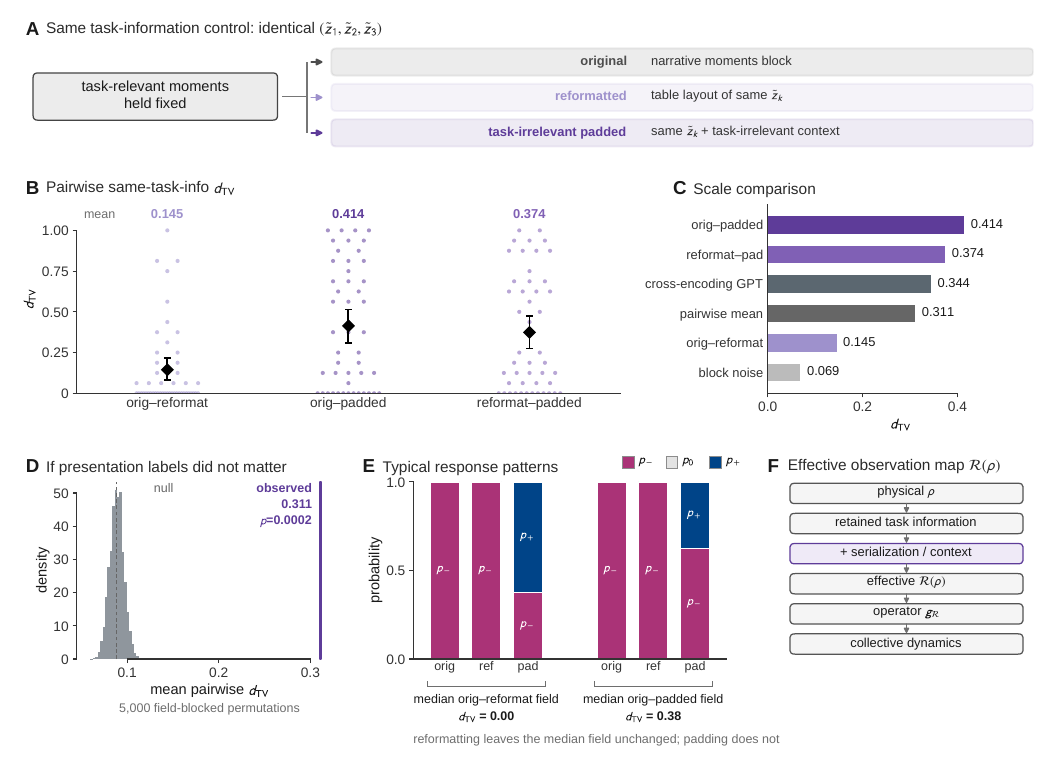}

\title{How a shared state is described determines whether AI agents synchronize}

\author{
Takahiro Ezaki$^{1,*}$, Naoto Imura$^{1}$, Katsuhiro Nishinari$^{1,2}$\\[1mm]
\small $^{1}$Research Center for Advanced Science and Technology,\\
\small The University of Tokyo, Tokyo, Japan.\\[0.6mm]
\small $^{2}$Department of Aeronautics and Astronautics, School of Engineering,\\
\small The University of Tokyo, Tokyo, Japan.\\[0.6mm]
\small $^{*}$Correspondence: \href{mailto:tkezaki@g.ecc.u-tokyo.ac.jp}{tkezaki@g.ecc.u-tokyo.ac.jp}
}

\date{}

\usepackage{float}
\usepackage{accsupp}
\newcommand{\pageofn}{}
\DeclareCaptionLabelFormat{sipaged}{#1~#2\pageofn}

\newcommand{\SIpage}[1]{\centering
  \includegraphics[width=180mm]{#1}}

\newcommand{\SIpageT}[3]{\centering
  \includegraphics[width=180mm,trim=0 #2 0 #3,clip]{#1}}

\makeatletter
\newcommand{\ident}{\begingroup
  \catcode`\_=12 \catcode`\#=12 \catcode`\&=12 \catcode`\%=12 \catcode`\$=12
  \catcode`\~=12 \catcode`\^=12 \ident@aux}
\newcommand{\ident@aux}[1]{%
  \BeginAccSupp{method=escape,ActualText={#1}}%
  \texttt{\ident@break#1\@nil}%
  \EndAccSupp{}\endgroup}
\def\ident@break#1\@nil{\ident@loop#1\@nil}
\def\ident@loop#1{%
  \ifx#1\@nil\else
    #1%
    \expandafter\ident@maybebreak\expandafter#1%
  \fi}
\def\ident@maybebreak#1{%
  \if#1_\allowbreak\else\if#1/\allowbreak\else
  \if#1.\allowbreak\else\if#1-\allowbreak\fi\fi\fi\fi
  \ident@loop}
\makeatother
\begin{document}
\maketitle

\begin{abstract}
Language-model agents increasingly act in populations, where the outcome that matters is collective: whether they align, split or fail to coordinate. Each acts not on the world but on a text description of it, a choice usually fixed in software. Using synchronization, the canonical probe of how interaction rules produce collective order, we show that this choice can decide the outcome. Agents on a circle chose to advance, stay or move back after reading the others' relative positions, in 507,112 valid responses across matched populations, controlled inputs and three model families. In GPT, numerical summaries aligned every matched population at both positive couplings, whereas histograms aligned none; Claude showed the reverse at the stronger coupling. Re-describing identical states shifted action probabilities in all three families, even between histograms carrying the same information. No single directional coefficient explained the outcome: state descriptions are part of the interaction rule that turns individual responses into collective order.
\end{abstract}

\section*{Introduction}

Language-model agents increasingly act in populations: pricing agents that share a market \cite{fish2024algorithmic,calvano2020artificial,ezaki2026shippers}, logistics agents that share a supply network \cite{ezaki2026shippers}, and software agents that negotiate, vote and coordinate \cite{park,tom,du,agentverse}. In such populations the outcome that matters is collective: whether the agents align, split or fail to coordinate. Every one of these agents acts not on the world but on a text description of it. The same state can be written as a few summary statistics or as a full distribution, as prose or as a table, and the choice is usually fixed inside the software rather than tested as an experimental variable. We ask whether this choice alone can change what a population does.

Language-model responses are sensitive to prompt formatting \cite{sclar}, the order \cite{lu} and position \cite{liu_lost} of information, and the labels assigned to choices \cite{zheng}. But a change in one response does not tell us what will happen in an interacting population. Each agent's action changes what the others observe next. Through this feedback, a small change in action probabilities may disappear, accumulate or redirect the population. Previous work has shown that language-model populations form social conventions and amplify biases \cite{ashery}, respond to network structure \cite{zomer}, produce collective outcomes in social dilemmas \cite{willis} and drive market concentration \cite{ezaki2026shippers}. Critical analyses ask for systematic measurement of such emergent behavior \cite{lamalfa}. None of these studies varied the description of the state while holding the model, the dynamics and the initial conditions fixed; that is the test reported here.

We call the rule that converts a physical state into the text an agent reads its \emph{observation map}. In sequential decision-making, actions depend on observations rather than directly on the underlying state \cite{kaelbling,zhang_dbc}; the resulting actions then change the states encountered later \cite{perdomo,ross}. For interacting agents, the observation map can therefore shape both individual responses and collective dynamics. We test these two effects separately: first by changing descriptions in an interacting population, and then by measuring responses to exactly the same fixed states.

Our test system is deliberately simple. Each agent has a position, or phase, on a circle. Such systems provide a standard way to study how individual interactions produce synchronization \cite{kuramoto1975,winfree,acebron}, with applications to chemical oscillators \cite{kiss}, power grids \cite{dorfler} and circadian clocks \cite{yamaguchi}. Here, each agent reads the other agents' positions relative to its own and chooses to advance, stay or move back. A shared simulation engine applies the chosen action. Unlike the standard Kuramoto model, the interaction rule is not a prescribed sine function: it comes from the language model's responses. The agents receive no instruction to synchronize.

We describe the same distribution of relative positions in three ways: a compact numerical summary using the first three circular moments, or a 24-bin histogram whose bins are labelled by their centers or their intervals. Moments summarize broad angular patterns while discarding finer detail; the two histograms retain the same bin masses under different labels. GPT populations aligned under the summary description but not the histograms, whereas Claude showed the reverse at the stronger coupling. Responses to fixed states also differed across descriptions, including the two information-matched histograms. Controlled inputs exposed differences in the response curves but showed that a positive directional component alone could not explain synchronization. A further GPT control changed responses even when the numerical summary was unchanged and only its presentation varied.

\section*{Results}

\subsection*{State descriptions reverse which populations synchronize}

We simulated agents on a circle, following the phase-oscillator framework \cite{acebron,kuramoto1975}. Agent $i$ has a phase $x_i(t)$ that accumulates over time and a position $\theta_i(t)=x_i(t)\bmod 2\pi$ on the circle. At each step, it receives a text description of the other agents' relative positions and returns one action, $f_i(t)\in\{-1,0,+1\}$: retard, stay or advance. All agents then update together according to
\begin{equation}
x_i(t+1)=x_i(t)+\omega_i+Kf_i(t).
\label{eq:update}
\end{equation}
The fixed increment $\omega_i$ gives each agent its own motion in the absence of interaction. The coupling $K$ sets how far a chosen action moves it, in radians per step. The model sees neither of these quantities, nor its absolute position, identity, time or previous actions.

For each agent, we summarize its peers' relative positions in a 24-bin histogram, which we call the \emph{relative-phase field}. We then encode that same field as \emph{moments}, \emph{centers} or \emph{intervals} (Fig.~1A). Moments reports six numbers: the real and imaginary components of the first three circular moments, which capture broad one-, two- and three-group angular patterns. Centers lists the mass at each bin center; intervals lists the same masses with interval labels to six decimal places. All three descriptions come from the same binned data. The comparison between moments and histograms changes both information and presentation; the comparison between the two histograms changes the labels while retaining the bin masses.

Within each model, initial conditions, the update rule and the action set were matched across encodings. Only the description of the observation and its numerical content changed. Each call was stateless, with no memory or learning within a run. Thus any effect on the simulated population had to act through the probabilities of the three actions.

We first tested \modelid{gpt-5.4-mini} with $N=17$ agents over $T=100$ steps at $K=-0.15$, $0$, $0.08$ and $0.15$. Each coupling used six initial configurations, generated by six physical random seeds and matched across encodings. One fixed vector of natural increments was used throughout. All 122,400 calls returned valid actions. The six initial configurations, rather than agents or individual calls, provide the independent comparisons.

GPT populations synchronized under moments but not under either histogram. We measure alignment by the polar order parameter $r_1$: it equals 1 when all phases coincide and approaches 0 when they are spread evenly around the circle (Materials and Methods, Eq.~(\ref{eq:order})). At $K=0.08$, mean final order was $0.999$, $0.710$ and $0.382$ for moments, centers and intervals; at $K=0.15$, it was $0.996$, $0.727$ and $0.507$ (Fig.~1B,D,F). Moments reached the prespecified high-order threshold, $r_1(T)\geq0.9$, in all six runs at both positive couplings; neither histogram reached it in any run. All 12 moments runs also stayed above 0.9 for the final 68 observations (Supplementary Figs.~S3,S4). We use \emph{locking} for this finite-time alignment, without claiming permanent frequency synchronization. Every paired comparison favored moments over either histogram ($p=0.03125$ for each two-sided exact sign test; individual tests are not adjusted for multiplicity).

The histogram outcomes also differed in structure. Centers usually produced partial alignment, whereas intervals more often produced two opposed groups. Such groups can have low $r_1$ but high second-order alignment $r_2$; a single aligned group has high values of both. We therefore use $Q_2=r_2-r_1$ to distinguish two-group structure from ordinary alignment \cite{hansel,okuda}. For intervals, mean $Q_2$ was $0.128$ and $0.213$ at the two positive couplings, compared with $-0.004$ and $-0.013$ for moments (Fig.~1G; Supplementary Figs.~S3,S4).

Two controls help interpret this separation. At $K=0$, actions have no effect on motion, and all encodings gave exactly the same $r_1,r_2,r_3$ trajectories from each shared initial condition (Supplementary Fig.~S1). This verifies that the conditions used the same engine and initial states. At $K=-0.15$, all three encodings remained active but produced low polar order. Their mean signed actions, termed \emph{social torques}, were $-0.335$, $-0.033$ and $+0.062$ for moments, centers and intervals, respectively (Materials and Methods, Eq.~(\ref{eq:torque}); Supplementary Fig.~S5). Low alignment therefore did not simply reflect inactivity.

Claude showed the opposite ordering. We repeated the collective experiment with the same six initial configurations at $N=17$, $T=100$ and $K\in\{0,0.08,0.15\}$ (Fig.~1C,E,F; Fig.~4A). At $K=0.15$, both histograms locked in all six runs and moments in none. At $K=0.08$, the counts were $0/6$, $2/6$ and $5/6$ for moments, centers and intervals (Fig.~4C). At zero coupling, the trajectories again coincided exactly across encodings.

The Claude comparison combined the two positive couplings within each seed index rather than treating them as independent repetitions. Encoding affected both the terminal-lock score and mean final order ($p=0.00103$ and $p=0.000386$, respectively, over all 46,656 within-seed label permutations; Fig.~4D). Pairwise comparisons supported moments versus each histogram, but did not establish an ordering between the histograms. Four additional initial configurations reproduced the direction of the effect and were excluded from the confirmatory tests (Supplementary Figs.~S15--S17). The reversal therefore concerns moments versus histograms, a change in both retained information and presentation.

The reversal also persisted beyond a single endpoint. In a post hoc check at $K=0.15$, both Claude histograms stayed above $r_1=0.9$ for all of the final 20 observations in every run; moments did so in none. At $K=0.08$, the intervals count was $4/6$ over this window, compared with $5/6$ at the endpoint; centers remained at $2/6$. These descriptive checks leave the confirmatory tests unchanged (Supplementary Methods).

\subsection*{Controlled inputs reveal how the response rules differ}

In a running population, each encoding generates its own sequence of states. We therefore measured responses to fixed, synthetic fields to separate the effect of the description from the effect of the states encountered. Repeated queries estimated the three action probabilities for each field $\rho$ and encoding $\R$,
\begin{equation}
\bm p_{\R}(\rho)
=
\left[
p_{\R}(-1\mid\rho),
p_{\R}(0\mid\rho),
p_{\R}(+1\mid\rho)
\right].
\label{eq:operator}
\end{equation}
This distribution is the agent's \emph{response operator}. Its activity, $A_{\mathrm{op}}=1-p_{\R}(0\mid\rho)$, is the probability of moving rather than staying. Its mean signed action, $g_{\R}=p_{\R}(+1\mid\rho)-p_{\R}(-1\mid\rho)$, indicates whether responses favor advancing or moving back.

We first placed the peers in a single angular cluster and rotated that distribution around the focal agent (Fig.~2A). The resulting curve shows how the mean action changes as the cluster moves from behind the agent to ahead of it. We describe the curve with Fourier components, analogous to the phase-interaction functions used in oscillator theory \cite{daido} and measured in physical oscillators \cite{kiss_entrain}. The sine coefficient $a_1$ captures a directional component that changes sign between ahead and behind. The cosine coefficient $b_1$ shifts the angle of this first component, and the second-harmonic amplitude $R_2$ measures a component that repeats twice around the circle. These are descriptions of the response curve, not assumptions about collective stability.

In GPT, tightening the cluster changed the three response curves in different ways. We varied its concentration over $\kappa\in\{2,4,6,9,12\}$, where larger $\kappa$ means a narrower cluster, and measured 36 rotation angles in two separately acquired batches, or \emph{blocks} (25,920 valid calls). At high concentration, moments retained a strong directional component and always acted; centers reversed its first sine component and favored retarding overall; intervals developed a prominent second harmonic and little first sine component (Fig.~2B--G). All three patterns met the criteria specified before acquisition in both blocks (Supplementary Figs.~S6--S8).

The differences extended beyond single-cluster fields. For two-peaked, asymmetric, exactly opposed and sparse few-peer distributions, the encodings differed in activity, mean action and the full action probabilities (Fig.~2J; Supplementary Figs.~S7--S10). We compare distributions using total-variation distance, which ranges from 0 for identical probabilities to 1 for no overlap. When peers formed two balanced, opposite groups, the first sine component was near zero for all encodings, but their full response distributions still differed. Moments now mostly stayed still. Introducing even the smallest tested imbalance, $|\varepsilon|=0.02$, made it active again (Fig.~2H,I; Supplementary Fig.~S9). This is a sharp change at the measured resolution, not evidence of a mathematical discontinuity.

We next repeated the concentration sweep in Claude using the exact GPT prompts and rotation angles (25,920 valid responses in two blocks). This descriptive follow-up was designed after the collective reversal was known. Claude's histograms retained strong positive $a_1$ across all concentrations ($1.23$--$1.28$), unlike GPT's negative centers and weak intervals responses at high concentration (Fig.~2E,F). Yet Claude moments also had positive $a_1$ throughout ($0.94$--$1.03$), despite not synchronizing in the collective experiment. Its response curve was shifted in angle, with a much larger cosine component $b_1$ (Fig.~2D,F). At $\kappa=12$, the first-component angle was $37.9^\circ$ and $37.4^\circ$ in the two blocks, and stayed between $37.3^\circ$ and $38.5^\circ$ when the number of fitted Fourier components changed (Supplementary Fig.~S26). This describes the response to a rotated input, not a measured lag between interacting agents.

Thus the controlled inputs reveal differences that could contribute to the collective reversal, while ruling out a simple explanation based on the sign of $a_1$ alone. The shifted moments response is a candidate explanation for further testing. It has not been shown to cause the lack of synchronization: a running population visits changing distributions, often with more than one peak, and these response curves have not been validated as a model of that feedback.

\subsection*{Different descriptions of the same state elicit different actions}

We also tested states taken directly from the interacting populations. This \emph{replay} freezes a state and presents it in each encoding, so differences in the response are no longer confounded with differences in the input state. We selected 48 fields using a rule fixed before replay acquisition, balancing the encoding that generated each field (its \emph{source}) and six types of trajectory state. We then showed every field to GPT in all three \emph{presented} encodings, collecting 32 responses per field and encoding across two blocks (4,608 valid calls; Fig.~3A; Supplementary Figs.~S11--S13).

Changing the description changed the action probabilities. Mean pairwise total-variation distance was $0.344$; none of 5,000 within-field label permutations gave a difference this large ($p=0.0002$, the test's resolution limit; Fig.~3F). The difference was $3.76$ times the variation between repeated measurements of the same encoding and field ($0.092$; Fig.~3E). We call this repeat-measurement variation \emph{block noise}. All three encoding pairs exceeded it, with field-bootstrap intervals for the differences above zero. In particular, centers and intervals differed by $0.290$ (95\% CI $0.220$--$0.364$), or $3.2$ times the block noise, despite retaining the same bin masses (Fig.~3D). Information loss alone therefore cannot explain the response differences.

Two further prespecified tests asked whether the responses depended on the encoding that had originally generated a field. Neither the source effect on mean action ($p=0.078$) nor its interaction with presented encoding on activity ($p=0.072$) crossed the prespecified threshold (Fig.~3B,C,G; Supplementary Fig.~S13). These tests leave source dependence unresolved; they do not establish its absence.

\subsection*{Response differences extend to Claude and Gemini}

Because model families can behave differently on comparable tasks \cite{xie_trust,cemri}, we repeated the replay panel with \modelid{claude-haiku-4-5-20251001} and \modelid{gemini-3.5-flash}. Each used 16 responses per field and presented encoding across two blocks. Claude returned 2,304 valid responses out of 2,304 attempts; Gemini returned 2,281, with invalid responses not concentrated in one encoding. Both models changed their action distributions when the same field was re-encoded ($p=0.0002$ for each; Fig.~4F; Supplementary Fig.~S14). Their differences were larger relative to block noise than in GPT, mainly because their nearly deterministic responses varied less on repeat measurement, rather than because re-encoding caused larger absolute changes.

\subsection*{Presentation matters with the numerical summary held fixed}

To isolate presentation more directly, we gave GPT the same circular-moment values in three forms: the original text, a table, and the original text with added task-irrelevant material. The 48 replay fields each received 16 responses per version across two blocks (2,304 valid calls; Fig.~5A; Supplementary Fig.~S18).

Presentation changed responses even though the numbers were identical. Mean pairwise distance across the three versions was $0.311$ ($p=0.0002$), compared with block noise of $0.069$ (Fig.~5B--D). Reformatting as a table had a mean effect of $0.145$ (95\% CI $0.079$--$0.219$), concentrated in a subset of fields: 27 of 48 had an estimated distance of zero, as did the median. The block-noise interval was $0.049$--$0.090$; overlap between intervals is not a test of their difference. Adding irrelevant text had a larger mean effect, $0.414$, exceeding the $0.344$ average difference between the original encodings (Fig.~5C,E,F). These are fixed-input results; the reformatted versions were not tested in interacting populations.

The added text changed length, context and the position of the relevant numbers together \cite{liu_lost,sclar,lu,leidinger,min_icl}, so it does not isolate any one of these factors. Two further controls did not support a simple explanation based on character count: the histogram farther from a fixed moments prompt in length was closer in response at both tested anchors, and multiplying the character-count gap by $2.6$--$7.8$ caused little change in response distance (Supplementary Figs.~S19,S20). These comparisons were secondary analyses of the control designs.

\section*{Discussion}

Changing how agents read a shared state changed the collective state they produced. With the same physical dynamics and initial conditions, numerical summaries led GPT populations to align, whereas histograms did so in Claude at the stronger coupling. Replaying identical inputs isolated changes in action probabilities in all three tested models, including on states generated by the agents' own interactions \cite{perdomo}. Together, these experiments show that a state description is part of the interaction rule linking an individual model to the population's behavior.

The reversal between GPT and Claude means that no tested encoding consistently favored synchronization across models. Controlled inputs help explain where their responses differ: Claude's histograms retained strong positive directional components where GPT's did not, while Claude moments had both a positive directional component and a shifted response curve. Because moments did not synchronize in Claude, a positive first sine coefficient cannot by itself explain the collective outcome. What generalizes is sensitivity to the description, rather than one shared mechanism of synchronization. Evaluating an agent therefore requires specifying both the model and what it is shown \cite{cemri,lamalfa,xie_trust}.

The experiments separate three kinds of evidence. The strongest collective contrast, moments versus histograms, changes both retained information and presentation. Centers and intervals retain the same bin masses but give different responses to fixed fields. The GPT presentation control keeps the numerical moments unchanged and alters their layout or surrounding text \cite{sclar,lu}. A further, post hoc analysis linked response differences more closely to whether values were explicitly paired with coordinates than to character count (Supplementary Fig.~S20) \cite{leidinger,min_icl}. Information loss is therefore not the only source of response differences. Whether presentation alone can switch a population between synchronized and unsynchronized states still requires a collective experiment with retained information fixed.

Our simple system separates these effects from learning within a run and changes to the physical engine. Its simplicity also sets the scope of the result. The agents have no explicit goal, so synchronization measures the behavior elicited from the model, not success at a coordination task or similarity to human behavior. Applications to social and economic simulation \cite{park,li2024econagent,ashery}, pricing \cite{fish2024algorithmic,calvano2020artificial,ezaki2026shippers} and task-based coordination \cite{autogen,du,tom,moa} share the same feedback structure: software describes a state, models act on that description, and their actions change the next state. This motivates testing descriptions in those settings; it does not establish that they will show the same effects. The broader connection is to decision-making under partial observations \cite{kaelbling,givan,zhang_dbc,lesort} and to the way deployed policies change the states they encounter \cite{perdomo,hardt_pastfuture,narang_multiplayer}.

A quantitative explanation of the collective outcomes remains open. The measured curves describe responses to particular shapes and concentrations of peer distributions. They need not act as pairwise coupling functions for arbitrary population states. In Claude moments, even the fitted slope at zero rotation changes sign with Fourier order, so it cannot provide a robust stability test (Supplementary Fig.~S26). A stronger test would use a fixed response model to generate new collective trajectories under conditions not used to fit it, and compare their alignment and persistence with language-model populations. The shifted moments response provides one candidate mechanism to examine. Replaying a fixed sequence of states cannot test the feedback that such trajectories accumulate.

Fitted surrogate models offer one route to this test by replacing expensive language-model calls with learned action probabilities \cite{ha,dreamer,janner}. In our secondary analysis, all three surrogates predicted held-out controlled fields well enough to meet their prespecified criteria. Only moments had adequate training coverage of the states encountered by interacting populations. A revised moments surrogate improved average predictions on new collective runs, though not every measured outcome; the histogram surrogates were stopped before that evaluation. These results underline the need to test approximations on states generated through feedback \cite{ross,hgdagger}, as well as under externally imposed distribution shifts \cite{koh}.

The collective experiments used 17 agents, one fixed vector of natural increments, a small set of couplings and a fixed duration. Their locking labels describe finite-time alignment, not thermodynamic phases or permanent frequency entrainment \cite{acebron,strogatz_crawford}. Collective dynamics were tested only in GPT and Claude, and the fixed-number presentation control only in GPT. Each model used one prompt contract and a fixed temperature. The Claude concentration sweep was acquired in September, after the July GPT and collective measurements; the shared model identifier cannot exclude changes over time. Backend updates \cite{chen_drift}, different prompts, memory, goals or larger populations could alter the responses. Within the tested system, however, the result is clear: the choice of state description changes collective order, and its effect can reverse across models. State descriptions should therefore be reported and tested as part of the design of interacting language-model populations.

\section*{Materials and Methods}

\subsection*{Experimental design}

We tested whether changing the description of a shared physical state changes individual actions and collective synchronization. Within each model, matched populations shared initial conditions, natural increments and update rules across descriptions. Controlled-field and replay experiments compared responses to fixed inputs; separate presentation controls held the numerical information constant. Protocols and analysis rules were fixed in repository records before the corresponding acquisition or analysis, except for explicitly identified post hoc checks and amendments. The Claude controlled-field follow-up was designed after the collective reversal was observed. The sections below specify each design, sampling unit and exception; Supplementary Materials provide the protocol and decision records.

\subsection*{Circular-agent dynamics}

Agent $i\in\{1,\ldots,N\}$ had an accumulating phase $x_i(t)\in\mathbb R$ and a wrapped phase $\theta_i(t)=x_i(t)\bmod 2\pi$. All agents updated together according to Eq.~(\ref{eq:update}) \cite{kuramoto1975,winfree,acebron}. The natural increments were $N$ evenly spaced values on $[-0.05,+0.05]$ rad per step, assigned in fixed agent order and reused across seeds, encodings and models. Their mean was zero. This spread permits alignment at the positive couplings through differences in time-averaged actions. Without coupling, it produces a 10-rad difference in accumulated motion between the fastest and slowest agents over 100 steps.

Initial phases were drawn uniformly from $[-\pi,\pi)$ using a \emph{physical seed}, separate from the local seeds used to schedule model calls. Within each coupling, the initial phases matched across encodings and model families. Different couplings used separately generated initial phases through a $K$-dependent seed. Actions were mapped as \texttt{retard}$\mapsto-1$, \texttt{stay}$\mapsto0$ and \texttt{advance}$\mapsto+1$.

Negative coupling was implemented only through the sign of $K$ in the engine. The model received the same instruction at positive, zero and negative $K$. At $K=0$, the engine satisfies $x_i(t)=x_i(0)+t\omega_i$ independently of the sampled actions.

\subsection*{Relative-phase field}

For focal agent $i$, peer $j$ contributed
\begin{equation}
\delta_{ij}(t)=\operatorname{wrap}\left[\theta_j(t)-\theta_i(t)\right]
\in[-\pi,\pi).
\end{equation}
The focal agent was excluded, and the histogram masses were normalized by $N-1$. Prompts omitted the peer count, absolute phase, agent identity, time step, natural increment, coupling and history. Bin boundaries were half-open, and bins appeared in a fixed order. Reordering the peers therefore leaves the encoding unchanged. Numerical checks verified the encoder's rotation convention and invariance to peer ordering; these checks required no additional language-model queries.

\subsection*{Observation maps}

The first three circular moments compactly summarize one-lobed, two-lobed and three-lobed angular structure (a net direction, a pair of opposed groups, a three-group arrangement) while discarding finer bin-level detail. The three principal maps were:
\begin{enumerate}[leftmargin=1.6em,label=(\roman*)]
    \item \textbf{Moments}: the real and imaginary components of the binned circular moment
    \begin{equation}
    \tilde z_m=\sum_b h_b\exp(\mathrm{i}mc_b),
    \qquad m=1,2,3,
    \end{equation}
    written in a fixed text format. The supplied feature is the binned moment
    $\tilde z_m$, i.e.\ the $m$-th moment of the fixed 24-bin representation (bin masses $h_b$,
    bin centers $c_b$); for finite-peer fields it approximates the raw peer moment
    $z_m=\frac{1}{N-1}\sum_{j\neq i}\exp(\mathrm{i}m\delta_{ij})$. All three maps are therefore
    deterministic functions of the same binned field, and the moments agent received
    $\{\tilde z_m\}$ rather than the raw peer moments.
    \item \textbf{Centers}: 24 normalized bin masses $h_b$, each paired with its bin center $c_b$, in a fixed canonical order.
    \item \textbf{Intervals}: the same 24 masses paired with half-open bin intervals and serialized to six decimal places.
\end{enumerate}
Complete prompts, token counts under each provider tokenizer and boundary conventions are reported in Supplementary Fig.~S2 and Supplementary Table~S1.

\subsection*{Language-model response contract}

Each model call was stateless and contained one user message requesting a JSON object with one field, \texttt{social\_action}, set to \texttt{advance}, \texttt{stay} or \texttt{retard}. The instruction, named \texttt{response-law-v0.1} in the code, was fixed except for the description of the encoding and its payload heading (Supplementary Fig.~S2). For the original acquisitions, a deterministic parser first required this exact schema. If that failed, a fixed rescue rule read an action word from the response; it never invented or substituted an action. Unreadable responses were recorded as invalid and retried, with at most three attempts. None entered the retained collective trajectories, and action probabilities were normalized over valid calls. Parser outcomes by model and encoding are reported in Supplementary Table~S2. The later Claude sweep used the narrower fence-removal rule and documented retry exceptions described below.

The three model IDs were \modelid{gpt-5.4-mini}, \modelid{claude-haiku-4-5-20251001} and
\modelid{gemini-3.5-flash}. Generation parameters, provider and API endpoint, acquisition date range, SDK/API versions, sampling parameters (including parameters left unset, for which the provider default applied), retry policy and provider seed support are reported in Supplementary Table~S2.

\subsection*{Microscopic response acquisition}

Repeated calls on a fixed field $\rho$ estimated the three action probabilities in Eq.~(\ref{eq:operator}). We then rotated a single-peaked field by an angle $\delta$ relative to the focal agent and measured the mean signed action. This gives an empirical counterpart of the phase-interaction functions studied in oscillator theory \cite{daido} and experiments \cite{kiss_entrain}. In that framework, odd and even components describe different contributions to phase dynamics. Here we use them to compare response shapes, without assuming that they determine stability in this stochastic, discrete-time system. We report activity and the full action distribution as well, because the same mean can result from very different mixtures of advancing, staying and retarding. For the rotated field $\rho_\delta$, the response curve was
\begin{equation}
g_{\R}(\delta)=
p_{\R}(+1\mid\rho_\delta)-p_{\R}(-1\mid\rho_\delta).
\end{equation}
We fitted
\begin{equation}
g_{\R}(\delta)
=
a_0+
\sum_{m=1}^{M}
\left[
a_m\sin(m\delta)+b_m\cos(m\delta)
\right],
\end{equation}
Here $a_m$ and $b_m$ are the sine (odd) and cosine (even) coefficients. The intercept $a_0$ is the component independent of rotation angle. We also report $C_m=a_m+\mathrm{i}b_m$, its amplitude $R_m=|C_m|$ and angle $\phi_m=\arg C_m$. Fits used the 36 offset means, separately for each acquisition block. A multinomial bootstrap within each offset estimated uncertainty in $(a_m,b_m)$. Angles were reported only when the amplitude was sufficiently large and its uncertainty region excluded the origin.

The GPT concentration sweep crossed three encodings, five concentrations and 36 rotation offsets, with 24 responses per combination in each of two blocks. Offsets were paired around zero and slightly displaced from an evenly spaced grid to distinguish low-order components from higher harmonics. The primary fit used $M=6$; prespecified sensitivity fits used $M\in\{2,4,8,12\}$. Supplementary Fig.~S6 and the Supplementary Methods give the offset design, analytic checks and coefficient-recovery tests.

Additional fields contained two peaks, asymmetric distributions, exactly opposed groups, unequal masses in those groups, or a small number of peers. For the opposed groups, $\varepsilon$ measures the signed excess of mass on one side; $\varepsilon=0$ is balanced. We tested
\begin{equation}
\varepsilon\in\{-0.10,-0.05,-0.02,0,0.02,0.05,0.10,0.20\}.
\end{equation}
Activation scales inferred from this grid were treated as resolution-limited.

\subsection*{Measures of collective behavior}

Order parameters, the standard Kuramoto measures of collective coherence \cite{acebron,daido}, were
\begin{equation}
r_m(t)=
\left|
\frac{1}{N}\sum_{i=1}^{N}
\exp\left[\mathrm{i}m\theta_i(t)\right]
\right|,
\qquad m=1,2,3,
\label{eq:order}
\end{equation}
so that $r_1$ measures polar alignment on the circle. A high $r_2$ is compatible with both one-cluster alignment and two opposed clusters; it is interpreted jointly with $r_1$.
We further used $Q_2=r_2-r_1$, the realised collective activity $A_{\mathrm{run}}(t)=N^{-1}\sum_i\mathbf 1[f_i(t)\neq0]$ (distinct from the expected operator activity $A_{\mathrm{op}}$ above), and the full-run social torque
\begin{equation}
\tau_{\mathrm{run}}
=
\frac{1}{NT}\sum_{t=0}^{T-1}\sum_i f_i(t),
\label{eq:torque}
\end{equation}
the run-averaged net action per agent (positive when agents advance and negative when they retard). With phase states indexed $t=0,\ldots,T$ and actions sampled at $t=0,\ldots,T-1$, summing Eq.~(\ref{eq:update}) over $t$ yields the exact engine identity
\begin{equation}
\Omega_{\mathrm{coll}}
=
\frac{1}{N}\sum_i\frac{x_i(T)-x_i(0)}{T}
=
\overline{\omega}+K\tau_{\mathrm{run}},
\qquad
\overline{\omega}=\frac{1}{N}\sum_i\omega_i,
\end{equation}
The population's mean drift therefore departs from $\overline{\omega}$ by exactly $K$ times the social torque. Windowed torques (for example the final-20-step average $\tau_{20}$) are named explicitly; reported torques are run-averaged unless stated otherwise.

The prespecified endpoint labels were:
\begin{itemize}
    \item polar locked: $r_1(T)\geq0.9$;
    \item high-$r_2$ non-polar: $r_2(T)\geq0.5$ and $r_2(T)>r_1(T)$;
    \item partial polar order: $0.35\leq r_1(T)<0.9$, excluding high-$r_2$;
    \item low-polar active: $r_1(T)<0.35$.
\end{itemize}
These labels describe outcomes in a finite population, with reference to higher-harmonic oscillator dynamics \cite{hansel,okuda,daido}; they are not stability classifications or thermodynamic phases. The endpoint label uses the non-strict criterion $r_1(T)\ge0.9$, whereas \emph{terminal lock} uses the strict criterion $r_1(s)>0.9$: a run showed terminal lock if there existed a time $t^{\ast}$ with $r_1(s)>0.9$ for every $s\in[t^{\ast},T]$, and the terminal-lock time was the smallest such $t^{\ast}$. It admits $t^{\ast}=T$ (no minimum duration); behaviour under an imposed minimum duration is reported in Supplementary Fig.~S4. Both thresholds were fixed before confirmatory analysis. In a post hoc descriptive check, the same positive-coupling trajectories were also summarized by the number of consecutive terminal observations with $r_1>0.9$, the number satisfying a 20-observation trailing window, and mean $r_1$ over those 20 observations (Supplementary Methods). No threshold was selected to maximize a contrast, and these summaries do not enter the confirmatory tests.

\subsection*{GPT matched collective experiment}

The GPT experiment in Fig.~1 used 17 agents, 100 steps, four couplings and six seed indices per coupling. Initial phases matched across encodings within each coupling. Each encoding contributed 40,800 valid calls, for 122,400 in total. Seeds for call ordering and retries were deterministically derived and encoding-specific. Run order and task submission were fixed before acquisition to avoid associating an encoding with a particular acquisition period.

The primary comparison paired runs by physical seed. Acquisition cost limited the design to six seeds, for which a two-sided exact sign test reaches its minimum $p=0.03125$ only when all six differences have the same sign. This occurred for moments versus each histogram at both positive couplings: $2/2^6=0.03125$. We report the four tests separately, without multiplicity adjustment, alongside effect sizes; they do not establish family-wise significance at 0.05. For each coupling and contrast, percentile bootstrap intervals for continuous effects resampled the six paired seed differences 2,000 times. Figure~1 shows individual trajectories and arithmetic means without confidence bands; Supplementary Fig.~S4 and the released tables report endpoint uncertainty. Neither agents nor time points were treated as independent replicates.

\subsection*{Claude controlled-field follow-up}

We repeated the concentration sweep in Claude on 18 September 2026, after observing the collective reversal. The model was \modelid{claude-haiku-4-5-20251001}, with temperature $0.7$ and a 64-token output cap. Three encodings, five concentrations, 36 offsets, 24 responses per combination and two blocks yielded 25,920 retained responses. Prompts and hashes exactly matched the recorded GPT inputs. Within each block, concurrent requests followed the original shuffled task order; blocks were acquired sequentially. We fitted each block separately at $M=6$, with sensitivity fits at $M=2,4,8,12$ and 2,000 within-offset multinomial bootstrap replicates. The cross-model comparison and phase interpretation are descriptive: GPT's prespecified phenotype tests were not reused as confirmatory Claude tests. The July GPT and Claude collective experiments were not acquired at the same time as this sweep.

At $\kappa=12$, Claude moments had $b_1=0.775$ and $0.766$ in the two blocks, compared with $0.063$ and $0.057$ for centers and $0.296$ and $0.294$ for intervals. The moments angle remained $37.3^\circ$--$38.5^\circ$ across $M=2,4,6,8,12$ (Supplementary Fig.~S26).

All nine responses from an initial check were enclosed in Markdown JSON fences. Before the full acquisition, we amended the parser to remove one complete outer fence and then require the original one-field JSON schema. We retained the first received response for each of the four initial tasks; five duplicate retries were excluded but kept in the original journal. All 25,920 retained responses passed after fence removal; none was bare JSON. Communication failures and interrupted requests were retried on the same task, with no fallback action. The complete amendment and retry accounting appear in the Supplementary Information with Figs.~S26,S27. Neither prompts nor response selection depended on the action returned.

\subsection*{Identical-field replay}

Replay fields were selected before their responses to new encodings were acquired. Candidate trajectory frames were assigned to six prespecified state categories, deduplicated by their 24-bin histogram hashes and sampled using quotas for source encoding and category. The panel contained 48 fields: 16 from each source encoding and eight from each category, with every category represented by at least one source encoding.

Each field was encoded under all three presented maps and queried 32 times in two blocks. The primary presented-encoding statistic was the mean, over physical fields, of the three pairwise total-variation distances between action operators,
\begin{equation}
\TV(\bm p,\bm q)=
\frac{1}{2}\sum_{f\in\{-1,0,+1\}}
|p_f-q_f|.
\end{equation}
For the primary test, we shuffled encoding labels among individual responses within each physical field and re-estimated the three action distributions. Labels never moved between fields. A parallel test used multinomial deviance: twice the log-likelihood difference between separate encoding-specific action distributions and one pooled distribution per field, summed over fields. It gave the same decision under the same permutations. A further check preserved acquisition block during shuffling (Supplementary Fig.~S13).

Two other prespecified tests examined the source encoding. The source main effect used the field's mean signed action averaged over presented encodings, with source labels shuffled within state-category$\times$coupling groups. The interaction test used the corresponding field-level activity contrast. Each test asked whether any levels differed, rather than testing individual pairs. All three used 5,000 Monte Carlo permutations at $\alpha=0.05$, without correction across tests. Pairwise comparisons are reported as effect sizes with field-bootstrap intervals, not additional tests. With the plus-one estimator, the smallest reportable $p$ is $1/5001\approx0.0002$; none is zero.

Each field and presented encoding was measured in two acquisition blocks. Their total-variation distance measures repeat-measurement variation, labelled \emph{block noise} in the figures. We compared this with the distance between encodings on the same field. Confidence intervals for both means and their paired difference resampled whole fields, keeping all responses and encodings for a field together.

\subsection*{Cross-family microscopic replication}

Claude and Gemini received the same 48 fields in the same three encodings, with 16 responses per field and encoding across two blocks (2,304 attempted calls per family). Replication required five prespecified conditions: an overall encoding effect with permutation $p<0.05$; greater distance between encodings than between repeat measurements; a positive field-bootstrap interval for that difference; at least two of the three encoding pairs above repeat-measurement variation; and an acceptable invalid-response rate without concentration in one encoding. Replication did not require the encoding pairs to have the same ranking across models.

The $n=16$ design was a prespecified conservative response count, not a per-family power optimum.

\subsection*{Claude collective replication}

The confirmatory Claude experiment reused the six GPT physical seeds at $N=17$, $T=100$ and $K\in\{0,0.08,0.15\}$. Four additional seeds were measured only at positive coupling to check the direction of the result. They were kept separate from all confirmatory tests.

For each core seed $s$ and encoding $\R$, the two positive couplings were combined into a terminal-lock score
\begin{equation}
L_{s\R}=\frac{1}{2}\sum_{K\in\{0.08,0.15\}}
\mathbf{1}\{\text{terminal polar lock}\}
\label{eq:lockscore}
\end{equation}
and a continuous final-order score
\begin{equation}
Y_{s\R}=\frac{r_1(T;0.08)+r_1(T;0.15)}{2}.
\label{eq:orderscore}
\end{equation}
For each core seed, the three encoding labels were permuted jointly across the two positive couplings. We enumerated all 46,656 within-seed permutations and computed the sum of squared deviations of encoding means for $L$ and $Y$. Pairwise contrasts used seed-level bootstrap intervals and exact sign-flip tests (flipping the sign of each seed's paired difference in all $2^6$ ways); the bootstrap resampled the paired differences of the six core seeds 5{,}000 times and reported the percentile interval. The error bars drawn in Fig.~4D are seed-bootstrap percentile intervals of the same construction; the four held-out seeds are shown but enter neither the intervals nor the reported $p$-values.

Three criteria were fixed before Claude acquisition. Gate A required an overall encoding effect on both terminal lock and final order. Gate B required a separation between locking and nonlocking outcomes. Gate C required the same encoding-to-outcome pattern as GPT. Claude passed A and B but failed C: it showed the reverse pattern, which had been specified as a separate possible outcome. Thus the encoding effect replicated, while its direction reversed.

\subsection*{Presentation control with fixed numerical information}

The control used the 48 replay fields in three moments formats: original text, a reformatted table and text with added irrelevant context. All numerical moment values were identical. Each field and version had 16 responses across two blocks. The statistic and within-field permutation followed the replay analysis; pairwise and block-noise intervals resampled whole fields.

Fixed-width formatting made character and token counts constant within each version under the backend tokenizer. Adding irrelevant text changed context, number position and length together. We therefore interpret it as a combined presentation change, not a test of length alone.

\subsection*{Additional presentation controls}

Two controls kept numerical information fixed while changing length, layout and whether each mass was explicitly paired with a bin index or coordinate. Both used the 48 replay fields and returned 3,072 valid responses from 3,072 calls. The prespecified contrasts did not distinguish the proposed explanations; interpretation rests on secondary, post hoc comparisons within the crossed designs (Supplementary Figs.~S19,S20).

\subsection*{Predicting responses and dynamics with surrogate models}

A 17-agent population over 100 steps requires 1,700 language-model calls. In a secondary analysis, we asked whether a fitted surrogate could replace those calls, as learned models replace expensive dynamics in other settings \cite{ha,dreamer,janner}. The surrogate predicts the probabilities of retard, stay and advance from numerical descriptors of the peer distribution. It receives no text, coupling, time or source-encoding label. Separate surrogates were trained for the responses elicited by each encoding. Centers and intervals used identical descriptors, so their differences reflect the measured responses rather than different input features.

For field $\rho$, the input vector $x(\rho)$ was computed from the same 24-bin record in every branch. Moments used 15 shared descriptors; the histogram branches also received the 24 bin masses, giving 39 inputs. The training target was the measured action distribution $\hat{\bm p}(\rho)=(\hat p_{-},\hat p_{0},\hat p_{+})$, estimated from $n(\rho)$ calls with counts $c_f(\rho)$ for action $f\in\{-1,0,+1\}$. We fitted predicted probabilities $\bm q_\theta(x(\rho))$ by minimizing the multinomial log loss over training fields $\rho_1,\ldots,\rho_M$,
\begin{equation}
\ell
=
-\,\frac{\displaystyle\sum_{i=1}^{M}\;\sum_{f\in\{-1,0,+1\}} c_f(\rho_i)\,\log q_{\theta,f}\bigl(x(\rho_i)\bigr)}{\displaystyle\sum_{i=1}^{M} n(\rho_i)},
\label{eq:surrogateloss}
\end{equation}
where $\theta$ contains the fitted model parameters. The loss is the mean negative log-likelihood per call, in nats; lower values indicate better predictions. We evaluated the same loss on held-out data and compared it with a baseline that used only peer count: $\bar{\bm q}^{(m)}$ is the mean training response for $m$ peers, ignoring their distribution. Improvement, $\Delta\ell=\ell_{\mathrm{baseline}}-\ell_{\mathrm{model}}>0$, therefore indicates that the distribution's shape provides predictive information.

We distinguished three tests. \emph{Compressibility} asks whether a surrogate predicts controlled fields excluded from training. Cross-validation held out whole fields, keeping repeated measurements of each field in the same fold, and compared held-out loss with the peer-count baseline. \emph{Closed-loop support} asks whether the training data cover fields encountered by interacting populations. For each such field, nearest-neighbor distance $d_{\mathrm{NN}}$ measures proximity to the training data, while local response dispersion $V_{\mathrm{local}}$ measures whether nearby fields elicited consistent responses. Both had to meet prespecified thresholds. \emph{Transportability} asks whether a frozen surrogate can replace the language model in new collective runs. Starting from matched initial conditions, it generates its own trajectory under the same engine and coupling. We compare action rates and final order using absolute endpoint error $|E_{\mathrm{pred}}-E_{\mathrm{obs}}|$. This last test is reported only where new language-model runs were acquired. The Supplementary Methods specify model selection, folds, descriptors, thresholds and endpoints.

All three surrogates met their prespecified criteria on held-out controlled fields. Centers and intervals failed the training-coverage checks and were stopped before prospective evaluation. For moments, replay exposed a specific error: the surrogate predicted staying on collective fields where the language model acted, although it correctly predicted staying on symmetric fields. We revised the surrogate to distinguish these cases using field descriptors, froze it and evaluated it against 44 new collective runs. Average predictions of actions and final order improved, but not every endpoint did. These conclusions apply to the tested descriptors, candidate models and acquisition domains; Supplementary Figs.~S21--S25 give the detailed comparisons.

\subsection*{Statistical analysis}

The reported experiments comprise 507,112 valid model responses in total, across the collective runs in GPT and Claude, the controlled-field sweeps, the identical-field replays, the presentation controls and the prospective surrogate validation; acquisition parameters for each are listed in Supplementary Table~S2. The total counts each retained collective action once and excludes invalid responses, mock calls and superseded or exploratory collective runs (Supplementary Methods, response-count accounting). The independent comparisons differ between experiments. Collective tests compare initial configurations (physical seeds); replay tests compare responses within the same physical field. Repeated model calls estimate action probabilities and are not independent population replicates. Table~1 summarizes these distinctions.

Repeated calls estimate an action distribution at a fixed field; they do not provide additional physical seeds or fields. Collective comparisons were blocked by physical seed, response labels were permuted within fixed fields, and confidence intervals across fields used field-cluster bootstrap resampling. Unless stated otherwise, reported confidence intervals are percentile cluster bootstraps with 5{,}000 resamples, drawing all of a unit's values together.

Unless stated otherwise, confidence intervals describe variation across the sampled seeds, fields or runs. They do not include uncertainty from future provider updates, untested prompts or other model families. Confirmatory collective tests use exact sign tests or exhaustive within-seed permutations; their decisions do not depend on bootstrap or normal-approximation intervals.

We versioned and hash-locked protocols, encodings, prompts, model IDs, physical seeds, outcome measures and stopping rules before the corresponding acquisitions. Throughout, \emph{prespecified} means fixed in these internal records, not registered publicly. Deviations, including the Claude follow-up's parser and retry amendments, are disclosed below and in the Supplementary Information. We retained raw responses, parser decisions and retry histories. No failed response was silently assigned a valid action. The repository provides the hash manifest and acquisition records; the Supplementary Information documents their timing and the replay-field selection rule.

\section*{Acknowledgments}

\noindent\textbf{Funding:} This research received no specific grant from any funding agency in the public, commercial or not-for-profit sectors.

\noindent\textbf{Author contributions:}\\
Conceptualization: TE\\
Methodology: TE\\
Software: TE\\
Validation: TE\\
Formal analysis: TE\\
Investigation: TE\\
Data curation: TE\\
Visualization: TE\\
Supervision: NI, KN\\
Writing---original draft: TE\\
Writing---review and editing: TE, NI, KN

\noindent\textbf{Competing interests:} The authors declare no competing interests.

\noindent\textbf{Data and materials availability:} Physical seeds, fixed field hashes, serialized prompts, raw model outputs, parsed actions, trajectories, analysis tables and figure-generation scripts are publicly available in version \texttt{v1.2.0} of the code repository, \url{https://github.com/tkEzaki/observation-maps-llm-agents}, archived on Zenodo at \url{https://doi.org/10.5281/zenodo.22899385}~\cite{ezaki2026data}. This release includes all 15 original acquisition groups and the additional September Claude sweep, including raw attempt journals, accepted traces, protocol and analysis tables under \nolinkurl{runs/claude_kappa_v2/anthropic/}; its preflight journal is retained separately. It also includes the revised figure scripts and the audit of 507,112 valid responses. Access to proprietary provider services is not required to reproduce the deterministic engine and statistical analyses, but exact re-acquisition may depend on continued availability of the specific provider model IDs and backend behaviour.

\noindent\textbf{Code availability:} Code for the engine, encoders, acquisition runners, statistical analyses and figure generation is available at \url{https://github.com/tkEzaki/observation-maps-llm-agents}, together with the hash manifest, the fixed model identifiers and generation parameters, and the per-call records of the response contract.

\noindent\textbf{Use of generative artificial intelligence:} Claude (Opus / Fable 5.0 and 5.1) and OpenAI Codex were used to assist with language editing of the manuscript, and with drafting and refactoring parts of the simulation, analysis and figure-generation code. All scientific claims, numerical values, statistical procedures, analyses and final text were reviewed and verified by the authors, who take full responsibility for them. Within the research pipeline itself, language models appear only as the experimental subject: they generate the agent actions that constitute the data. No language-model output was used as an automated rule to select, filter, impute or statistically adjudicate results; all scientific interpretations and decisions were made and verified by the authors.

\subsection*{Supplementary Materials}
Supplementary figures, tables and methods follow the main text in this PDF.

Supplementary Text and Methods\\
Figs. S1 to S27\\
Tables S1 and S2

\clearpage
\newgeometry{left=1in,right=1in,top=0.5in,bottom=0.5in}
\begin{figure}[p]
\centering
\draftfigure{\FigOnePath}{GPT matched collective observation-map intervention.}
\caption{\textbf{The description that promotes synchronization differs between GPT and Claude.}
\textbf{(A)} Experimental loop. A focal agent reads the relative positions of the other agents, represented by the same 24-bin histogram and encoded as moments, centers or intervals. The example uses a recorded initial state ($N=17$, $K=0.15$, seed 0); two payload lines are shown per encoding. The model chooses an action and a shared engine updates the population. Within each model, initial conditions, natural increments and update rules match across encodings. \textbf{(B, C)} The same seed at four times under each encoding in GPT and Claude, respectively, at $K=0.15$. Numbers inside the circles give $r_1$. \textbf{(D, E)} Alignment over time for six seeds (thin lines) and their mean (thick lines); the dotted line marks $r_1=0.9$. \textbf{(F)} Final alignment at both positive couplings (M, C, I: moments, centers, intervals). Grey lines connect matched seeds; counts give runs with $r_1(T)\geq0.9$. At $K=0.15$, moments reaches this threshold in $6/6$ GPT runs and $0/6$ Claude runs; each histogram shows the reverse. \textbf{(G)} Final $Q_2=r_2-r_1$ at $K=0.15$; positive values indicate two-group structure, as seen in GPT intervals. These are finite-time outcomes, not stability classifications. Supplementary Fig.~S2 gives complete prompts, S4 the GPT paired and persistence analyses, and S1 the zero-coupling control. Fig.~4 and Supplementary Figs.~S15--S17 report Claude's additional seeds and inference.}
\label{fig:gptmacro}
\end{figure}

\begin{figure}[p]
\centering
\draftfigure{\FigTwoPath}{Controlled microscopic response operators.}
\caption{\textbf{Controlled inputs reveal differences in the agents' response rules.}
\textbf{(A)} A single-peaked (von Mises) peer distribution is rotated by $\delta$ around the focal agent (open circle) and shown in each encoding. Repeated queries give the mean action $g(\delta)$. \textbf{(B)} Action probabilities for one field ($\kappa=12$, $\delta=-0.25$; peers just behind the agent), pooled across two blocks (M, C, I: moments, centers, intervals). \textbf{(C, D)} GPT and Claude response curves at $\kappa=12$. Points show means at the 36 offsets in each block. Curves are unconstrained order-6 Fourier fits, clipped to $[-1,1]$ only for display. Triangles mark the block-1 fitted switch from retard to advance nearest zero. \textbf{(E, F)} First-component coefficients $a_1$ (sine) and $b_1$ (cosine) for both blocks, with marker size increasing from $\kappa=2$ to 12. Positive $a_1$ describes a component directed toward the peers; $b_1$ changes its angular shift. GPT centers crosses to negative $a_1$, whereas all Claude encodings retain $a_1\geq0.94$. \textbf{(G)} Second-harmonic amplitude $R_2$; lines average the two blocks and bands span them, not confidence intervals. \textbf{(H, I)} GPT activity and $a_1$ for two opposite groups with signed imbalance $\varepsilon$. Moments mostly stays at balance but acts at $|\varepsilon|=0.02$. \textbf{(J)} Pairwise total-variation distances between encodings across six GPT field families. These conditional responses do not establish collective stability. Supplementary Figs.~S6--S10 give GPT uncertainty and extensions; S26,S27 give details of the later, descriptive Claude follow-up.}
\label{fig:micro}
\end{figure}

\begin{figure}[p]
\centering
\draftfigure{\FigThreePath}{Identical endogenous-field cross-encoding replay.}
\caption{\textbf{Describing the same fixed state differently changes GPT's actions.}
\textbf{(A)} Forty-eight fields were selected before replay acquisition, balanced across the encoding that generated them (source) and six trajectory-state categories. Every field was shown in all three presented encodings, with 32 responses per field and encoding across two blocks. \textbf{(B, C)} Activity and mean action, respectively. Rows identify the source encoding; columns identify the presented encoding (M, C, I: moments, centers, intervals). \textbf{(D)} Pairwise total-variation distances with 95\% intervals from resampling whole fields. \textbf{(E)} Distances between encodings compared with repeat-measurement variation (block noise) on the same fields. \textbf{(F)} Within-field permutation test: mean pairwise distance $\TV=0.344$, $p=0.0002$. \textbf{(G)} The three prespecified tests support a presented-encoding effect. Neither the source effect ($p=0.078$) nor its interaction with presented encoding ($p=0.072$) crossed the threshold; this does not establish their absence. The 48 fields, not the 4,608 calls, are the units for field-level inference. Supplementary Figs.~S11--S13 give field selection, individual response distributions and the additional tests.}
\label{fig:replay}
\end{figure}

\begin{figure}[p]
\centering
\draftfigure{\FigFourPath}{Claude macro replication and three-family microscopic replication.}
\caption{\textbf{Encoding affects Claude's collective behavior, but in the opposite direction to GPT.}
\textbf{(A)} Design: six seed indices matched to GPT within each coupling ($K=0,0.08,0.15$), plus four additional seeds at positive coupling. \textbf{(B)} Claude alignment trajectories; at $K=0.15$, histograms lock while moments remains partially ordered. \textbf{(C)} Outcomes for confirmatory and additional seeds, shown separately. \textbf{(D)} Confirmatory tests. Mean terminal-lock scores are $0.00,0.67,0.92$ and mean final-order scores are $0.598,0.921,0.973$ for moments, centers and intervals. Exact within-seed permutation tests give $p=0.00103$ and $p=0.000386$, respectively. \textbf{(E)} Matched GPT--Claude comparison for the six core seed indices: the effect of encoding replicates, but the locking pattern reverses. \textbf{(F)} Fixed-field response differences in GPT, Claude and Gemini. Collective dynamics were not tested in Gemini. Supplementary Fig.~S14 gives the three-model replay comparison; S15--S17 give all Claude trajectories, exact tests, replication criteria and seed-level comparisons.}
\label{fig:family}
\end{figure}

\begin{figure}[p]
\centering
\draftfigure{\FigFivePath}{Same-task-information presentation control.}
\caption{\textbf{Presentation changes GPT's responses even when the numerical information is fixed.}
\textbf{(A)} Identical moment values shown as the original text, a table or text with added irrelevant context. \textbf{(B)} Pairwise total-variation distances: means are $0.145$ for original--table, $0.414$ for original--added context and $0.374$ for table--added context. \textbf{(C)} Comparison with block noise ($0.069$) and the GPT cross-encoding mean ($0.344$). \textbf{(D)} Within-field permutation test: mean pairwise $\TV=0.311$, $p=0.0002$. \textbf{(E)} For each contrast, the example field closest to its median distance. The printed value is the median across all 48 fields, not the selected field's distance. \textbf{(F)} The observation map includes both retained information and presentation. Added text changes context, number position and length together, so the effect cannot be attributed to length alone. Supplementary Fig.~S18 gives the full control; S19,S20 examine length and the explicit pairing of values with coordinates.}
\label{fig:omap}
\end{figure}

\clearpage
\restoregeometry
\begin{table}[htbp]
\centering
\caption{\textbf{Units of inference by analysis.} For each experiment, the
question asked, the unit that is permuted or resampled, what the repeated model
calls contribute and the comparison that carries the result.}
\label{tab:units}
\small
\setstretch{1.1}
\begin{tabular}{@{}>{\raggedright\arraybackslash}p{0.16\textwidth}>{\raggedright\arraybackslash}p{0.20\textwidth}>{\raggedright\arraybackslash}p{0.15\textwidth}>{\raggedright\arraybackslash}p{0.17\textwidth}>{\raggedright\arraybackslash}p{0.20\textwidth}@{}}
\toprule
Analysis & Question & Unit of inference & Role of repeated calls & Main comparison \\
\midrule
GPT collective
  & Does the encoding change the trajectory?
  & physical seed ($n=6$)
  & generate the trajectory
  & paired within seed \\
\addlinespace
Controlled fields (GPT; Claude follow-up)
  & Does the response rule on a fixed field change?
  & field and acquisition block
  & estimate action probabilities
  & response curve and its harmonics; Claude follow-up descriptive \\
\addlinespace
Identical-field replay
  & Does the operator change on the same physical field?
  & physical field ($n=48$)
  & estimate the trinomial per field
  & response labels permuted within field \\
\addlinespace
Claude collective
  & Is there a macroscopic effect in Claude?
    & confirmatory seed-index block ($n=6$)
  & generate the trajectory
  & all label permutations within seed \\
\addlinespace
Same-information control
  & Does presentation matter at fixed numbers?
  & physical field ($n=48$)
  & estimate probabilities per variant
  & variant labels swapped within field \\
\addlinespace
Surrogate
  & Can responses be predicted in and out of distribution?
  & stimulus profile or complete run
  & provide training labels
  & out-of-fold and prospective \\
\bottomrule
\end{tabular}
\end{table}

\clearpage
\appendix
\newgeometry{left=12mm,right=12mm,top=14mm,bottom=16mm}
\setstretch{1}
\makeatletter
\renewcommand\normalsize{\@setfontsize\normalsize{10}{12}}
\makeatother
\normalsize
\setcounter{secnumdepth}{0}
\setcounter{tocdepth}{1}
\setcounter{figure}{0}
\setcounter{table}{0}
\setcounter{equation}{0}
\renewcommand{\thefigure}{S\arabic{figure}}
\renewcommand{\thetable}{S\arabic{table}}
\renewcommand{\theequation}{S\arabic{equation}}
\renewcommand{\theHfigure}{S\arabic{figure}}
\renewcommand{\theHtable}{S\arabic{table}}
\renewcommand{\theHequation}{S\arabic{equation}}
\renewcommand{\figurename}{Supplementary Figure}
\renewcommand{\tablename}{Supplementary Table}
\captionsetup{font=footnotesize,labelfont=bf,labelsep=period,skip=4pt}
\captionsetup[figure]{font=footnotesize,labelfont=bf,labelsep=period,skip=4pt}

\part*{Supplementary Information}

{\small\tableofcontents}

\section*{Claim-to-evidence map}
\addcontentsline{toc}{section}{Claim-to-evidence map}

Changing the state description changed collective alignment, with opposite effects in GPT and Claude. The evidence comes from three comparisons. Moments versus histograms changes both retained information and presentation. Centers versus intervals keeps the same bin masses but changes their labels. The GPT presentation control keeps the numerical moments fixed and changes their layout or surrounding text. The last comparison measures responses to fixed inputs, not collective dynamics. Controlled-field curves help identify possible explanations for the collective results but do not yet predict stability. The table below links each conclusion to its supporting figures.

\begin{table}[H]
\centering
\small
\begin{tabular}{@{}p{0.44\textwidth}ll@{}}
\toprule
Main claim & Main figure & Supplementary Figures \\
\midrule
Shared-engine validation ($K=0$ negative control)
  & Supplementary Information & S1, S2 \\
Encoding changes GPT collective order
  & Fig.~1 & S3--S5 \\
Each description gives the agent a different interaction rule
  & Fig.~2 & S6--S10, S26,S27 \\
The same state, re-described, elicits different actions
  & Fig.~3 & S11--S13 \\
Microscopic replication in GPT, Claude and Gemini
  & Fig.~4 & S14 \\
Collective replication and reversal in Claude
  & Figs.~1,4 & S15--S17 \\
Presentation changes the rule at fixed numbers
  & Fig.~5 & S18 \\
Serialization length versus retained information
  & Supplementary Information only & S19 \\
Which serialization feature is involved
  & Supplementary Information only & S20 \\
Surrogate compressibility and transportability
  & Supplementary Information only & S21--S25 \\
\bottomrule
\end{tabular}
\end{table}

\section{Analysis guide: what each statistical operation does}

The following guide explains the statistical terms used in the main text and this supplement. A \emph{field} is one distribution of peers' relative positions; an \emph{acquisition block} is a separate batch of queries. Repeated queries estimate the probabilities of the three actions. They are not independent repetitions of a collective experiment.

\begin{table}[H]
\centering
\small
\begin{tabular}{@{}p{0.26\textwidth}p{0.68\textwidth}@{}}
\toprule
Term & What it does \\
\midrule
Permutation test
  & Swaps the labels between conditions many times and asks how often chance
    alone produces a difference as large as the observed one. \\
\addlinespace
Exact sign test
  & Counts only the direction of the paired differences, not their size. \\
\addlinespace
Bootstrap confidence interval
  & Resamples the independent units specified for the analysis and uses the resulting estimates to construct an uncertainty interval. \\
\addlinespace
Total-variation distance
  & The difference between two advance/stay/retard probability distributions,
    on a scale from 0 (identical) to 1 (no overlap). \\
\addlinespace
Block variation (``block noise'')
  & The test-retest difference obtained when the same condition is measured
    again in a separate acquisition batch. \\
\addlinespace
Omnibus test
  & Asks whether any of several groups differ, without identifying which pair
    does. \\
\addlinespace
Permutation $p$ floor
  & With 5,000 resamples the observed statistic is counted among them
    (plus-one), so the smallest reportable $p$ is $1/5001\approx0.0002$. A $p$
    at the floor means ``none of 5,000 relabellings was as extreme'', not a
    measured zero. \\
\addlinespace
Out-of-fold prediction
  & Predicts a held-out fold using a model fitted without it. What is placed in
    the held-out fold decides what is being tested, so the holdout scheme is
    named wherever an out-of-fold number is reported. \\
\addlinespace
Support
  & Whether the training set contains fields similar to the test field, and
    whether those similar fields elicited consistent actions. \\
\bottomrule
\end{tabular}
\end{table}

\subsection*{Four levels of surrogate evaluation}
A surrogate is a fitted model that predicts action probabilities without calling the language model. We evaluate it at four levels (Supplementary Figs.~S21--S25). Accurate predictions on a test set do not by themselves show that the surrogate can replace an agent in a running population: its own actions will change the inputs it encounters.
\begin{enumerate}\setlength{\itemsep}{1pt}
\item \emph{Cross-block stability.} Does the response law measured on a field in
one acquisition block reproduce on a second, independent block of the
\emph{same} field? This is the \texttt{acquisition\_block\_holdout} scheme, and
it is the criterion by which the production model class was selected. It is a
reproducibility check: the same field appears in training and in test.
\item \emph{In-domain field generalization}, referred to as
\emph{compressibility}. Can the surrogate predict controlled fields that were
excluded from training altogether? This is what the four field-preserving
schemes measure (leave-one-profile-out, offset-group, sparse-realization and
source-family holdout).
\item \emph{Closed-loop support.} Do the fields an interacting population
generates for itself fall inside the region the training set covers, and do
their near neighbours in that set agree with one another about what the model
does?
\item \emph{Prospective closed-loop accuracy}, referred to as
\emph{transportability}. Does the surrogate predict new language-model
collective runs it played no part in producing?
\end{enumerate}

\noindent
We use \emph{transportability} only for level~4, and ``grouped out-of-fold, by
field'' refers to level~2 only. Failure at an earlier level stopped a branch
before the later, more expensive level was attempted: only moments reached
level~4.

\subsection*{Names used in the released code}
In legacy artifact names and analysis tables, \emph{representation} denotes
the observation map of the main text and \emph{target} denotes the presented
encoding; source encoding is \emph{source} in both.

\subsection*{Figure numbers and released file names}
The supplementary figures were reordered into narrative order after the
analysis was frozen, and the released scripts and PDFs keep their original
numbers. The complete mapping between manuscript figure numbers and released
file names is provided in the repository README.

\subsection*{Multiplicity across the three replay tests}
The three global replay tests (presented encoding, source, source $\times$
presented encoding) are reported without multiplicity correction, each carrying
its own prespecified $\alpha=0.05$, because they answer separate prespecified
questions rather than dividing one. The reported decisions do not depend on
that choice: a Bonferroni correction across the three would leave the
presented-encoding effect supported ($p=0.0002$ against $\alpha/3=0.0167$) and
would leave the source and interaction tests non-significant either way.

\noindent
The unit that is shuffled or resampled differs between experiments and is stated with each analysis: permutations are carried out within a field or within a seed, and fields and seeds are never mixed with one another.

\noindent
Throughout the captions, panels marked descriptive or secondary are post hoc
and not used for inference, and all conclusions are limited to the tested maps,
descriptors and acquisition domains.

\clearpage

\phantomsection\addcontentsline{toc}{section}{Supplementary Figure S1: Deterministic engine validation and causal isolation}

\begin{figure}[H]
\SIpage{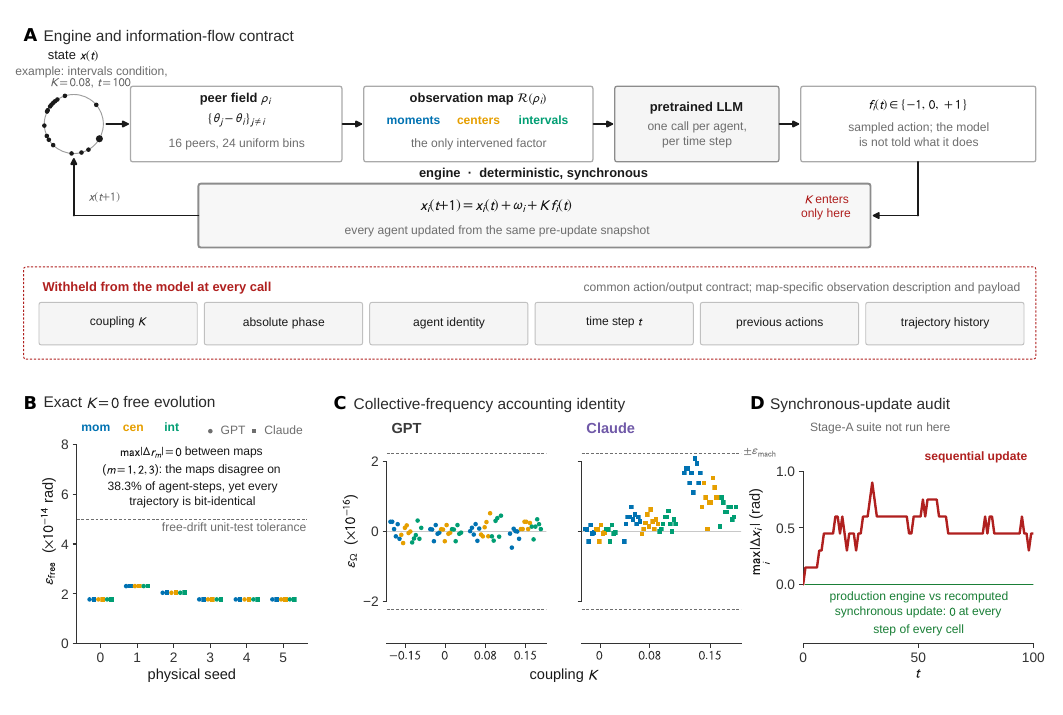}
\caption{\textbf{Deterministic engine validation and causal isolation.}
\textbf{(A)} Shared update rule and information available to the model. The three maps share one
action/output contract and differ only in the observation description and
payload. The relative-phase multiset $\{\theta_j-\theta_i\}_{j\neq i}$ is
reduced to the field $\rho_i$, encoded by $\R$ and passed to the language
model, which returns one action $f_i\in\{-1,0,+1\}$. All agents update
synchronously and the engine alone applies $Kf_i$ in
$x_i(t+1)=x_i(t)+\omega_i+Kf_i(t)$. Information never given to the model is
marked: $K$, absolute phase, agent identity, time step, previous actions and
history. The circle shows a final configuration (GPT intervals, $K=0.08$,
$t=100$), the enlarged dot marking focal agent $i=0$.
\textbf{(B)} Exact $K=0$ free evolution. The maximum residual
$\epsilon_{\mathrm{free}}=\max_{i,t}\lvert x_i(t)-x_i(0)-t\omega_i\rvert$ over
all 36 $K=0$ cells (2 families $\times$ 3 maps $\times$ 6 seeds), one point per
physical seed and observation map. The coupling term vanishes, so the sampled actions cannot change the trajectory. The maximum residual, $2.31\times10^{-14}$ rad, is due to floating-point accumulation.
\textbf{(C)} Collective-frequency accounting identity. The residual
$\epsilon_{\Omega}=\Omega_{\mathrm{coll}}-(\bar\omega+K\tau_{\mathrm{social}})$
across all GPT and Claude collective runs, coloured by encoding;
$\max\lvert\epsilon_{\Omega}\rvert=2.08\times10^{-16}$.
\textbf{(D)} Synchronous-update audit. A deterministic toy trajectory at
$K=0.15$ under a peer-mass policy is run with the correct synchronous rule and
with an incorrect sequential rule, and the production engine is checked
against an independently recomputed synchronous update over all 150 cells.
These checks validate the implementation, rather than the model responses; encoder rotation and peer-permutation invariance were
verified at the serialization level only.}
\end{figure}
\clearpage

\phantomsection\addcontentsline{toc}{section}{Supplementary Figure S2: Complete observation maps, prompt contracts and serialization metadata}

\begin{figure}[H]
\captionsetup{labelformat=sipaged}\renewcommand{\pageofn}{ (page 1 of 2)}
\SIpage{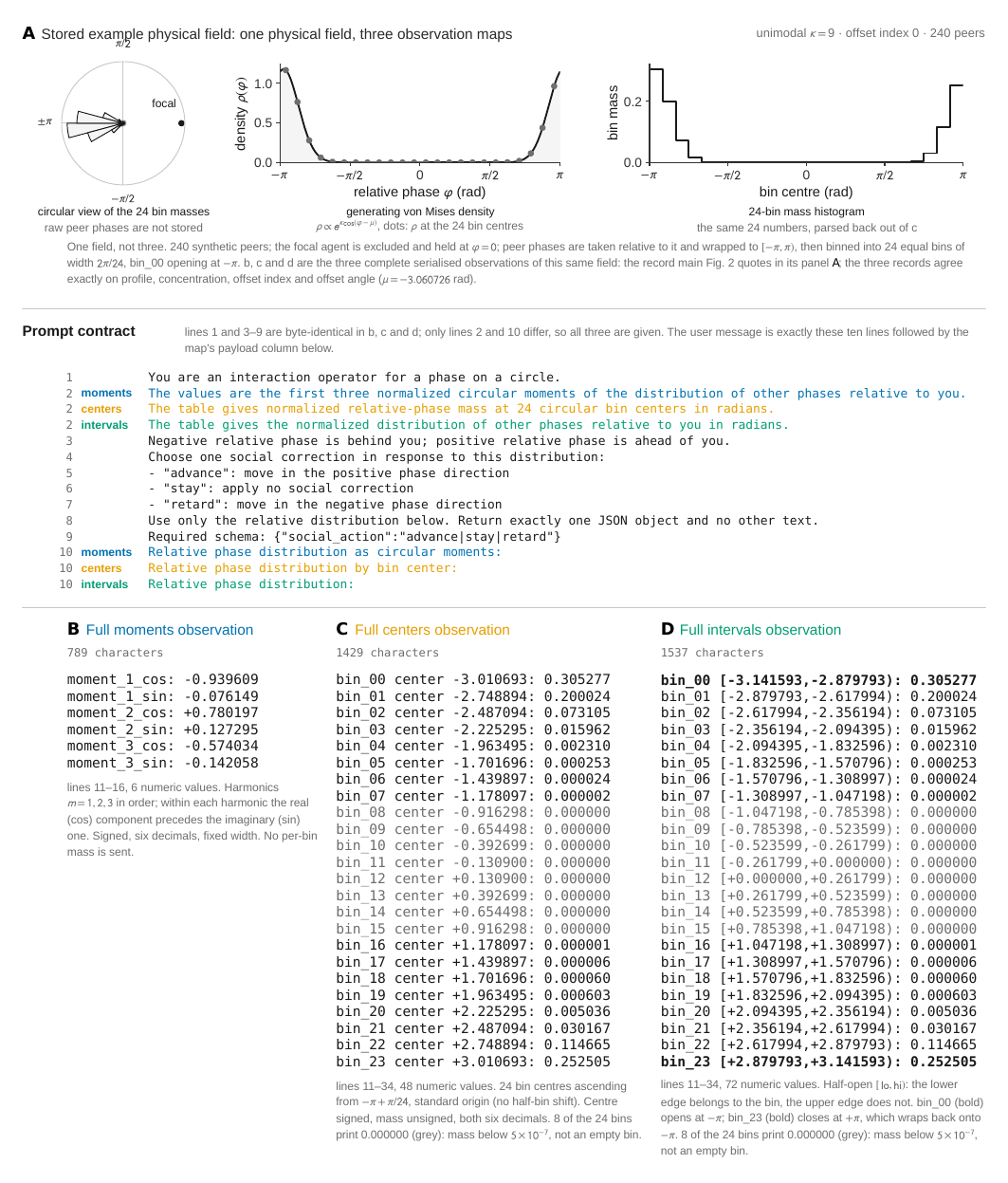}
\caption{\textbf{Complete observation maps, prompt contracts and serialization
metadata (the three complete payloads of one fixed field).}
\textbf{(A)} The example field (record: profile \texttt{unimodal\_k9}, offset index 0, 240 peers, focal agent
excluded and held at $\varphi=0$, wrapping to $[-\pi,\pi)$ with 24 bins of
width $2\pi/24$), shown in three ways: the serialized 24-bin masses
on the circle, the generating von Mises density, and the 24-bin mass histogram
parsed back out of the centers payload. Raw peer phases were not stored and
none are synthesised; the moments agent received only the fixed moment
features shown in panel~B.
\textbf{(B--D)} The complete user messages generated from this one field under the moments (\textbf{B}), centers (\textbf{C}) and
intervals (\textbf{D}) observation maps. Each payload is computed deterministically from the 24-bin histogram in \textbf{(A)} so the text shown is what the model
received; boundary conventions, bin origin, ordering and the fixed-width
six-decimal formatting are annotated on the panels. Concatenating the prompt
contract, using each map's own lines 2 and 10, with that map's payload column
below reproduces the user message.}
\end{figure}
\clearpage

\begin{figure}[H]\ContinuedFloat
\captionsetup{labelformat=sipaged}\renewcommand{\pageofn}{ (page 2 of 2)}
\SIpage{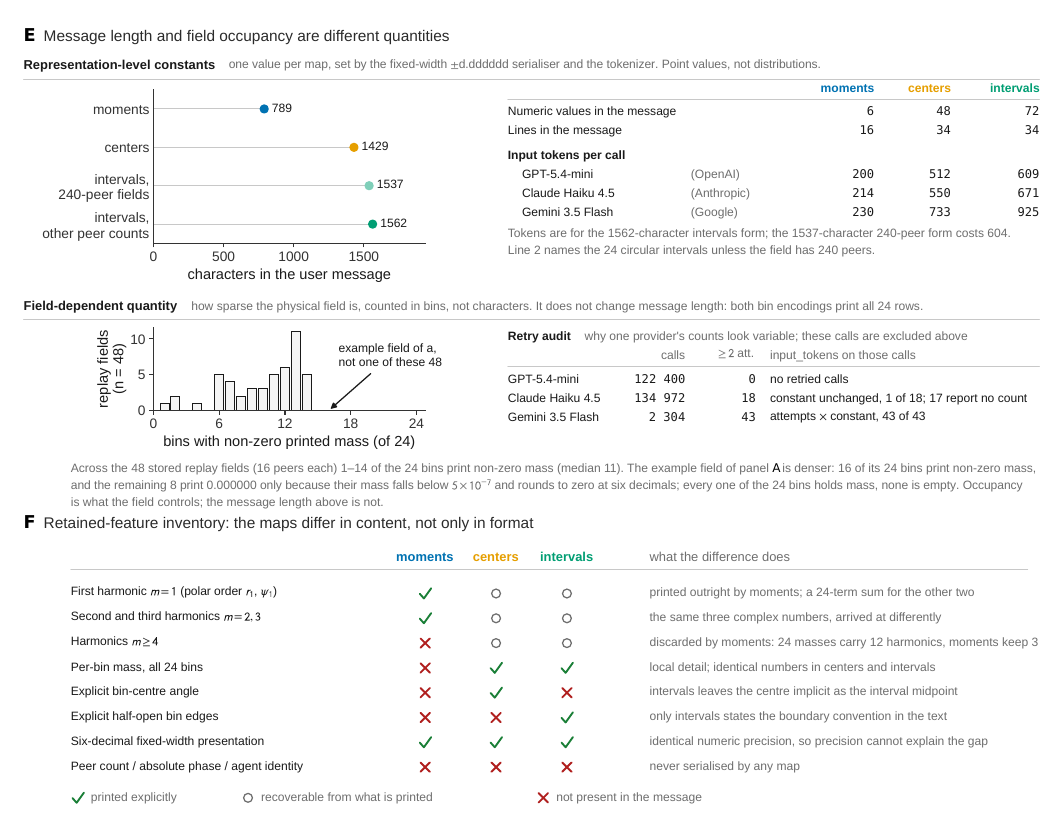}
\caption{\textbf{(continued: serialization metadata).}
\textbf{(E)} Character and token counts. Because the serializers use
fixed-width numeric formatting, character and token counts are
encoding-level \emph{constants}, not distributions, and are drawn as
point values (Supplementary Table~S1). The only field-dependent quantity is
the number of occupied bins (1--14 of 24, median 11), a sparsity descriptor of
the field shown on its own axis and explicitly not a proxy for prompt length. Retried calls are excluded from the token display.
\textbf{(F)} Retained-feature inventory: which physical information components
(low-order polar information, higher-harmonic information, 24-bin local
detail, explicit bin centers, explicit bin boundaries, six-decimal
presentation) each map retains. The maps differ in retained information as well as presentation.
}
\end{figure}
\clearpage

\phantomsection\addcontentsline{toc}{section}{Supplementary Figure S3: Complete GPT matched-collective trajectories}

\begin{figure}[H]
\captionsetup{labelformat=sipaged}\renewcommand{\pageofn}{ (page 1 of 3)}
\SIpage{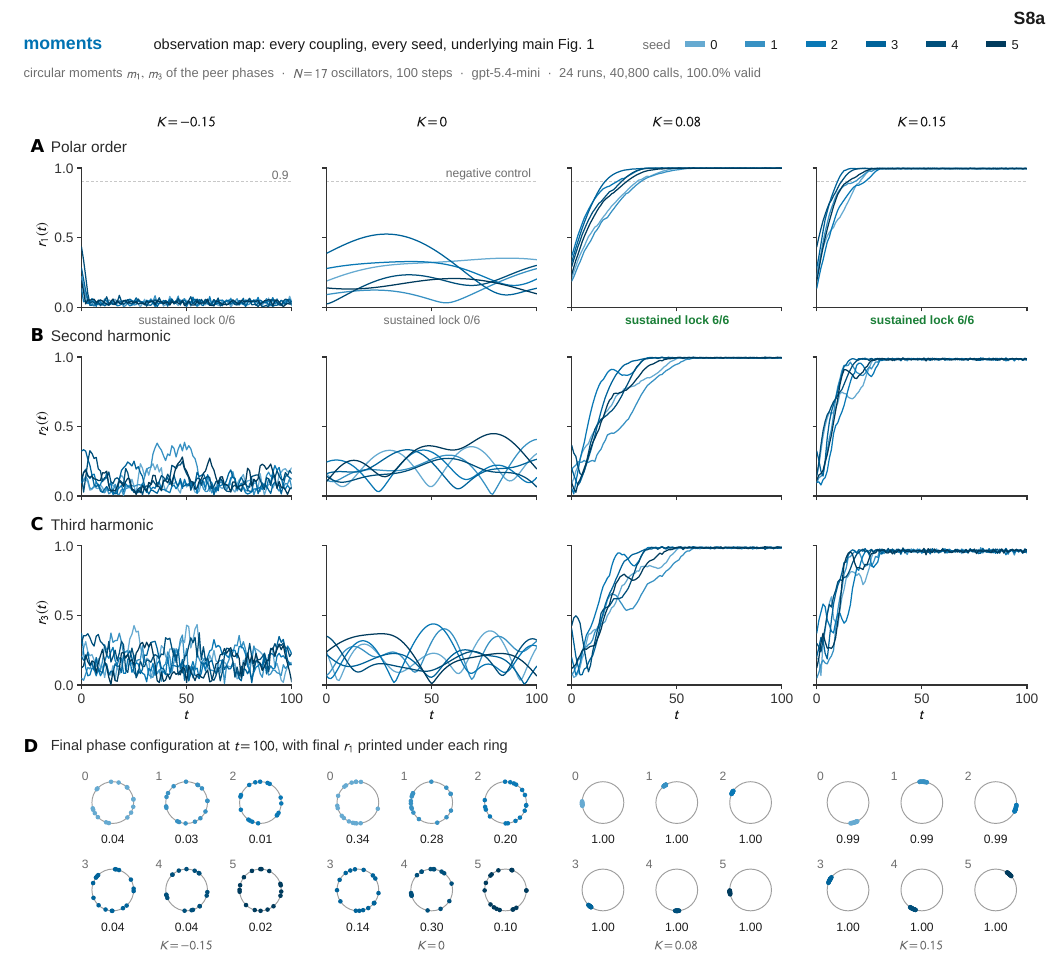}
\caption{\textbf{Complete GPT matched-collective trajectories (one page per
observation map; this page: moments).}
Complete trajectory sets underlying main Fig.~1, from the matched GPT
experiment (\texttt{gpt-5.4-mini}, $N=17$, $T=100$,
$K\in\{-0.15,0,0.08,0.15\}$, six paired physical seeds, 122{,}400 valid
calls). Each page shows, for one observation map, a $3\times4$ grid of
(harmonic, coupling) cells with all six seed trajectories of $r_m(t)$ for
$m=1,2,3$ (nothing is averaged across seeds, so run-to-run timing
variation remains visible), plus the 24 final phase configurations at
$t=100$. The $y$ range of each harmonic is identical on all three pages.
The dashed rule in \textbf{A} is the prespecified polar-locking threshold
$r_1=0.9$. At $K=0$ the order-parameter trajectories of the
three maps coincide exactly (Supplementary Fig.~S1B); at positive $K$ the
moments condition shows terminal lock in 6/6 seeds (the endpoint criterion
$r_1(T)\geq0.9$ gives the same counts).}
\end{figure}
\clearpage

\begin{figure}[H]\ContinuedFloat
\captionsetup{labelformat=sipaged}\renewcommand{\pageofn}{ (page 2 of 3)}
\SIpage{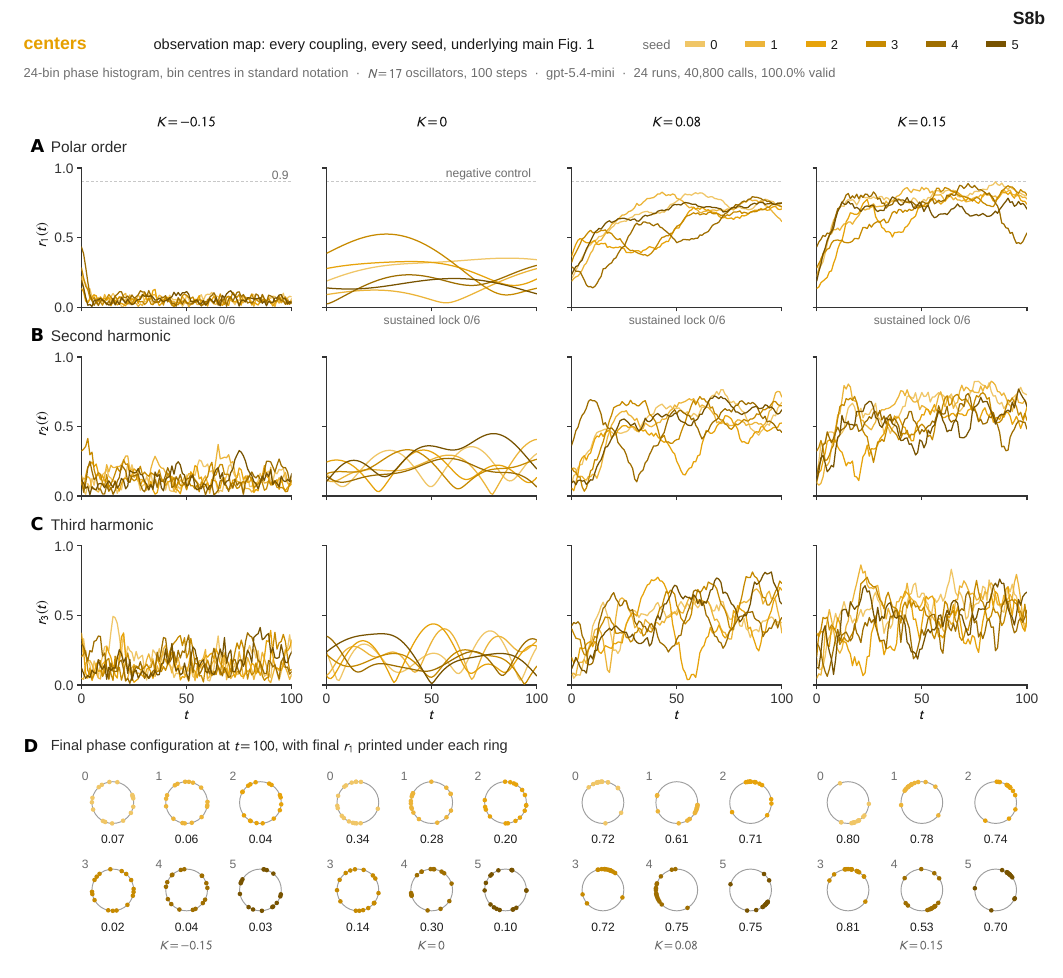}
\caption{\textbf{(continued: centers).} As on page 1 of Supplementary Fig.~S3, for the
centers observation map, which locks in 0/6 seeds at each positive coupling
and predominantly produces partial polar order.}
\end{figure}
\clearpage

\begin{figure}[H]\ContinuedFloat
\captionsetup{labelformat=sipaged}\renewcommand{\pageofn}{ (page 3 of 3)}
\SIpage{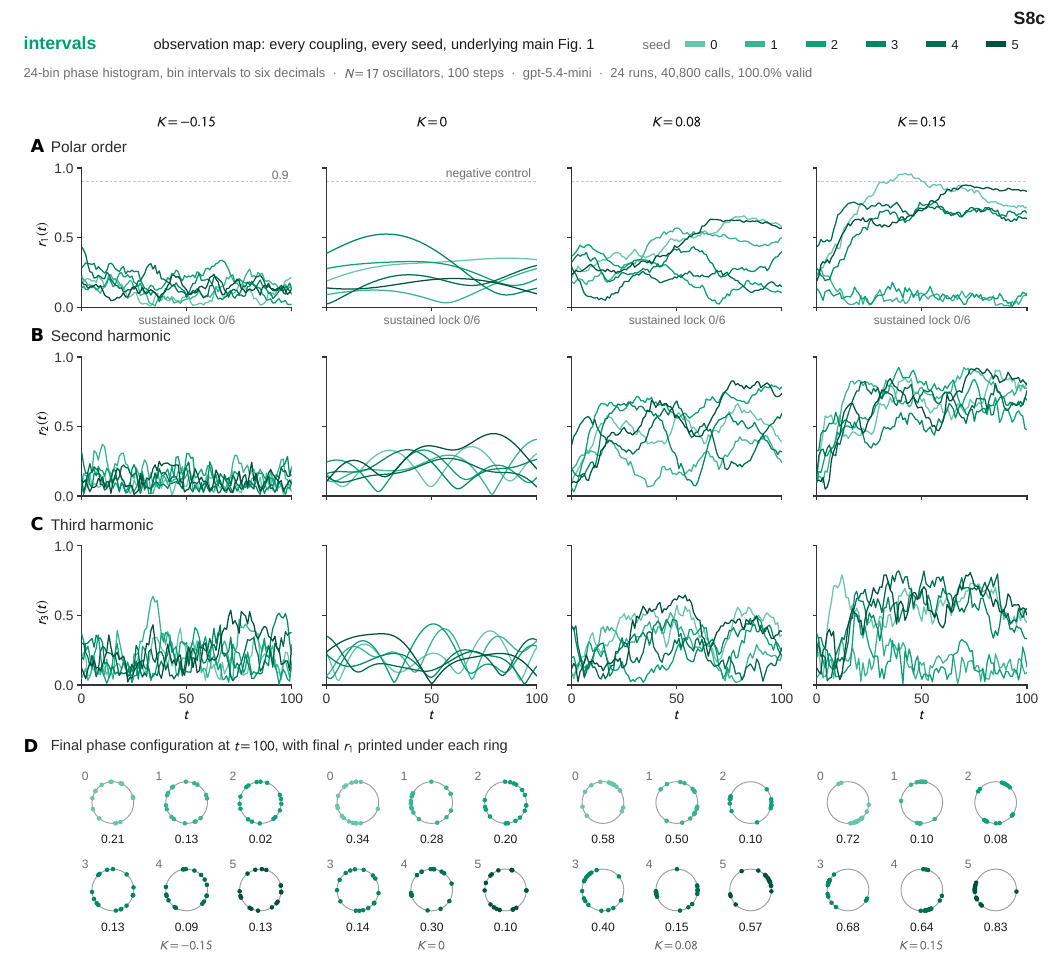}
\caption{\textbf{(continued: intervals).} As on page 1 of Supplementary Fig.~S3, for
the intervals observation map, which locks in 0/6 seeds at each positive
coupling and more often retains second-harmonic order relative to polar order
($Q_2=r_2-r_1$).}
\end{figure}
\clearpage

\phantomsection\addcontentsline{toc}{section}{Supplementary Figure S4: Robustness of the GPT collective phenotype classification}

\begin{figure}[H]
\SIpage{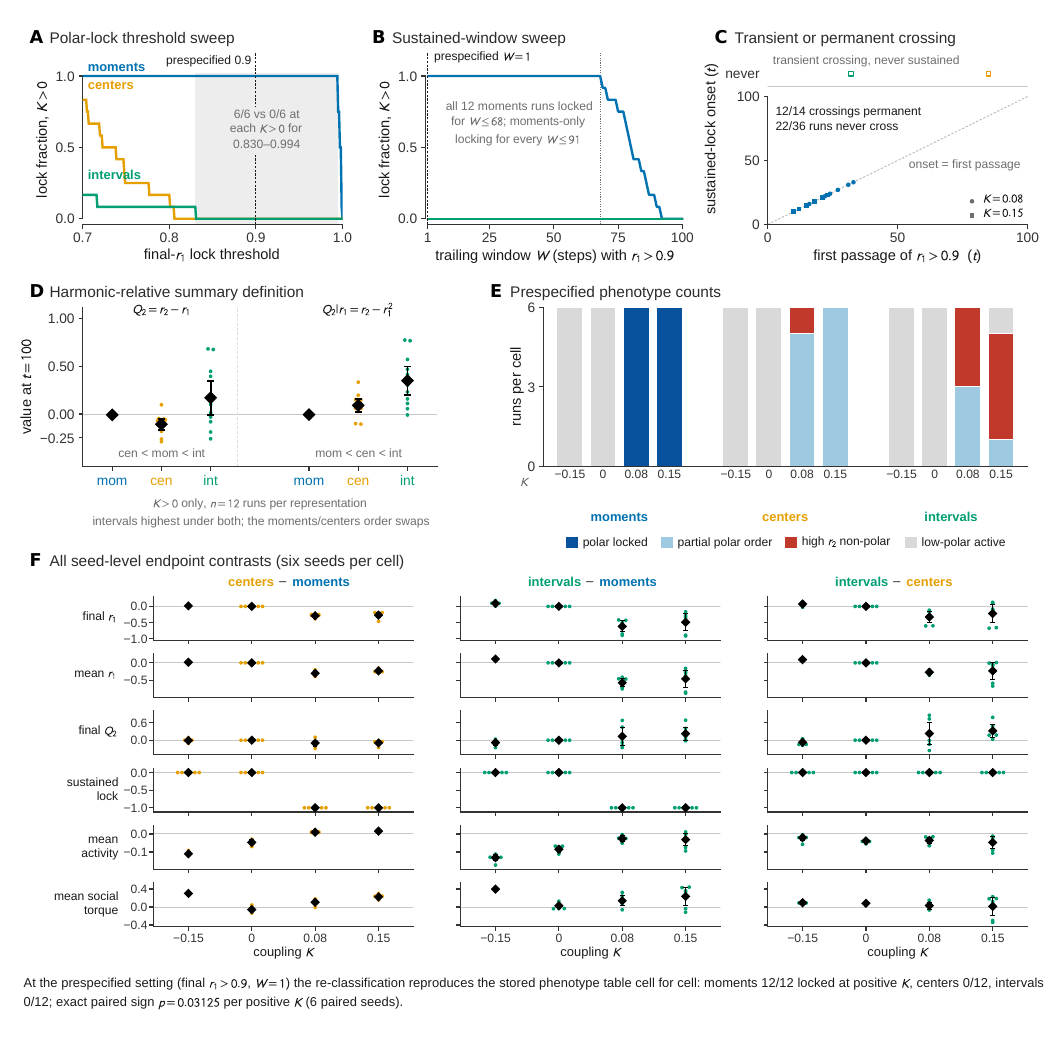}
\caption{\textbf{Robustness of the GPT collective phenotype classification.}
Sensitivity of the locking result (moments 6/6, centers and intervals 0/6; exact paired sign test $p=2/2^{6}=0.03125$) to the threshold and duration used to define locking. Panels \textbf{A}--\textbf{D}
apply the prespecified analysis functions to classify the same trajectories again.
\textbf{(A)} Lock fraction against the final-$r_1$ threshold, over a band
around the prespecified 0.9.
\textbf{(B)} Lock fraction against a minimum duration: the length of the
trailing window over which $r_1$ must remain above threshold. All 12 moments
runs (six seeds at each positive coupling) remain locked for trailing windows of up to 68 observations, and for every tested window of up to 91 observations locking occurs
only under moments, so the prespecified rule's admission of $t^{\ast}=T$ is not
what produces the $6/6$ versus $0/6$ separation.
\textbf{(C)} First passage versus terminal-lock onset: whether each crossing of 0.9 lasts to the end of the run or is followed by a fall below threshold.
\textbf{(D)} The harmonic-relative summary $Q_2=r_2-r_1$ beside the one
alternative already present in the artifacts, $r_2-r_1^{2}$.
\textbf{(E)} Phenotype counts for every encoding and coupling.
\textbf{(F)} All seed-level endpoint contrasts (final $r_1$, mean $r_1$, final
$Q_2$, terminal lock, activity, torque), one row per endpoint and one column
per encoding contrast.
This analysis checks sensitivity around the prespecified rule; no panel reports an optimised threshold, window or summary
($n=72$ runs).}
\end{figure}
\clearpage

\phantomsection\addcontentsline{toc}{section}{Supplementary Figure S5: Negative coupling and nonlocking dynamical pathways}

\begin{figure}[H]
\SIpage{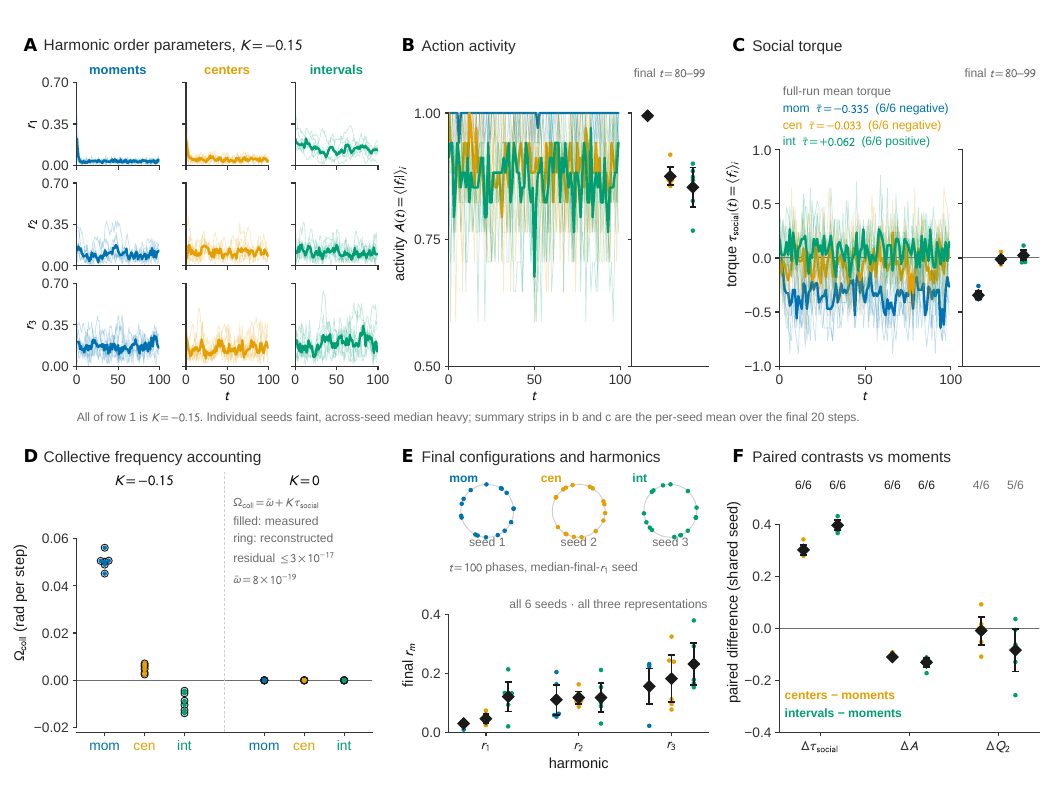}
\caption{\textbf{Negative coupling and nonlocking dynamical pathways.}
The GPT $K=-0.15$ condition.
\textbf{(A)} All six seed trajectories of $r_1,r_2,r_3$, faceted by
encoding: polar order stays low for every map.
\textbf{(B)} Activity trajectories and final-window means: near-zero polar
order coexists with near-unit activity, that is, almost no abstention.
\textbf{(C)} Social torque trajectories and final-window estimates; the
full-run mean torques differ in sign and magnitude across maps ($-0.335$
moments, $-0.033$ centers, $+0.062$ intervals).
\textbf{(D)} Collective-frequency accounting: measured $\Omega_{\mathrm{coll}}$
against $\bar\omega+K\tau_{\mathrm{social}}$; the drift is coupling-mediated
and vanishes at $K=0$.
\textbf{(E)} Mechanically selected final phase configurations (one per
encoding, not the maximum-difference seed) with all final order
parameters.
\textbf{(F)} Shared-seed paired contrasts against moments in torque, activity
and $Q_2$; counts give how many of the six shared seeds agree in sign. Torque
and activity separate the encodings at every seed, $Q_2$ does not.
The agents remain active despite low alignment. Panels \textbf{C} and
\textbf{F} do not test how responses depend on the states generated by an encoding.}
\end{figure}
\clearpage

\phantomsection\addcontentsline{toc}{section}{Supplementary Figure S6: Prespecified offset design and aliasing control for response harmonics}

\begin{figure}[H]
\SIpage{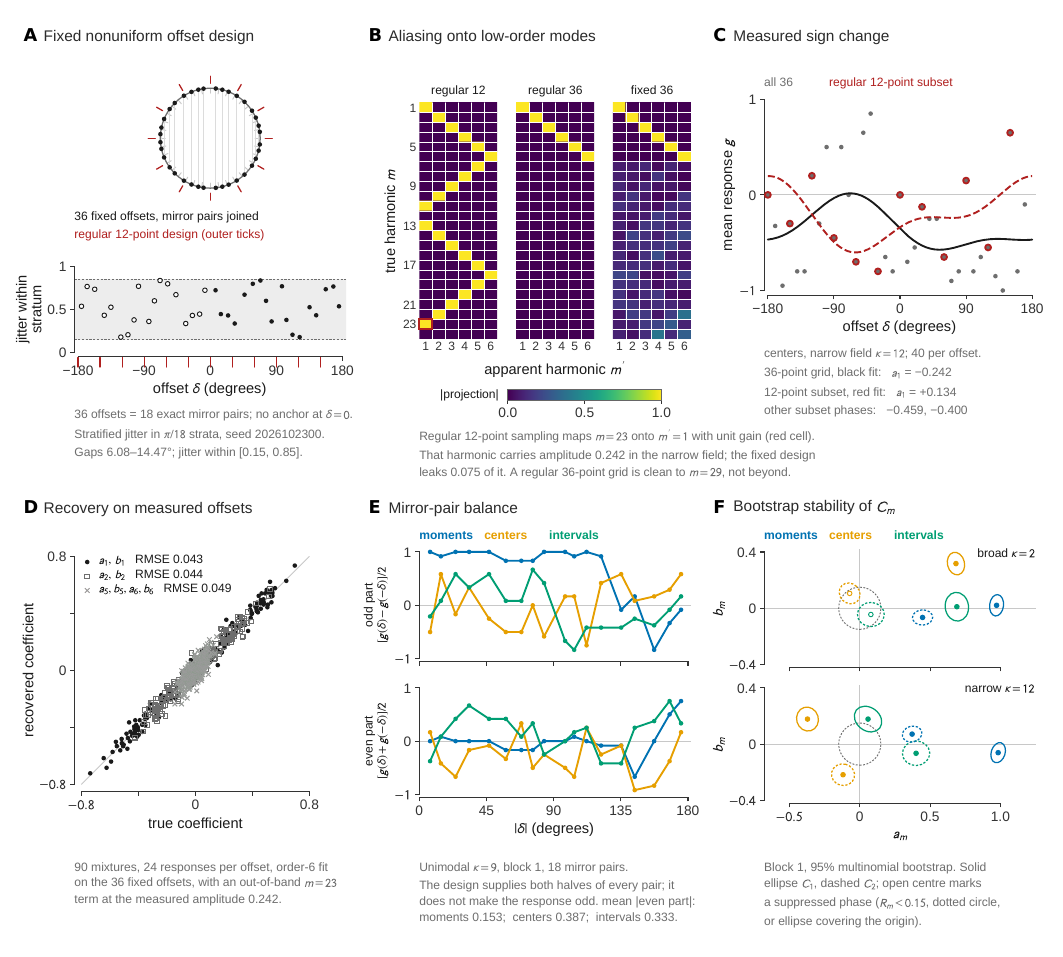}
\caption{\textbf{Prespecified offset design and aliasing control for response
harmonics.}
The harmonic coefficients reported throughout the study are meaningful only if
the sampled angles can separate the harmonics being fitted.
\textbf{(A)} The 36 prespecified nonuniform offsets (mirror-paired, stratified
jitter) shown on the circle and on the linear $[-\pi,\pi)$ axis.
\textbf{(B)} Analytic aliasing map for a regular 12-point design. With evenly
spaced angles, a fast harmonic can masquerade as a slower one when fitted
(aliasing); the map shows how much of each unit-amplitude true harmonic would
leak into the fitted low-order modes, highlighting the
$m=23\rightarrow m=1$ contamination the nonuniform design prevents.
\textbf{(C)} The same leak in practice: one measured response, from a separate
acquisition sampled on a regular 36-point grid, fitted twice, once with all 36
points and once with only its regular 12-point first block; only the 12-point
fit changes the fitted low-order coefficients.
\textbf{(D)} Ground-truth check on simulated data: harmonic mixtures with known
coefficients, including an out-of-band $m=23$ nuisance component, are sampled
on the measured offsets and refitted; recovered coefficients fall on the
identity line.
\textbf{(E)} Mirror-pair balance: sums and differences of responses at
$\pm\delta$, documenting odd/even decomposition stability.
\textbf{(F)} Bootstrap confidence ellipses for the complex kernels
$C_1$ and $C_2$ of representative broad and narrow fields, with the prespecified
phase-reporting gate (phase is reported only when $R_m\geq0.15$ and the
conditional bootstrap ellipse excludes the origin).
Panels \textbf{(A)} \textbf{B} and \textbf{D}--\textbf{F} use the block-1
design; \textbf{C} refits a separate regular 36-point grid acquisition and its
12-point first block.}
\end{figure}
\clearpage

\phantomsection\addcontentsline{toc}{section}{Supplementary Figure S7: Complete encoding-invariance screen}

\begin{figure}[H]
\SIpage{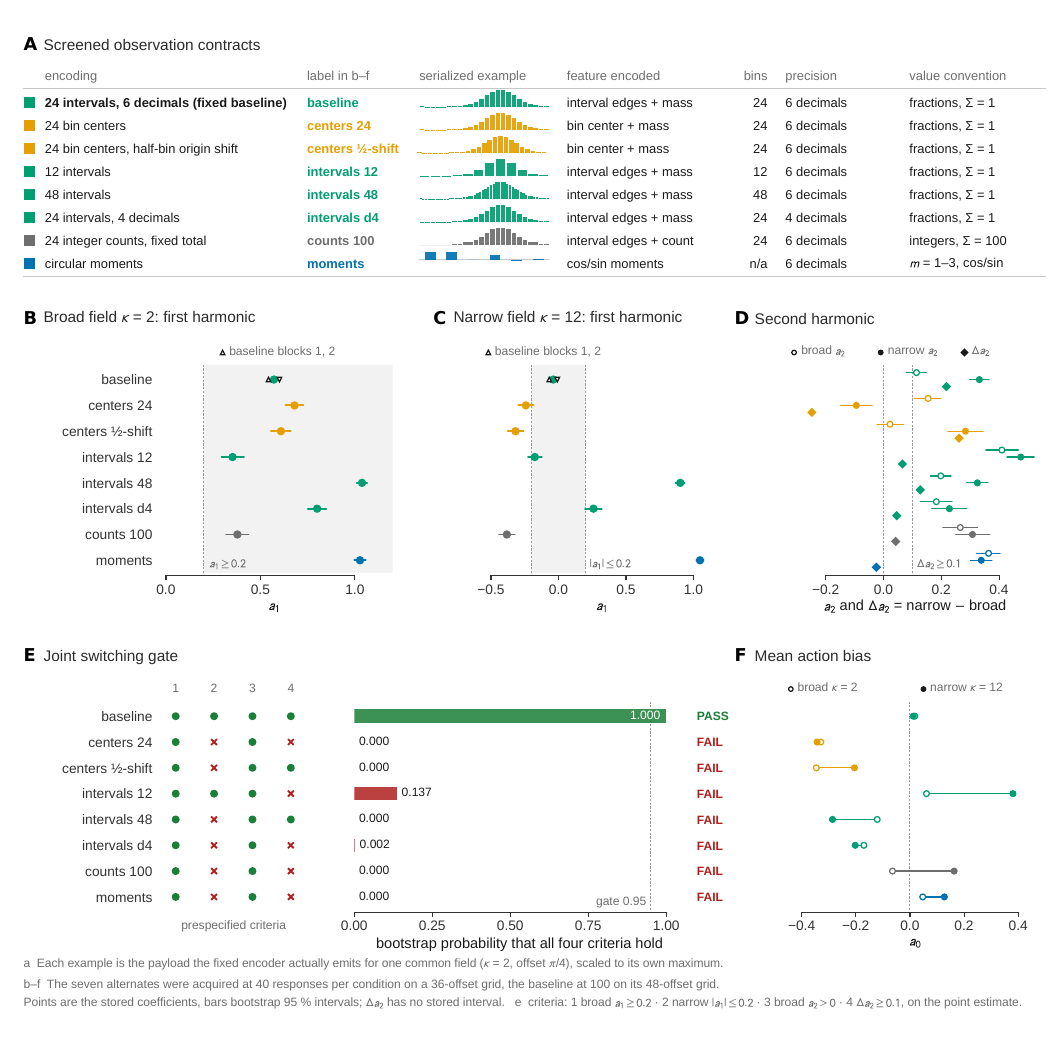}
\caption{\textbf{Complete encoding-invariance screen.}
This preliminary screen informed the choice of the three encodings used in the subsequent experiments. Eight observation contracts were compared on a common pair of fields before
any collective experiment, asking
whether changing the representation preserved the action probabilities. It did not.
\textbf{(A)} Screened encoding inventory (baseline intervals, centers, shifted
bin origin, 12 bins, 48 bins, four-decimal precision, integer counts,
moments); each example strip is the serialized payload for one common field.
\textbf{(B)}\textbf{(C)} First-harmonic estimates $a_1$ with bootstrap intervals
for the broad (\textbf{B}) and narrow (\textbf{C}) field, with the prespecified
thresholds.
\textbf{(D)} Second harmonic $a_2$ for both fields and the switching contrast
$\Delta a_2$, with all prespecified gate thresholds.
\textbf{(E)} Joint probability of satisfying every prespecified switching criterion,
per encoding, against the 0.95 gate, with the overall PASS/FAIL.
\textbf{(F)} Offset-averaged action bias $a_0$ per encoding: encoding
changes shift average directional bias, not only harmonic shape.
Scope: a motivating measurement, not a confirmatory result; no main-text claim
rests on it. Only the baseline encoding was acquired in two blocks (open
triangles in \textbf{(B)} \textbf{C}), so no between-block strip exists for the
seven alternates, and the three screened maps are not
information-equivalent.}
\end{figure}
\clearpage

\phantomsection\addcontentsline{toc}{section}{Supplementary Figure S8: Full concentration-dependent microscopic transmutation map}

\begin{figure}[H]
\SIpage{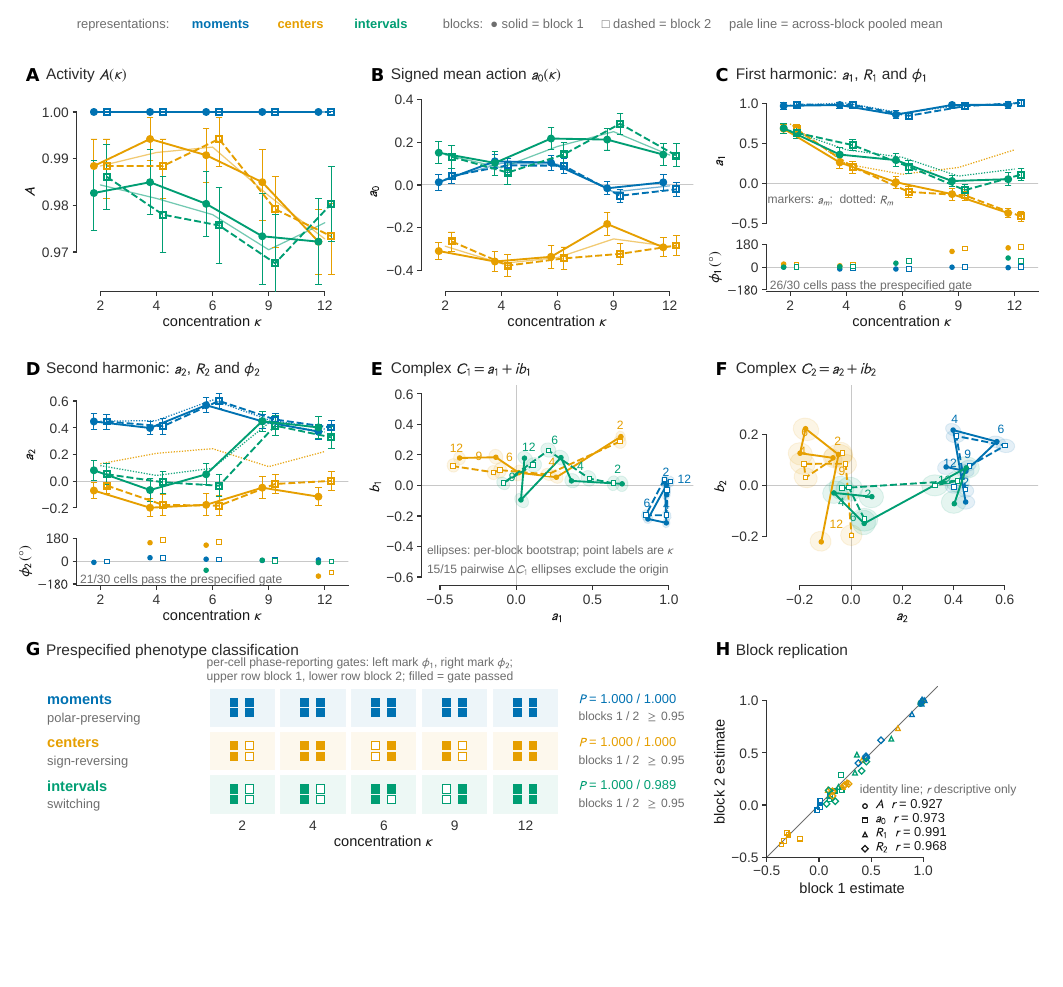}
\caption{\textbf{How concentration changes the response to each encoding.}
Response measurements for a single-peaked relative-phase distribution at five
prespecified concentrations $\kappa\in\{2,4,6,9,12\}$, three observation maps,
36 offsets, 24 responses per condition and two independent acquisition blocks
(25{,}920 valid responses at $M=6$).
\textbf{(A)} Activity $A(\kappa)$ with block-specific estimates, pooled means
and bootstrap intervals.
\textbf{(B)} Offset-independent response bias $a_0(\kappa)$.
\textbf{(C)}\textbf{(D)} First- and second-harmonic summaries: signed
projections $a_m$, magnitudes $R_m=\lvert C_m\rvert$, and a phase strip
showing $\phi_m$ only for cells passing the prespecified phase-reporting gate
(Supplementary Fig.~S6F).
\textbf{(E)}\textbf{(F)} Complex-plane trajectories of $C_1$ (\textbf{E}) and
$C_2$ (\textbf{F}) as $\kappa$ increases, one connected path per
encoding, with bootstrap ellipses.
\textbf{(G)} Prespecified microscopic phenotype classification per encoding and
concentration (polar-preserving, switching, sign-reversing); all three rules
passed in both blocks. The gate in \textbf{(C)}\textbf{D} is a display rule for
phase legibility; the phenotype rules are trajectory-level inference criteria
reading the $\kappa=2$ and $\kappa=12$ endpoints only.
\textbf{(H)} Between-block replication: block-1 against block-2 estimates of
$A$, $a_0$, $R_1$ and $R_2$ on identity lines; trajectory correlations are
$r(a_1)=0.939$--$0.994$ and $r(a_2)=0.868$--$0.986$, above the prespecified
$0.8$ threshold, and 39/45 block-difference intervals include zero. Error
bars are 95\% bootstrap intervals.
Correlations, uncertainty ellipses and exclusions of the origin are conditional on these two acquisition blocks.}
\end{figure}
\clearpage

\phantomsection\addcontentsline{toc}{section}{Supplementary Figure S9: Antipodal symmetry and dense signed-im\-balance response}

\begin{figure}[H]
\SIpage{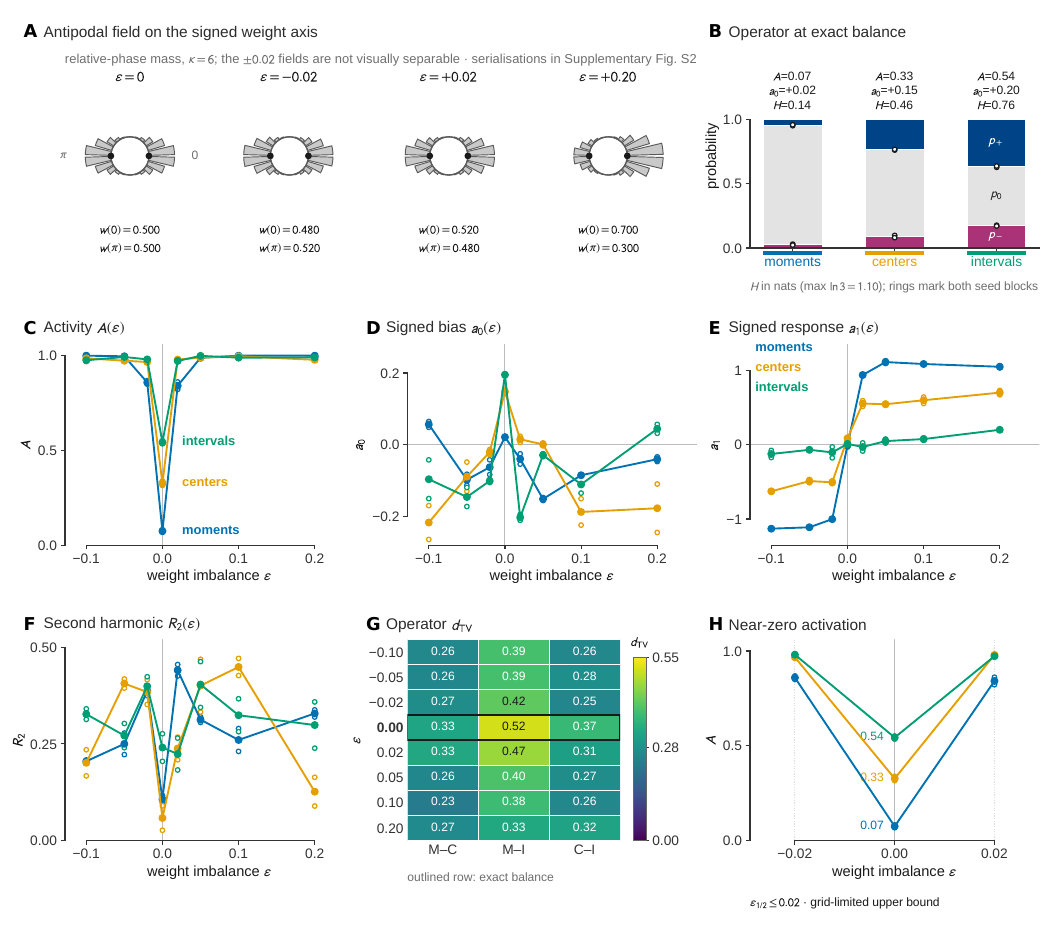}
\caption{\textbf{Antipodal symmetry and dense signed-imbalance response.}
Two von Mises modes at $0$ and $\pi$ with weights $0.5\pm\varepsilon$ at
$\kappa=6$, swept over
$\varepsilon\in\{-0.10,-0.05,-0.02,0,0.02,0.05,0.10,0.20\}$ for three
encodings, 36 offsets, 12 samples and two independent blocks
(20{,}736 calls). Activity is $A=p_-+p_+$; $a_0$ and $a_1$ are the order-0 and
order-1 Fourier coefficients of the signed action over the 36 offsets, and
$R_2$ is the second-harmonic amplitude.
\textbf{(A)} The physical field at exact balance, at the two smallest signed
imbalances and at the largest grid value; mode weights are printed because the
$\pm0.02$ fields are not visually separable.
\textbf{(B)} Full trinomial operator at exact balance, both blocks shown: the first sine component is near zero for all three encodings, while moments mostly returns stay.
\textbf{(C)} Activity $A(\varepsilon)$ over the full signed sweep, block by
block. \textbf{(D)} Signed action bias $a_0(\varepsilon)$.
\textbf{(E)} Signed first-harmonic response $a_1(\varepsilon)$.
\textbf{(F)} Second-harmonic amplitude $R_2(\varepsilon)$.
\textbf{(G)} Pairwise operator total-variation distance at every
$\varepsilon$: a vanishing first sine component does not imply identical action probabilities.
\textbf{(H)} The near-zero window with the unmeasured band marked: moments has
low activity at exact balance but activates at the smallest measured nonzero
imbalance, $\lvert\varepsilon\rvert=0.02$.
The grid contains $+0.20$ but no matching negative point; the
activation is sharp and grid-limited, with the shape between grid points
unresolved, and no mathematical discontinuity or derivative at
$\varepsilon=0$ is claimed.}
\end{figure}
\clearpage

\phantomsection\addcontentsline{toc}{section}{Supplementary Figure S10: Sparse finite-peer fields and variance decomposition}

\begin{figure}[H]
\SIpage{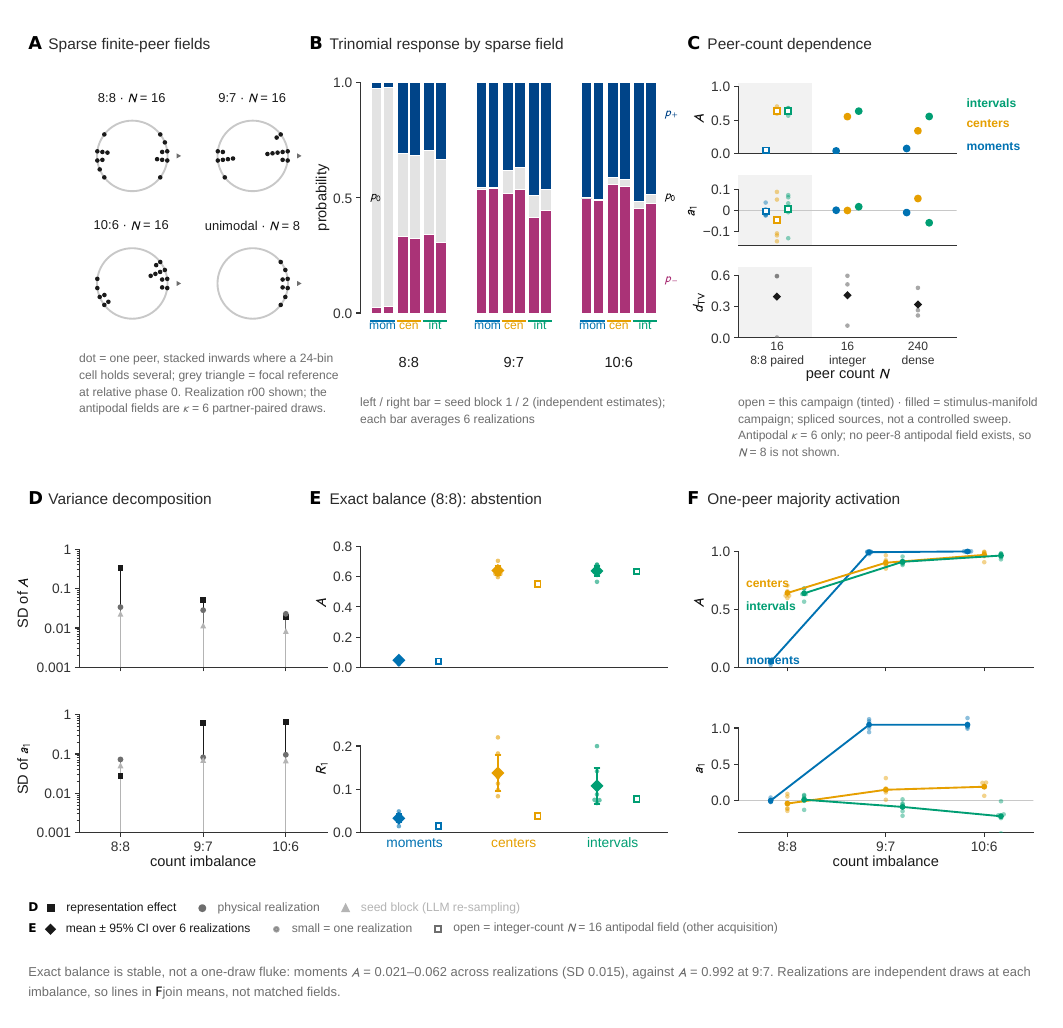}
\caption{\textbf{Sparse finite-peer fields and variance decomposition.}
The sparse-peer experiment used sparse antipodal fields with $N=16$ peers at
$\kappa=6$ and partner-paired draws, three count imbalances (8:8, 9:7, 10:6),
three observation maps, and 6 independent physical realizations $\times$ 2
seed blocks per cell.
Panels \textbf{C}--\textbf{F} use block-averaged realization means.
\textbf{(A)} The discrete stimulus inventory, with peer counts indicated.
\textbf{(B)} Trinomial responses per encoding and sparse field, with
independent-block estimates.
\textbf{(C)} Peer-count dependence, comparing the sparse $N=16$ campaign with
the dense 240-peer stimulus families where matched field families exist;
$\TV$ is the aggregate-trinomial distance between the encoding-specific aggregate action distributions.
\textbf{(D)} Variance decomposition separating physical-realization variance,
seed-block variance (which absorbs LLM re-sampling; the two are not resolved
separately) and the encoding effect; physical realizations, not repeated
calls, are the replication unit.
\textbf{(E)} Exact-balance abstention for moments across all balanced sparse
realizations: showing that abstention recurs across physical realizations.
\textbf{(F)} One-peer-majority activation: paired 8:8 versus 9:7 and 10:6
comparisons; a single-peer majority restores a strong signed polar channel
under each encoding.
The sparse experiment contains only $N=16$, so these data do not establish a trend toward the large-population limit; only
per-block aggregates are available, so within-block sampling cannot be
resolved apart from block variance.}
\end{figure}
\clearpage

\phantomsection\addcontentsline{toc}{section}{Supplementary Figure S11: Prespecified replay-field selection and audit}

\begin{figure}[H]
\SIpage{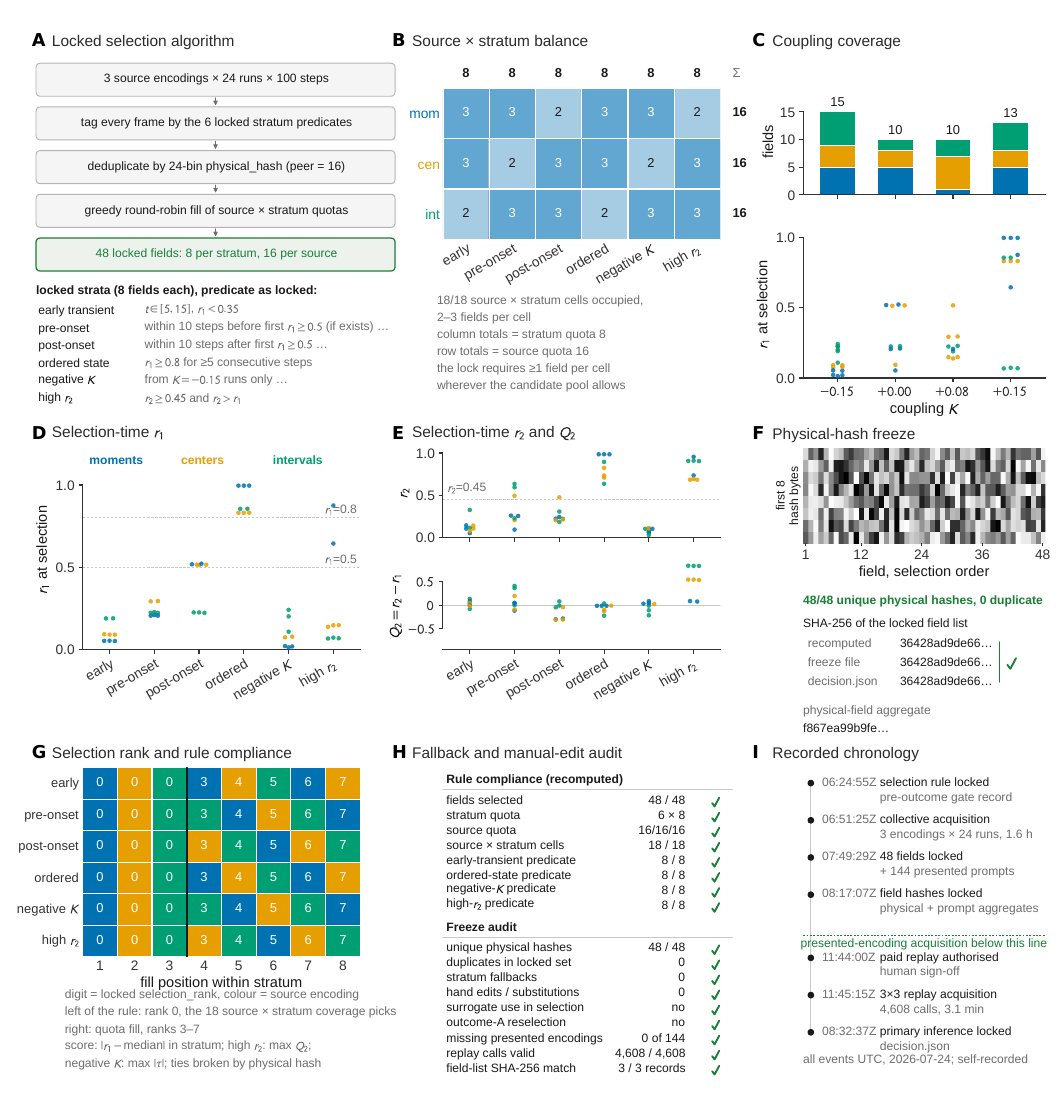}
\caption{\textbf{Prespecified replay-field selection and audit.}
The 48 physical fields replayed under all three presented encodings were chosen
by the mechanical rule prespecified in the replay selection record. Selection did not use the replay outcomes: the released selector reads only
the collective sessions' trajectories and metadata, and the stratum predicates
are functions of $r_1$, $r_2$ and $Q_2$ of the source trajectories alone, so no
presented-encoding response is an input to it.
\textbf{(A)} The prespecified selection algorithm, from complete collective
trajectories through stratum tagging, physical-hash deduplication and greedy
source-by-stratum quota fill, to the fixed 48-field list. The panel prints the
first clause of each predicate, with the full text in that record.
\textbf{(B)} Source $\times$ stratum balance: all 18 cells of the $3\times6$
matrix are populated, 16 fields per source and eight per stratum.
\textbf{(C)} Coupling coverage of the selected fields.
\textbf{(D)}\textbf{(E)} Selection-time $r_1$ (\textbf{D}) and $r_2$, $Q_2$
(\textbf{E}) per field, by stratum and source.
\textbf{(F)} Physical-hash freeze: 48 unique physical hashes, no duplicates.
\textbf{(G)} Selection rank and rule compliance for every field.
\textbf{(H)} Fallback and manual-edit audit: zero fallback selections, zero
manual substitutions. The duplicate-hash count is taken inside the selected
set; the number of candidate frames removed by the pre-selection
deduplication is not retained by the selector.
\textbf{(I)} The recorded order of events, from selection-rule lock through
replay acquisition to inference lock (timestamps in the repository audit); the selection code excludes replay responses from its inputs, as described above.}
\end{figure}
\clearpage

\phantomsection\addcontentsline{toc}{section}{Supplementary Figure S12: Complete fieldwise replay response panel}

\begin{figure}[H]
\captionsetup{labelformat=sipaged}\renewcommand{\pageofn}{ (page 1 of 4)}
\SIpage{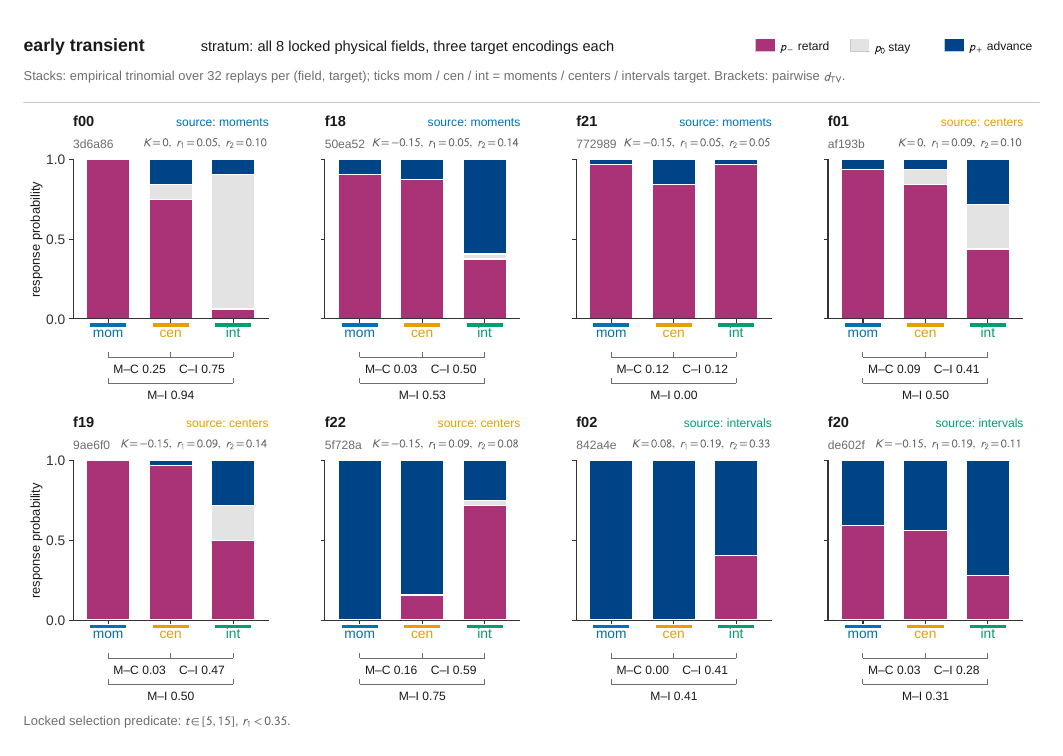}
\caption{\textbf{Complete fieldwise replay response panel (by replay stratum;
this page: early transient, defined as frames with
$t\in[5,15]$ and $r_1<0.35$).}
The evidence behind main Fig.~3: for each of the eight
fixed physical fields of the stratum, the empirical trinomial
$(p_-,p_0,p_+)$ stack under the moments, centers and intervals \emph{presented}
encodings, with the three pairwise total-variation distances drawn beneath
each field. The trinomials come from a single pass over the fixed replay trace (4{,}608 records $=48$ fields $\times$ 3 presented encodings $\times$ 32 responses in two blocks), and each stratum states its predicate. Panels are ordered by source encoding and are never averaged; all replays are GPT (Supplementary Table~S2). The hex string is the first six characters of the physical SHA-256, and the coloured tag names the source encoding.}
\end{figure}
\clearpage

\begin{figure}[H]\ContinuedFloat
\captionsetup{labelformat=sipaged}\renewcommand{\pageofn}{ (page 2 of 4)}
\SIpageT{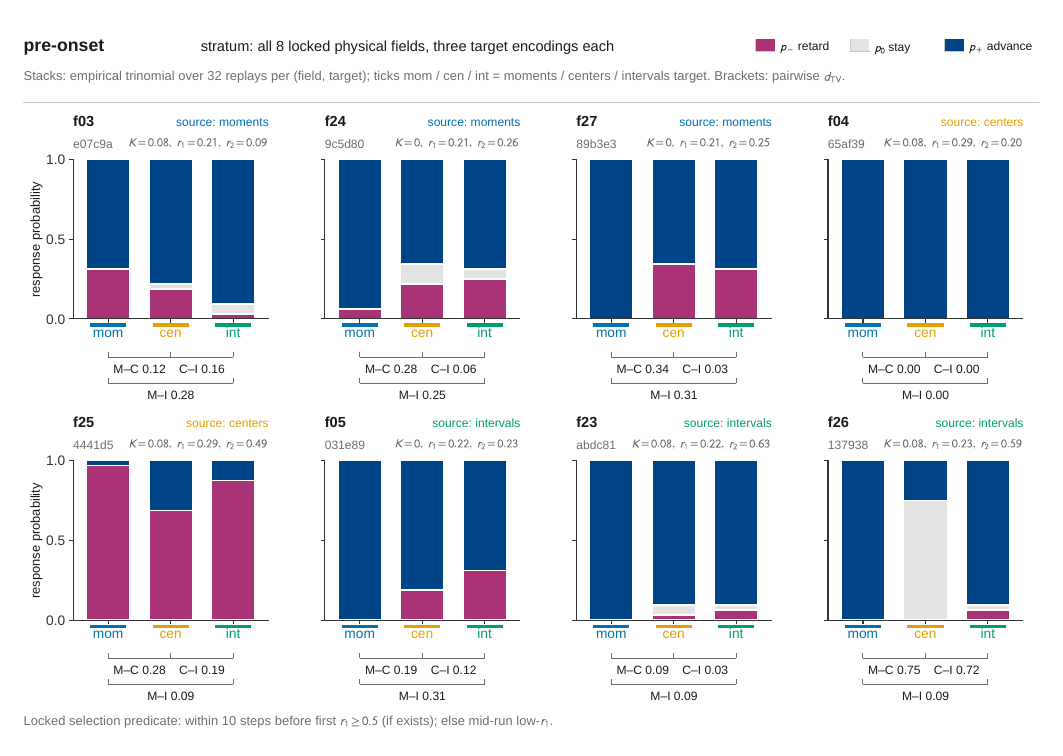}{4mm}{5mm}
\par\vspace{0.5mm}
\SIpageT{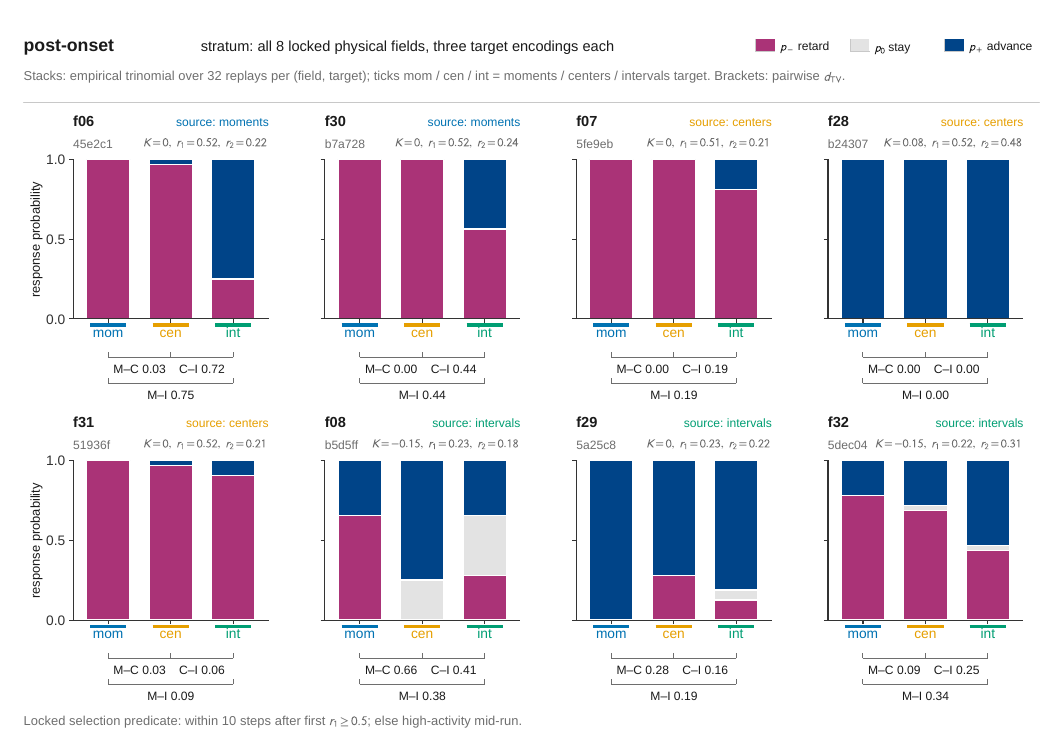}{4mm}{5mm}
\caption{\textbf{(continued: pre-onset and post-onset strata).} Top: pre-onset,
frames within 10 steps before the first crossing $r_1\geq0.5$ of their source
run (mid-run low-$r_1$ frames where no crossing exists). Bottom: post-onset,
frames within 10 steps after the first crossing (high-activity mid-run frames
where no crossing exists). Display as on page 1 of Supplementary Fig.~S12.}
\end{figure}
\clearpage

\begin{figure}[H]\ContinuedFloat
\captionsetup{labelformat=sipaged}\renewcommand{\pageofn}{ (page 3 of 4)}
\SIpageT{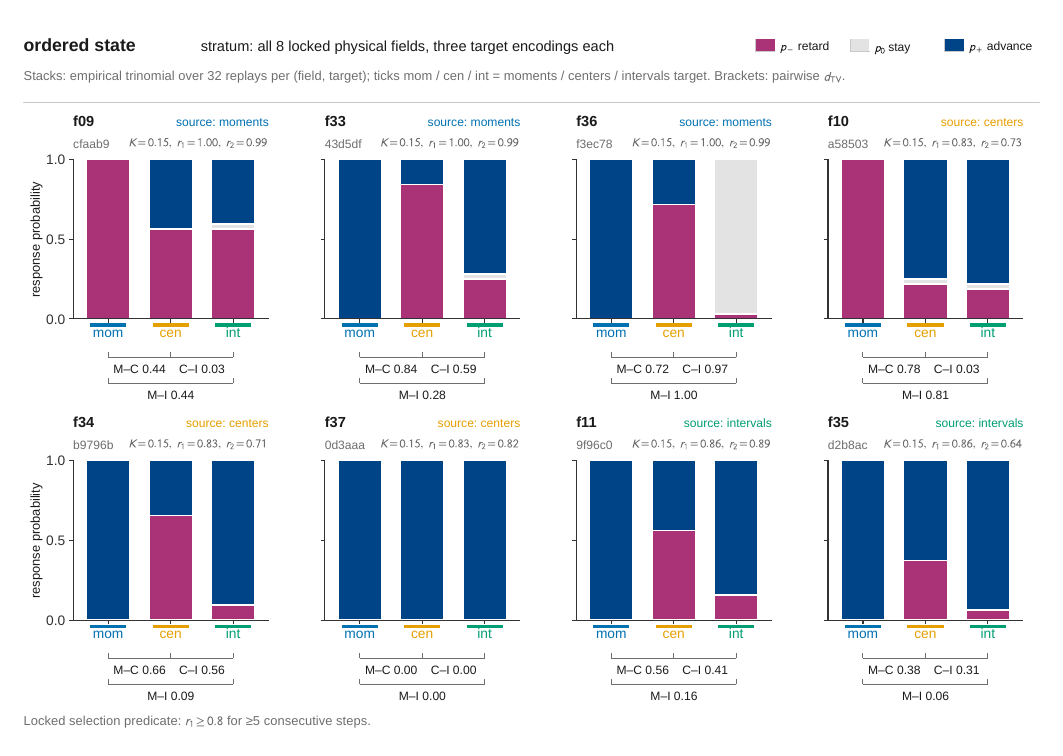}{4mm}{5mm}
\par\vspace{0.5mm}
\SIpageT{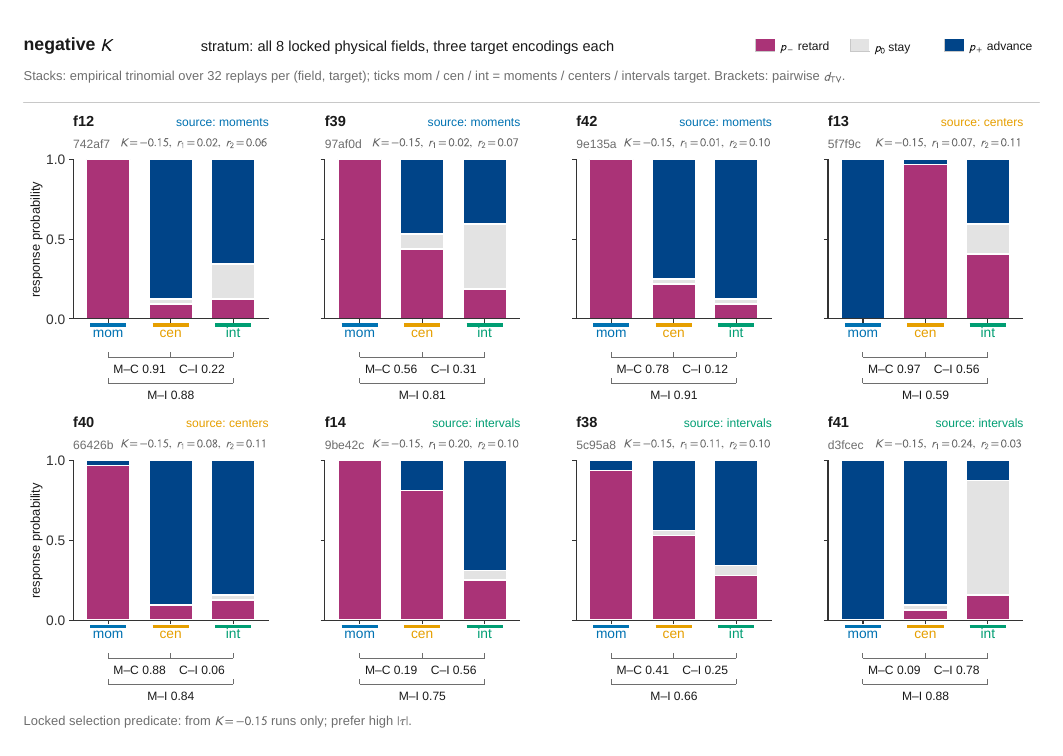}{4mm}{5mm}
\caption{\textbf{(continued: ordered-state and negative-$K$ strata).} Top:
ordered state, frames with $r_1\geq0.8$ sustained for at least five
consecutive steps. Bottom: negative $K$, frames drawn from $K=-0.15$ runs
only, preferring high $\lvert\tau_{\mathrm{social}}\rvert$. Display as on
page 1 of Supplementary Fig.~S12.}
\end{figure}
\clearpage

\begin{figure}[H]\ContinuedFloat
\captionsetup{labelformat=sipaged}\renewcommand{\pageofn}{ (page 4 of 4)}
\SIpage{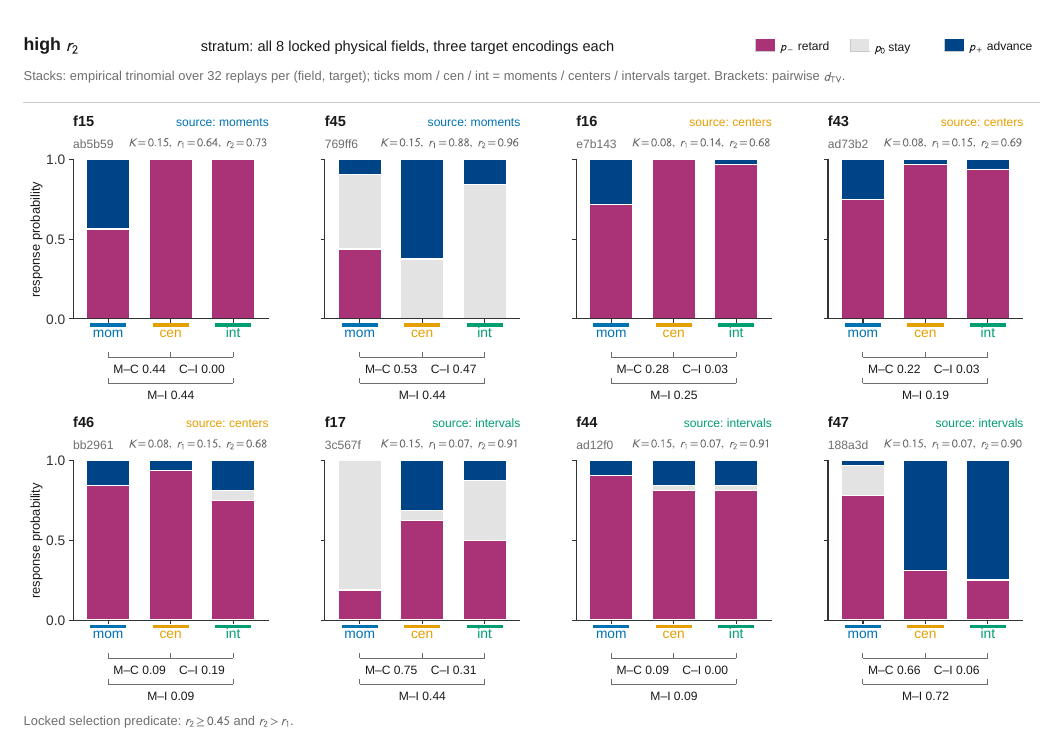}
\caption{\textbf{(continued: high-$r_2$ stratum).} Frames with $r_2\geq0.45$
and $r_2>r_1$. Display as on page 1 of Supplementary Fig.~S12.}
\end{figure}
\clearpage

\phantomsection\addcontentsline{toc}{section}{Supplementary Figure S13: Replay robustness, source and interaction tests}

\begin{figure}[H]
\SIpage{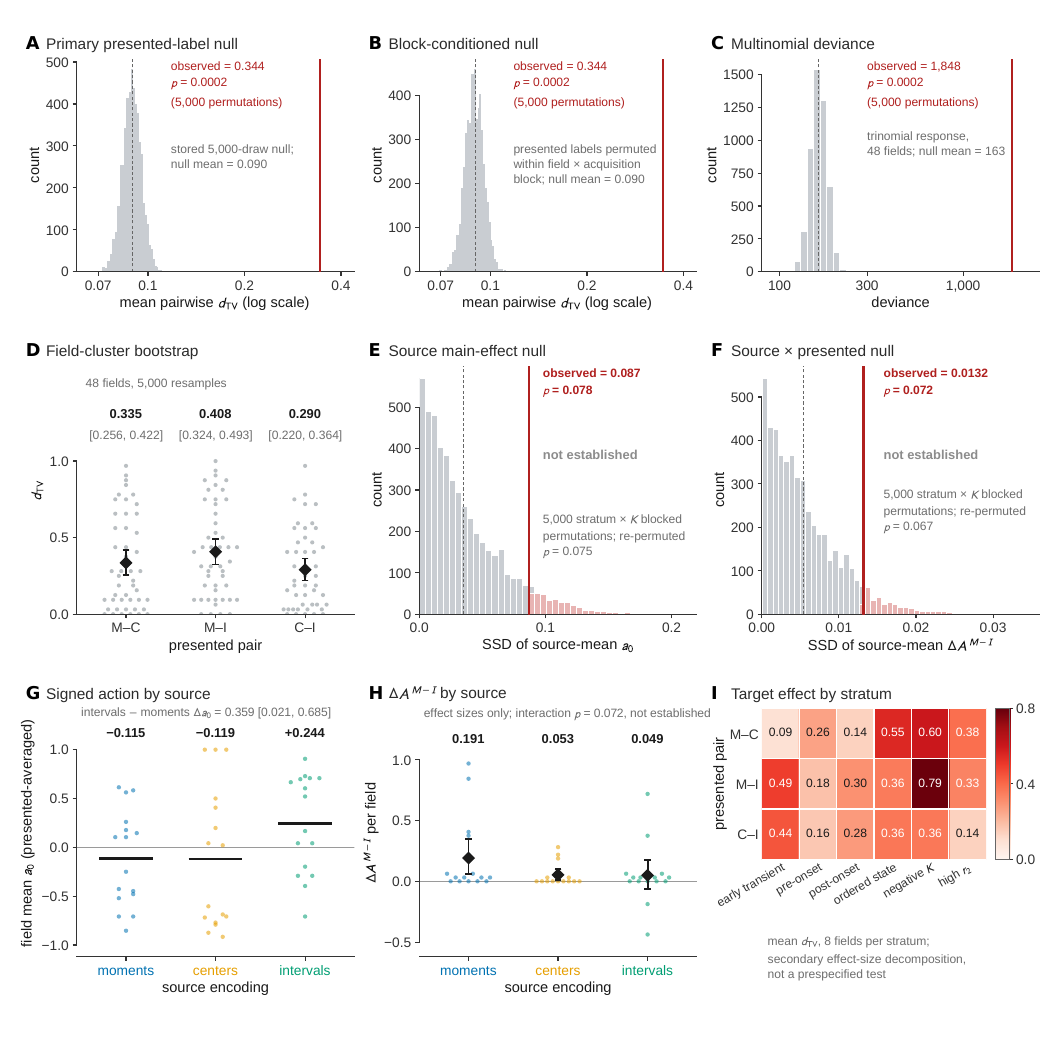}
\caption{\textbf{Replay robustness, source and interaction tests.}
\emph{How to read.} Panels \textbf{A}--\textbf{D} and the stratified
decomposition in \textbf{I} establish the presented-encoding effect on the same
physical fields; panels \textbf{E}--\textbf{H} examine the weaker question of
whether fields generated under different encodings also differ. Every
prespecified global test and secondary decomposition of the matched
$3\times3$ replay is shown.
Permutation distributions not stored in the original analysis (\textbf{(B)}\textbf{(C)}\textbf{(E)}\textbf{F}) were recomputed here using 5{,}000 permutations.
\textbf{(A)} Primary field-blocked presented-encoding-label permutation (5{,}000 draws):
observed mean pairwise $\TV=0.344$, $p=0.0002$.
\textbf{(B)} Block-conditioned presented-encoding permutation (robustness null preserving
block structure).
\textbf{(C)} Multinomial deviance test on the full trinomial response, giving
the same decision as the $\TV$ statistic.
\textbf{(D)} Field-cluster bootstrap distributions for the three pairwise
presented-encoding distances: moments--centers $0.335$ ($0.256$--$0.422$),
moments--intervals $0.408$ ($0.324$--$0.493$), centers--intervals $0.290$
($0.220$--$0.364$).
\textbf{(E)} Source main-effect null: $p=0.0784$.
\textbf{(F)} Source $\times$ presented encoding interaction null: $p=0.0716$ (reported as $0.072$ in the main text).
Neither meets the prespecified threshold of $p<0.05$, so neither effect is
established. Both are omnibus, between-field tests over 16 fields per source
condition, whereas the presented-encoding test in \textbf{A} is within-field; this design therefore provides less precision for the source and interaction effects, and these nonsignificant results are not evidence that no source or
interaction effect exists.
\textbf{(G)} Mean signed action $a_0$, averaged over presented encodings, by source encoding
($-0.115$ for moments, $-0.119$ for centers, $+0.244$ for intervals). The largest pairwise component, intervals minus
moments, has a field-cluster bootstrap interval excluding zero
($\Delta a_0=0.359$, unadjusted field-bootstrap 95\% CI $0.021$--$0.685$). This descriptive, uncorrected decomposition does not overturn the nonsignificant omnibus test in \textbf{E}; it illustrates one direction contributing to the observed source contrast.
\textbf{(H)} The prespecified presented-encoding contrast $\Delta A^{\mathrm{M-I}}$ shown
separately for each source ($0.191$, $0.053$, $0.049$ for moments, centers and
intervals). Effect sizes only; the interaction test in \textbf{F} is not established.
\textbf{(I)} Pairwise presented-encoding $\TV$ by the six prespecified replay strata
(secondary). The 48 physical fields, not the 4{,}608 calls, are the inference
units throughout.}
\end{figure}
\clearpage

\phantomsection\addcontentsline{toc}{section}{Supplementary Figure S14: Complete three-family microscopic operator replication}

\begin{figure}[H]
\SIpage{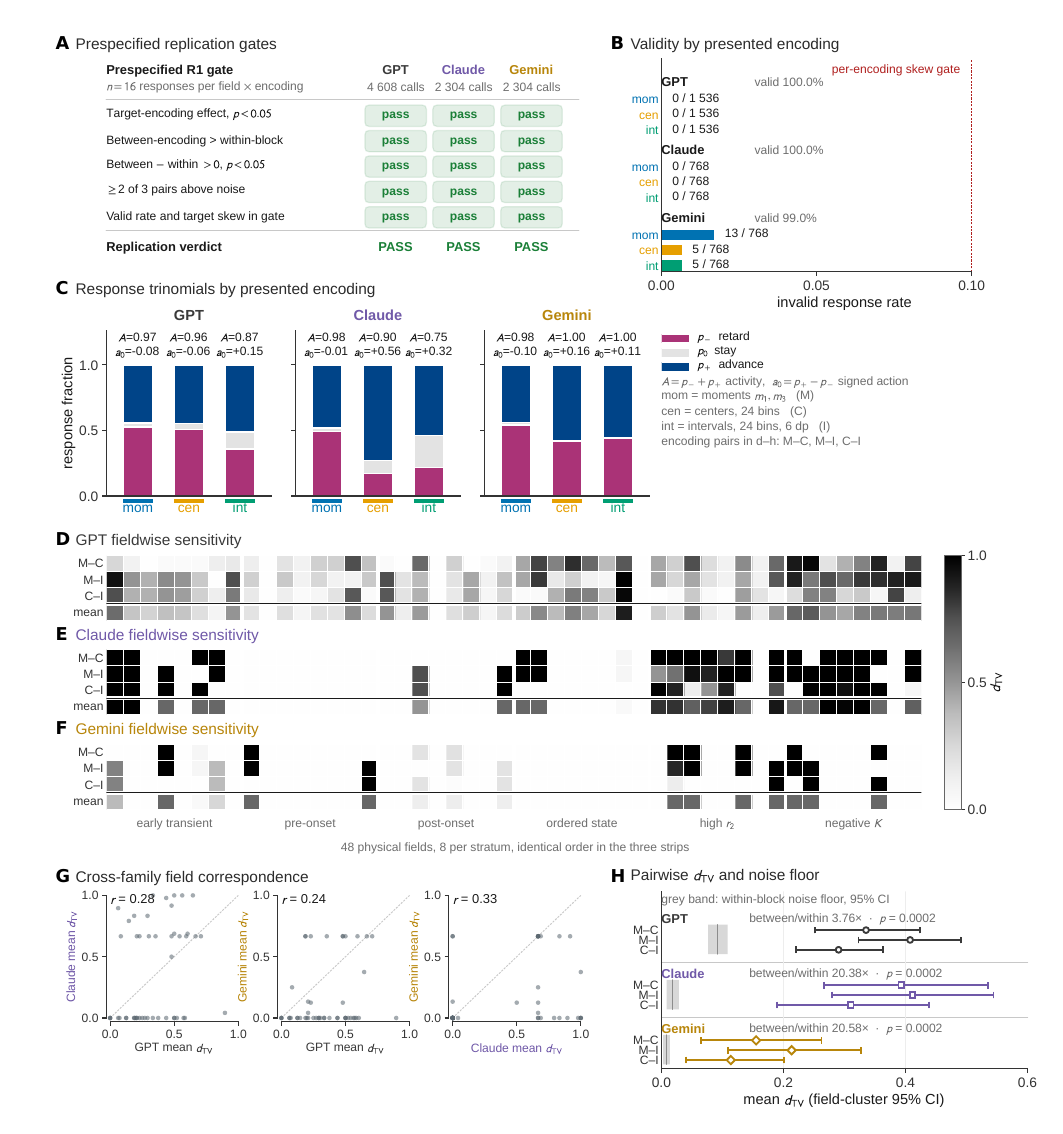}
\caption{\textbf{Complete three-family microscopic operator replication.}
The complete version of main Fig.~4F: GPT, Claude and Gemini (model IDs in
Supplementary Table~S2)
on the same 48 fixed physical-field hashes and presented encoders, 16 responses
per field and presented encoding in two blocks for the replication families.
\textbf{(A)} The prespecified replication gates and their outcomes for
all three families (global presented-encoding permutation $p<0.05$; between-presented-encoding
exceeding within-presented-encoding $\TV$; positive field-bootstrap interval for the
difference; at least two of three encoding pairs above the noise floor;
acceptable, non-encoding-skewed invalidity).
\textbf{(B)} Validity audit by presented encoding: Claude 2{,}304/2{,}304 and
Gemini 2{,}281/2{,}304 valid responses; invalidity is not sufficiently
encoding-skewed to explain the result.
\textbf{(C)} Response trinomials by family and presented encoding, reported as the
mean over the 48 fields of the per-field response distribution, with invalid
responses excluded.
\textbf{D}--\textbf{(F)} Fieldwise mean pairwise presented-encoding $\TV$ for all 48
fields in one shared field order: GPT (\textbf{D}), Claude (\textbf{E}),
Gemini (\textbf{F}).
\textbf{(G)} Cross-family field correspondence over the 48 shared fields;
exact pairwise ranking agreement was not a prespecified requirement, so the
panel is descriptive.
\textbf{(H)} Pairwise $\TV$ forest with within-block noise floors,
between/within ratios and the permutation $p$ per family ($p=0.0002$ for
each replication family). Permutation tests use 5{,}000 within-field
presented-encoding-label relabellings, so $p=0.0002$ is the resolution floor $1/(N+1)$
and not a family difference. The response count $n=16$ was fixed in advance rather than optimized per family.}
\end{figure}
\clearpage

\phantomsection\addcontentsline{toc}{section}{Supplementary Figure S15: Complete Claude matched-collective trajectories}

\begin{figure}[H]
\captionsetup{labelformat=sipaged}\renewcommand{\pageofn}{ (page 1 of 3)}
\SIpage{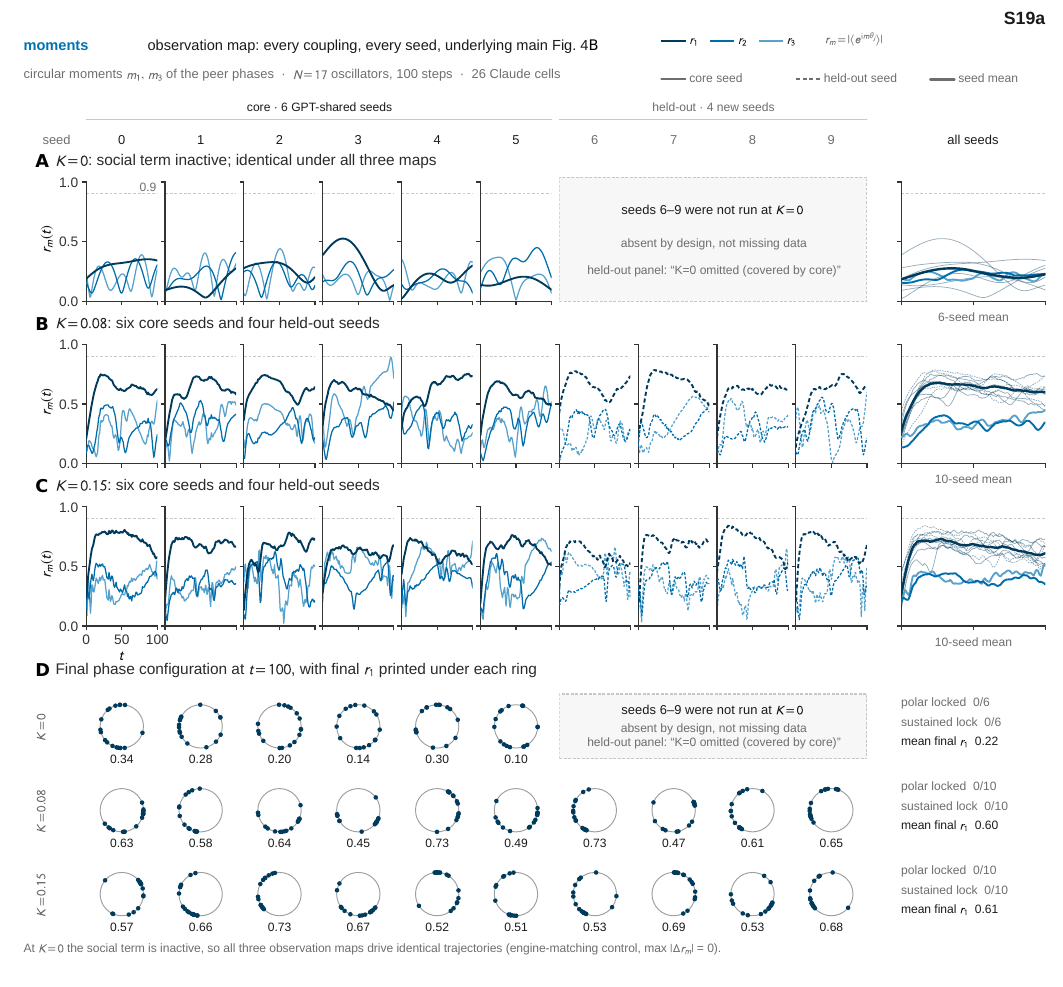}
\caption{\textbf{Complete Claude matched-collective trajectories (one page per
observation map; this page: moments).}
The complete Claude trajectory set as
small multiples over (coupling, seed): coupling rows $K\in\{0,0.08,0.15\}$
by ten seed columns, each cell holding $r_1$, $r_2$ and $r_3$ for one
physical seed, plus a seed-aggregate column and the final phase configuration
of every cell. Core seeds (the six physical seeds shared with GPT) and the
four held-out seeds are drawn distinctly throughout. The grid is not
rectangular: held-out seeds were acquired only at positive coupling and
therefore do not appear in the $K=0$ column; the four absent cells are
labelled as a design decision rather than left to read as missing data (26
cells per encoding, not 30). The dashed rule is the polar-lock
threshold $r_1=0.9$ (the \texttt{protocol.json} phenotype rule
\texttt{final\_r1}~$\geq0.9$), and final $r_1$ values set in bold in
\textbf{D} are polar locked. Every cell spans $t=0$--$100$ horizontally and $r_m=0$--$1$ vertically. The $K=0$ row is core-only by design, so its aggregate is a 6-seed mean
whereas the $K>0$ aggregates are 10-seed means, as in main Fig.~4B (main Fig.~1 uses only the six core seeds).}
\end{figure}
\clearpage

\begin{figure}[H]\ContinuedFloat
\captionsetup{labelformat=sipaged}\renewcommand{\pageofn}{ (page 2 of 3)}
\SIpage{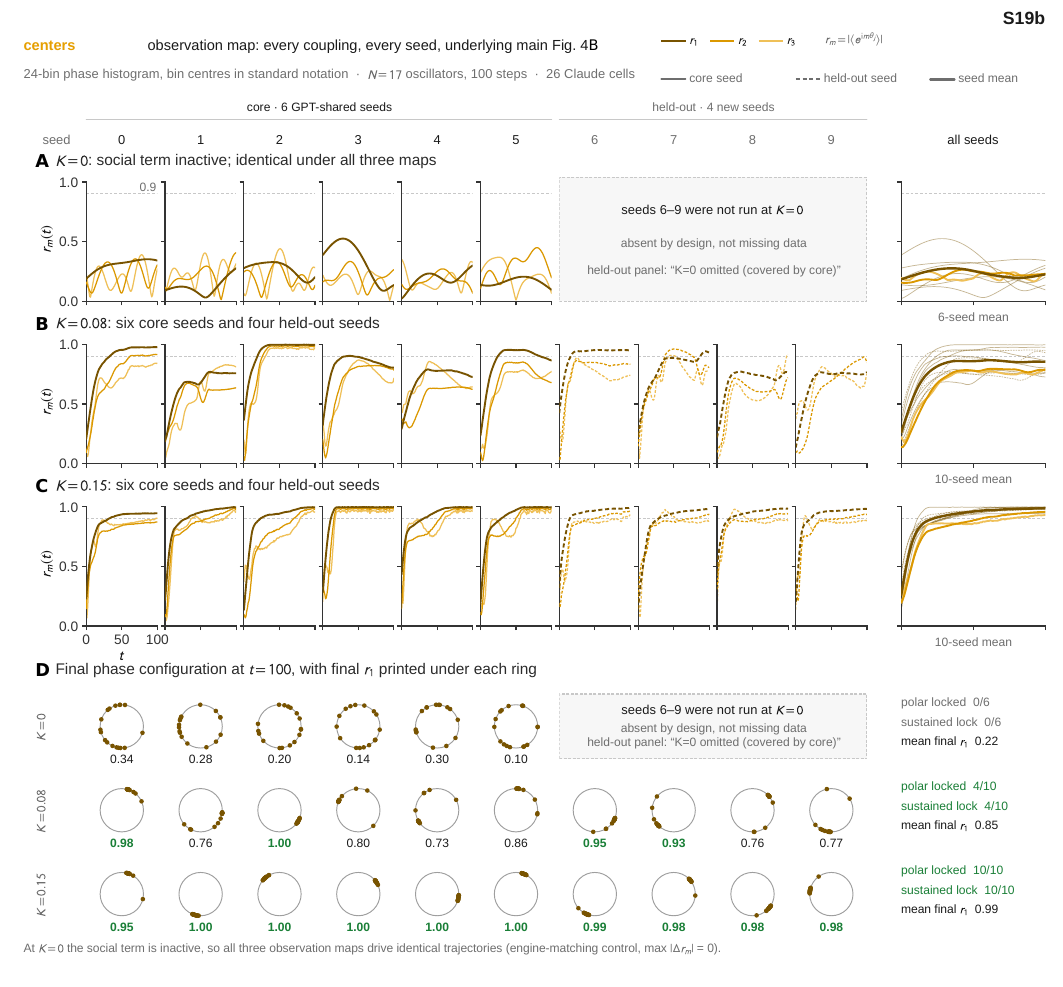}
\caption{\textbf{(continued: centers).} As on page 1 of Supplementary Fig.~S15, for
the centers observation map. Held-out seeds appear only at positive coupling,
by design.}
\end{figure}
\clearpage

\begin{figure}[H]\ContinuedFloat
\captionsetup{labelformat=sipaged}\renewcommand{\pageofn}{ (page 3 of 3)}
\SIpage{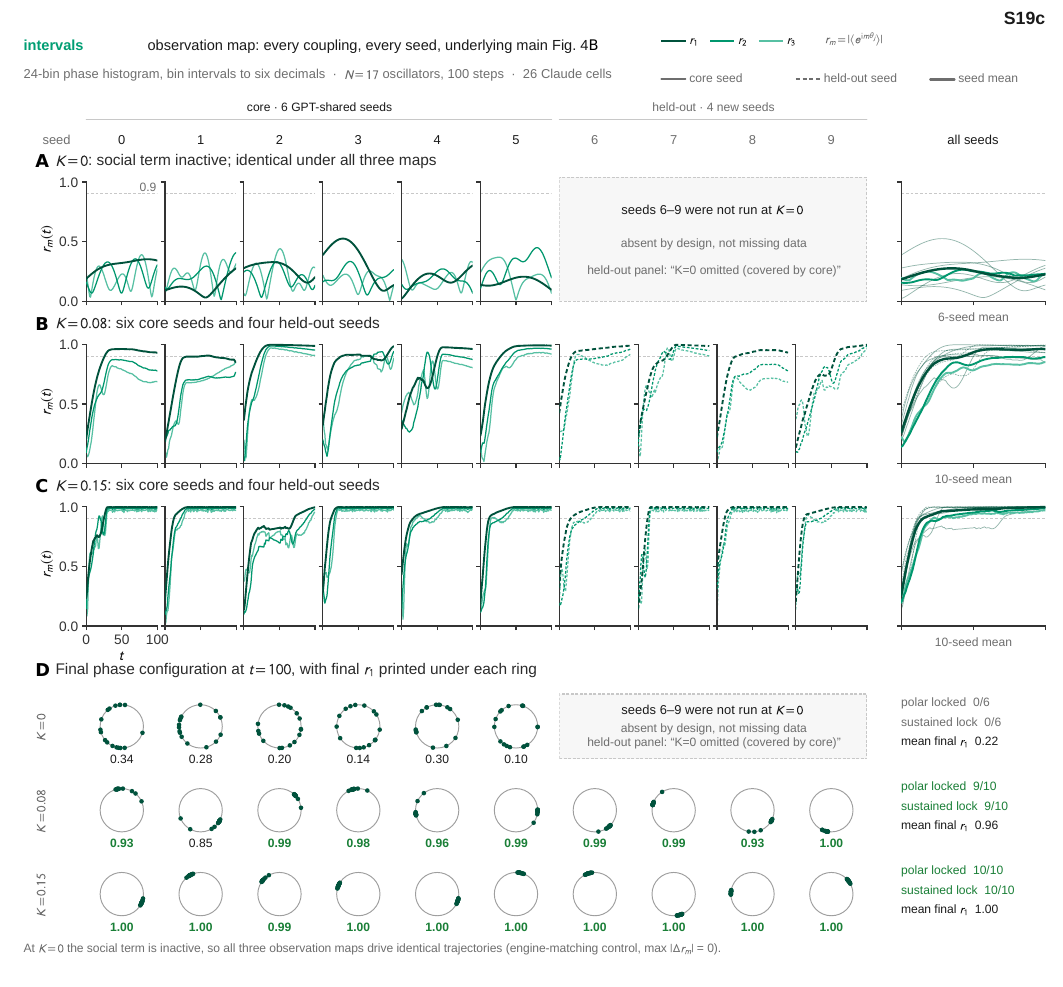}
\caption{\textbf{(continued: intervals).} As on page 1 of Supplementary Fig.~S15, for
the intervals observation map. Held-out seeds appear only at positive
coupling, by design.}
\end{figure}
\clearpage

\phantomsection\addcontentsline{toc}{section}{Supplementary Figure S16: Claude controls, exact inference and prespecified gates}

\begin{figure}[H]
\SIpage{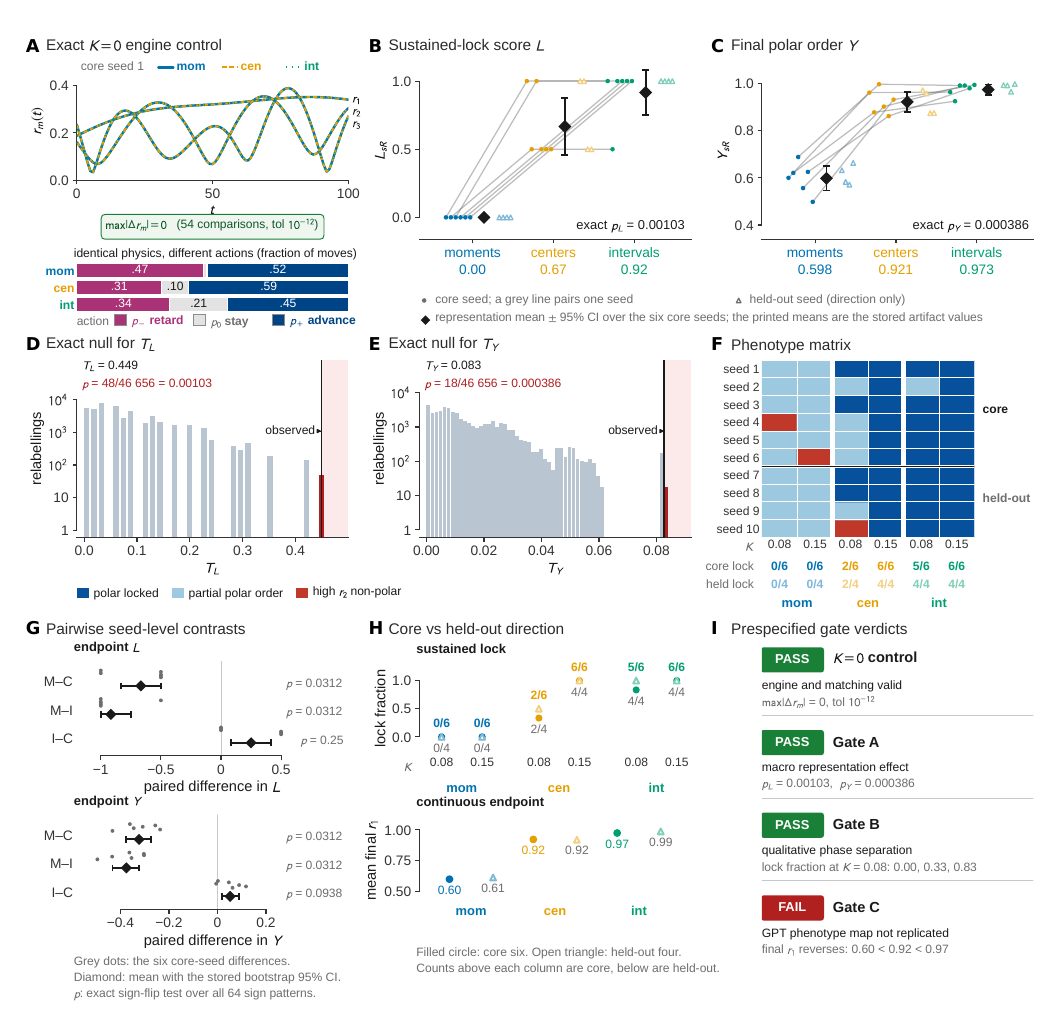}
\caption{\textbf{Claude controls, exact inference and prespecified
gates.}
The complete confirmatory analysis behind main Fig.~4, with $N=17$
oscillators and 100 steps throughout (Claude Haiku 4.5); the core seeds are
the six physical seeds shared with GPT and the held-out seeds are four new
seeds.
\textbf{(A)} Exact $K=0$ engine and matching control: the three observation
maps drive identical physical trajectories
($\max\lvert\Delta r_m\rvert=0$ for $m=1,2,3$ on every core seed) while the
actions they elicit differ.
\textbf{(B)} Terminal-lock score $L$ seed by seed (core means 0.00 moments,
0.67 centers, 0.92 intervals), with held-out seeds as separate marks.
\textbf{(C)} Final polar order $Y$ seed by seed (core means 0.598, 0.921,
0.973).
\textbf{(D)} Exact permutation null for $T_L$, with all $(3!)^{6}=46{,}656$
within-seed relabellings of the three encodings enumerated, so the null is
exact rather than sampled: observed $T_L=0.449$, $p=0.00103$. Red marks the
rejection tail $T\geq T_{\mathrm{obs}}$, where
$T=\sum_{\R}(\bar{x}_{\R}-\bar{x})^{2}$.
\textbf{(E)} The same for $T_Y$ over the same enumeration: observed
$T_Y=0.0827$, $p=0.000386$.
\textbf{(F)} Complete phenotype matrix, core seeds above the separator and
held-out seeds below, with lock counts ($K=0.08$: moments 0/6,
centers 2/6, intervals 5/6; $K=0.15$: 0/6, 6/6, 6/6).
\textbf{(G)} Pairwise seed-level contrasts for both endpoints with bootstrap
intervals and exact sign-flip $p$-values; the confirmatory distinction is moments versus the two histogram encodings, and intervals over centers is not established.
\textbf{(H)} Core versus held-out direction, shown separately; the held-out set is a directional replication only, and no pooled confirmatory test was prespecified.
\textbf{(I)} Prespecified gate verdicts: the $K=0$ control, Gate~A (global
encoding effect) and Gate~B (qualitative lock/nonlock separation) pass. Gate~C
was classified as FAIL because the GPT encoding-to-phenotype map was not
replicated; the observed ordering matched the prespecified reverse-map outcome,
which does not invalidate Gates A or B: encoding dependence generalized across
families while the mapping from encoding to phenotype was family-specific.}
\end{figure}
\clearpage

\phantomsection\addcontentsline{toc}{section}{Supplementary Figure S17: GPT--Claude reversal of the encoding-to-\hspace{0pt}phenotype map}

\begin{figure}[H]
\SIpage{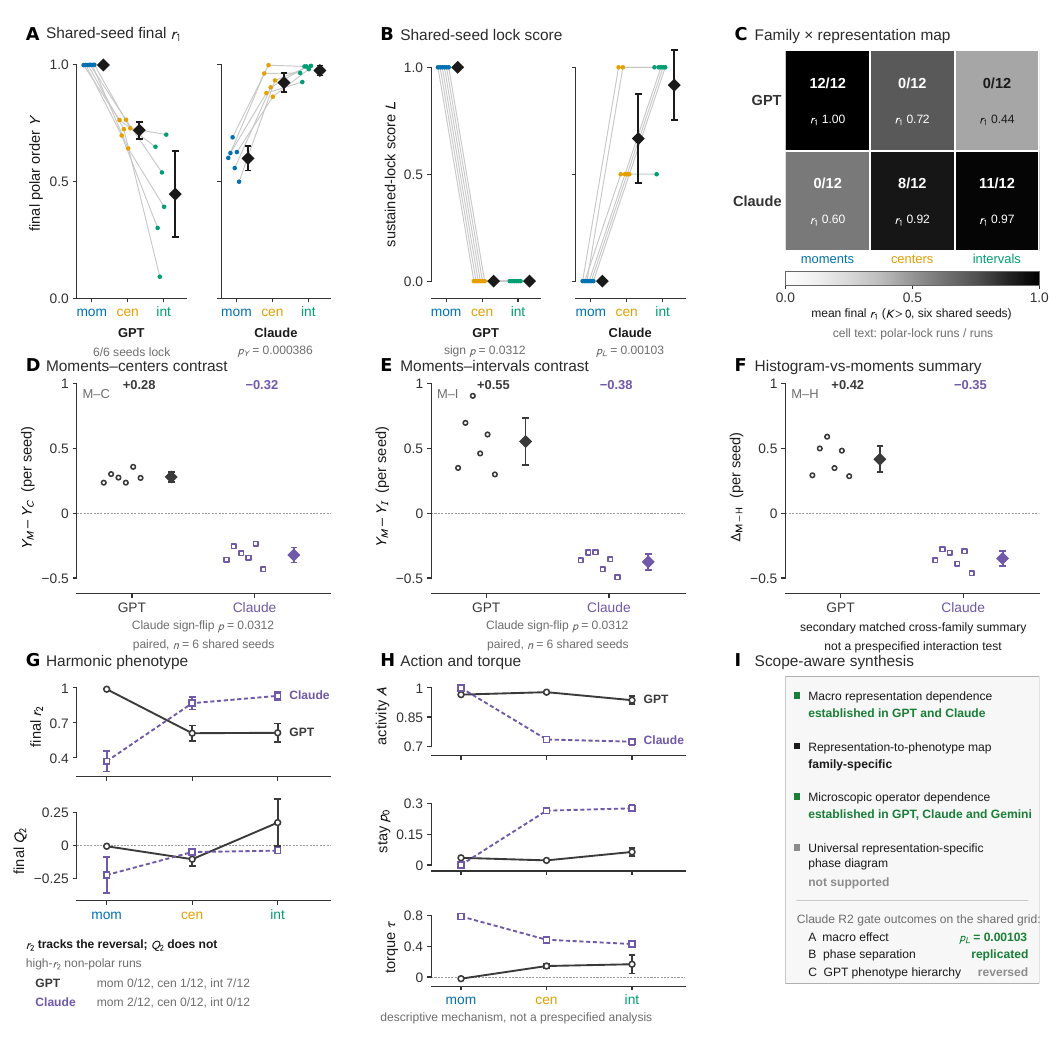}
\caption{\textbf{GPT--Claude reversal of the encoding-to-phenotype
map.}
All comparisons use the six physical seeds shared by GPT and Claude. Initial phases $\theta_i(0)$ and natural increments $\omega_i$ were checked and matched in all 18 core cells. Additional Claude seeds have no GPT counterpart and are omitted. The inference unit is one physical seed, each seed point averaging the two
positive couplings $K\in\{0.08,0.15\}$.
\textbf{(A)} Shared-seed final polar order $Y$ in GPT and Claude facets with
paired seed points on identical axes.
\textbf{(B)} Shared-seed terminal-lock score $L$, same structure. In \textbf{A} and \textbf{B} the grey lines join one seed across the three
encodings, diamonds are means and whiskers are seed-bootstrap percentile 95\% intervals;
the confirmatory inference for this comparison is the exact test in
Supplementary Fig.~S16, not these bars.
\textbf{(C)} Family $\times$ encoding macro map (shade: mean
positive-$K$ final $r_1$; cell text: locked runs), the principal visual summary of the reversal.
\textbf{(D)}\textbf{(E)} Seed-level contrasts $Y_{\mathrm M}-Y_{\mathrm C}$
(\textbf{D}) and $Y_{\mathrm M}-Y_{\mathrm I}$ (\textbf{E}) by family, with
means and intervals.
\textbf{(F)} The matched cross-family summary
$\Delta_{\mathrm{M-H}}=Y_{\mathrm M}-(Y_{\mathrm C}+Y_{\mathrm I})/2$ by
family; secondary, not a prespecified interaction test.
\textbf{(G)} Harmonic phenotype comparison (final $r_2$ and $Q_2$ by family
and encoding): the reversal concerns polar locking, and other harmonic
characteristics need not reverse identically.
\textbf{(H)} Descriptive action statistics for shared seeds at positive
coupling (activity, stay probability, social torque).
\textbf{(I)} Summary of the evidence: macroscopic encoding dependence is
established in GPT and Claude; the encoding-to-phenotype mapping is
family-specific; microscopic operator dependence is established in GPT,
Claude and Gemini; the results do not establish one encoding-to-outcome relationship that holds across models.}
\end{figure}
\clearpage

\phantomsection\addcontentsline{toc}{section}{Supplementary Figure S18: Complete same-task-information observation-map control}

\begin{figure}[H]
\captionsetup{labelformat=sipaged}\renewcommand{\pageofn}{ (page 1 of 2)}
\SIpage{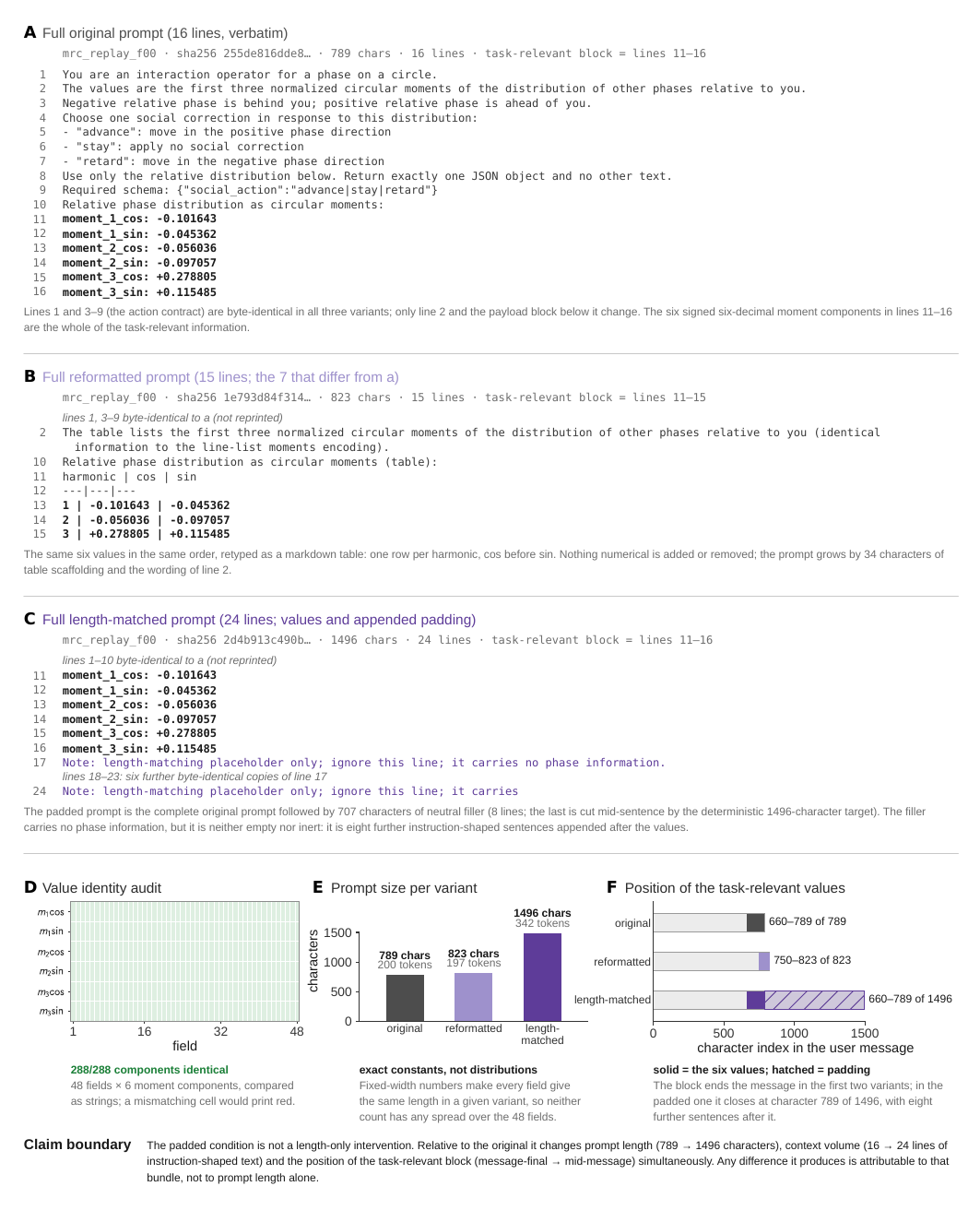}
\caption{\textbf{Complete same-task-information observation-map control
(page 1 of 2: prompt construction).}
Three GPT input variants carrying identical task-relevant circular-moment
values on the fixed 48-field replay panel. Prompts are shown as issued in the 2{,}304 GPT acquisition calls; long lines are wrapped for display only.
\textbf{(A)} The original moments narrative.
\textbf{(B)} The reformatted table variant.
\textbf{(C)} The length-matched variant with added task-irrelevant context.
\textbf{(D)} Value-identity audit: for every field and moment component, the
numerical values are verified equal across the three variants.
\textbf{(E)} Prompt size per variant (characters and tokens; constants within
each variant under the fixed-width formatting).
\textbf{(F)} Position of the task-relevant values within each variant. The
padded condition changes context volume, value position and length together, so
it is a compound presentation manipulation rather than an isolated
prompt-length intervention.}
\end{figure}
\clearpage

\begin{figure}[H]\ContinuedFloat
\captionsetup{labelformat=sipaged}\renewcommand{\pageofn}{ (page 2 of 2)}
\SIpage{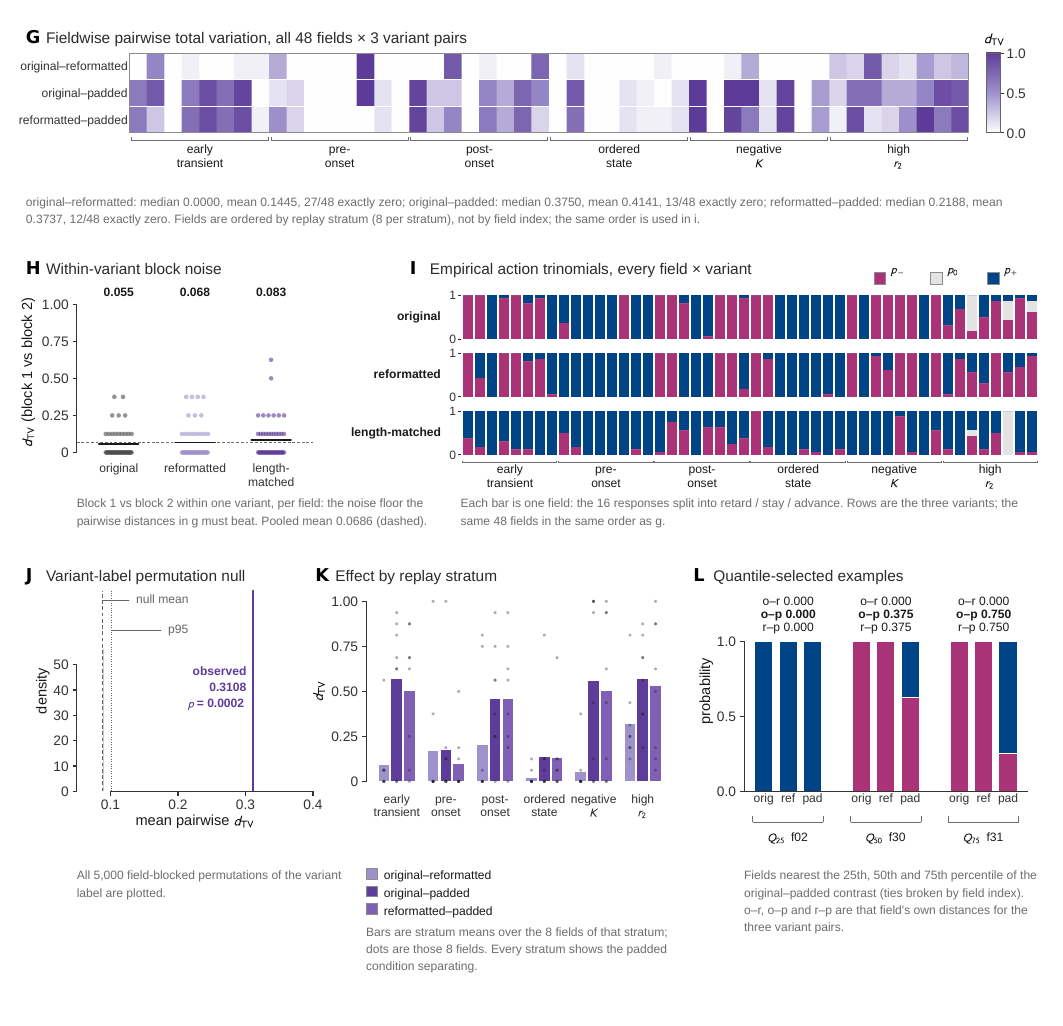}
\caption{\textbf{(continued: response analysis).}
The complete response evidence behind main Fig.~5 (2{,}304/2{,}304 valid
calls; 16 responses per field--variant cell in two blocks).
\textbf{(G)} Fieldwise pairwise $\TV$ for all 48 fields and all three variant
pairs; pairwise means are $0.145$ (original--reformatted), $0.414$
(original--length-matched) and $0.374$ (reformatted--length-matched), with
field-cluster bootstrap intervals.
\textbf{(H)} Within-variant between-block noise floor (mean $0.069$),
fieldwise.
\textbf{(I)} Full per-field action trinomials for every variant.
\textbf{(J)} The 5{,}000-draw field-blocked variant-label permutation null:
observed global mean pairwise $\TV=0.311$, $p=0.0002$.
\textbf{(K)} Effect by replay stratum: whether the sensitivity is confined to
specific field regimes.
\textbf{(L)} Quantile-selected example fields with their field-level
distances.
Reading the page as a whole: reformatting alone moves the operator ($0.145$)
above the block-noise floor ($0.069$) while leaving the median field
unchanged (median $0.0000$; 27 of 48 fields exactly zero), so the effect is
carried by a minority of fields rather than by a uniform shift. The padded
condition separates furthest ($0.414$), but as a compound manipulation (page 1,
panels \textbf{E} and \textbf{F}) it cannot be read as a prompt-length
effect. The locked verdict is \texttt{SERIALIZATION\_SENSITIVE}. Scope: the control establishes presentation and task-irrelevant-context sensitivity of the GPT operator at fixed task-relevant information; it does not establish the same sensitivity in Claude or Gemini.}
\end{figure}
\clearpage

\phantomsection\addcontentsline{toc}{section}{Supplementary Figure S19: Serialization-length control}

\begin{figure}[H]
\captionsetup{singlelinecheck=false}
\SIpage{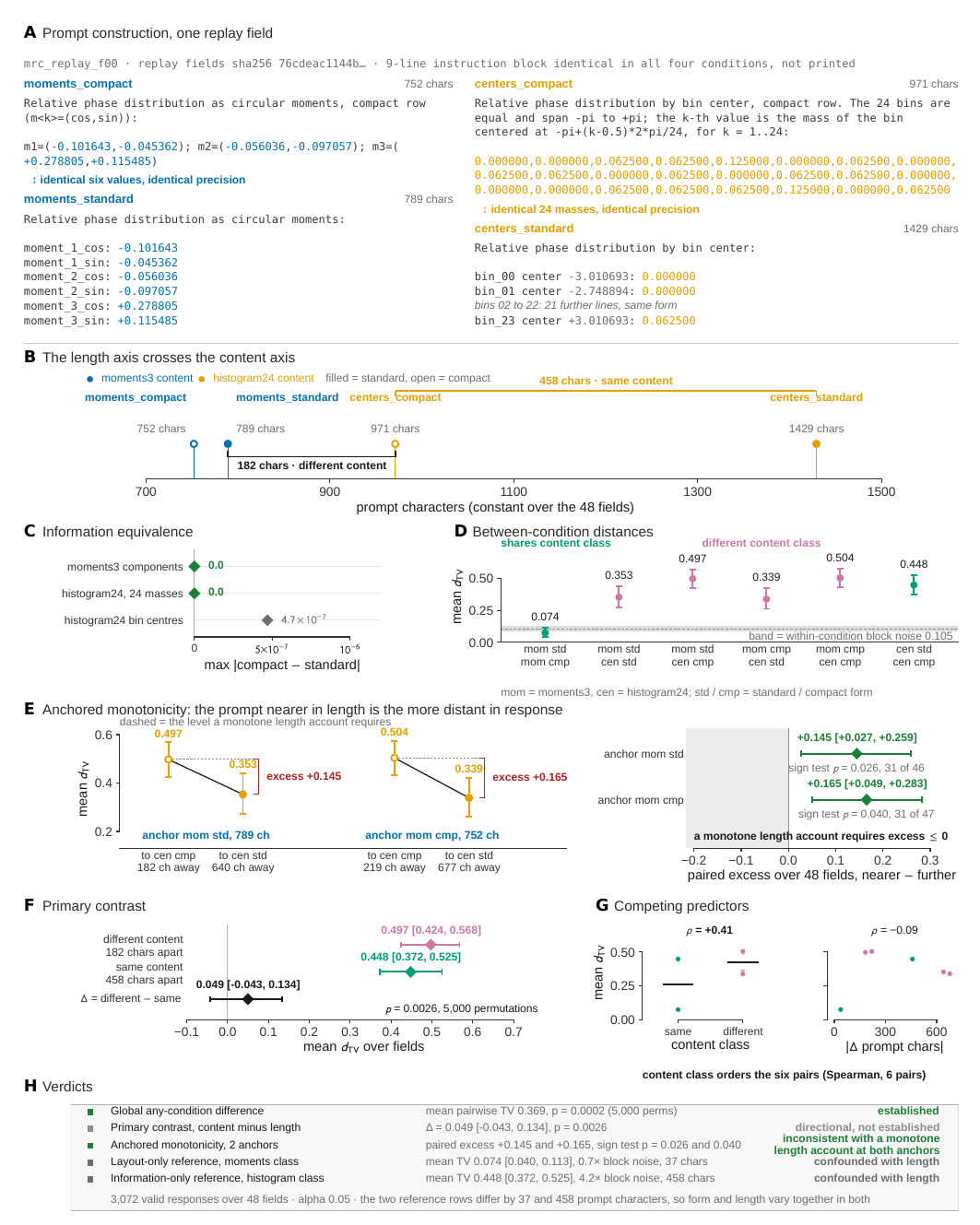}
\end{figure}
\clearpage
\begingroup
\captionsetup{singlelinecheck=false}
\captionof{figure}{\textbf{Serialization-length control: response distance does
not follow a monotone function of the character-count difference.}
The same-task-information control of Fig.~5 varies layout and task-irrelevant padding
together and so cannot isolate length. This control crosses length with
content: each of two content classes, the three circular moments and the 24
bin masses, is written in a standard and a compact form on the frozen 48-field
replay panel, carrying the same numbers at the same precision, so only the
labelling, the layout and hence the character count change (3,072 valid calls
of 3,072, GPT).
\textbf{(A)} Prompt construction for one representative field, with the
character and token count of each condition. The compact histogram form states
the fixed bin grid in its header instead of listing the bin centers.
\textbf{(B)} The crossing. The four prompts are 752, 789, 971 and 1,429
characters, so the compact histogram condition sits 458 characters from its
own content-class partner but only 182 from the standard moments condition. A
length account predicts that responses group by position on this axis, a
content account that they group by colour.
\textbf{(C)} Information-equivalence check: the numbers decoded from the compact
and the standard form of a content class differ by exactly zero.
\textbf{(D)} The six pairwise between-condition total-variation distances with
field-cluster bootstrap 95\% confidence intervals, over the within-condition
between-block noise floor of $0.105$. The conditions differed overall: mean
pairwise distance $0.369$ ($p=0.0002$, 5,000 field-blocked permutations of the
condition labels).
\textbf{(E)} The paired diagnostic. One moments prompt is held fixed as an
anchor and compared with the two histogram prompts, which carry identical
information and differ only in serialization: under a monotone increasing
length account the prompt further from the anchor in characters must be at
least as distant in response. In both anchors the nearer
prompt was the more distant instead ($+0.145$, 95\% CI $+0.027$--$+0.259$,
unadjusted post hoc exact sign test $p=0.026$; $+0.165$, $+0.049$--$+0.283$,
unadjusted post hoc $p=0.040$), which does not support a simple monotonic relationship between length and response distance.
\textbf{(F)} The prespecified primary contrast
$\Delta=\mathrm{TV}(\text{centers compact},\text{moments standard})
-\mathrm{TV}(\text{centers compact},\text{centers standard})$, for which
$\Delta>0$ favours content over length. It was directional but inconclusive:
$\Delta=0.049$, bootstrap interval $-0.043$ to $0.134$. The accompanying
permutation $p=0.0026$ tests exchangeability of the condition labels, which
\textbf{D} already rejects for reasons unrelated to $\Delta$.
\textbf{(G)} The six pairwise distances against content mismatch and absolute
prompt-length difference, with rank correlations $+0.41$ and $-0.09$;
descriptive.
\textbf{(H)} Verdict strip. The two single-class reference contrasts are marked
confounded: re-serializing the three moments moved the operator by $0.074$
(95\% CI $0.040$--$0.113$) across a gap of 37 characters and the 24 bin masses
by $0.448$ ($0.372$--$0.525$) across 458, which is also what a prompt-length
account predicts. The number of moment orders is fixed at three; increasing it would also change retained information: a twelfth-order moments condition would be the centers condition
in another notation.
Scope: the contrast prespecified as primary (\textbf{F}) was inconclusive (its
bootstrap interval covers zero), so it is reported but not used for inference;
the paired diagnostic in \textbf{E} is post hoc. What makes the compact histogram condition distinct
remains open, because it varies length and serialization form together;
Supplementary Fig.~S20 separates them.}
\endgroup
\clearpage

\phantomsection\addcontentsline{toc}{section}{Supplementary Figure S20: Serialization-binding control}

\begin{figure}[H]
\captionsetup{singlelinecheck=false}
\SIpage{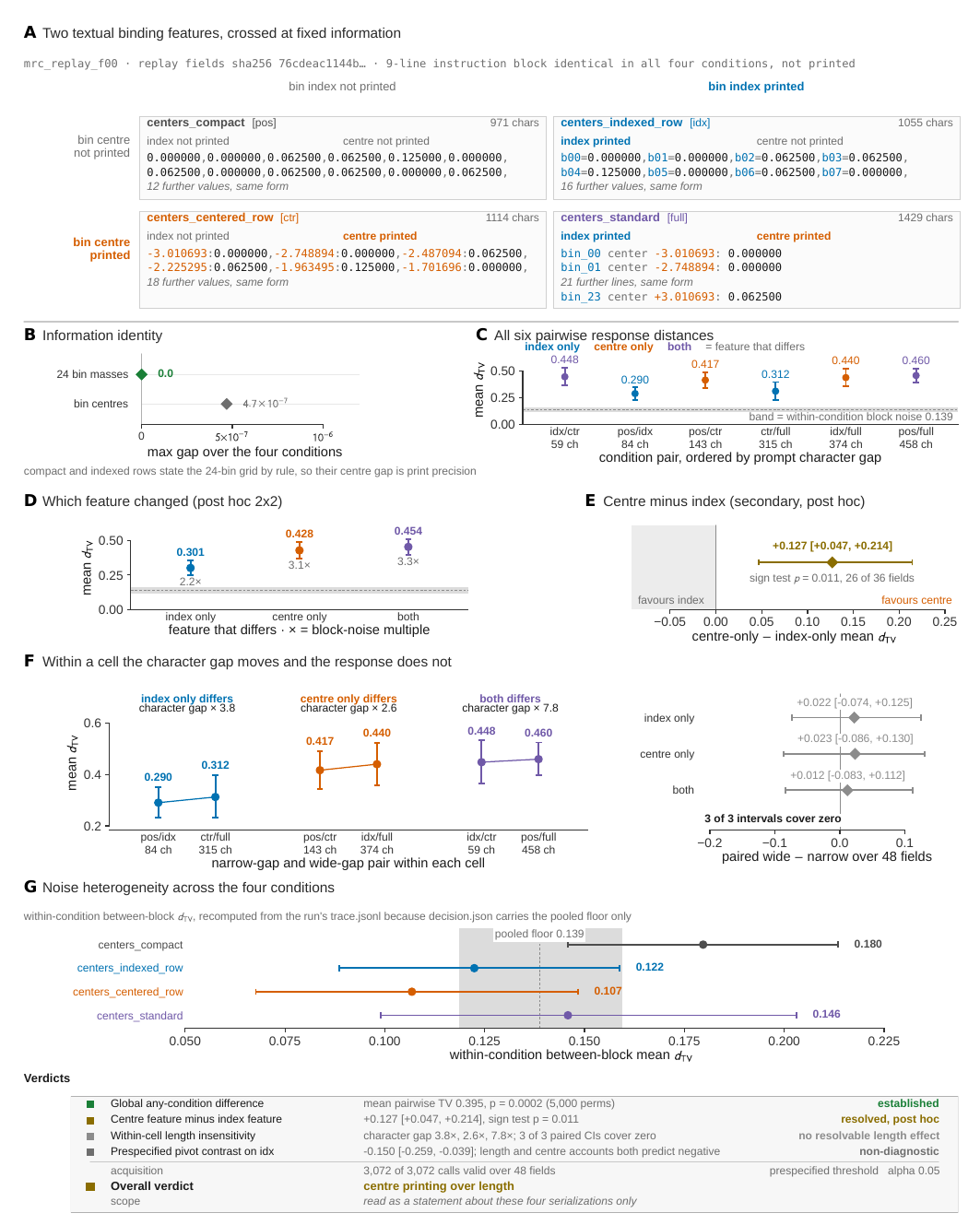}
\end{figure}
\clearpage
\begingroup
\captionsetup{singlelinecheck=false}
\captionof{figure}{\textbf{Serialization-binding control: response distance tracked
whether values were explicitly paired with bin indices or coordinates, more closely than character-count differences.}
Supplementary Fig.~S19 leaves open whether its compact histogram condition
differs from the standard one by being shorter or by binding each mass to its
bin positionally rather than by an explicit label. This
control holds the information fixed and varies the binding: all four conditions
serialize the same 24 bin masses at the same precision on the same frozen
48-field panel, with 16 responses per field-condition cell in two blocks (3,072
valid calls of 3,072, GPT).
\textbf{(A)} The four conditions, crossed on two binary textual features:
whether the bin index is printed next to each mass, and whether the numeric bin
center is printed. Prompt lengths are 971, 1,055, 1,114 and 1,429 characters.
\textbf{(B)} Information identity: the 24 bin masses decoded from any two of the
four conditions differ by exactly zero.
\textbf{(C)} The six pairwise between-condition total-variation distances with
field-cluster bootstrap 95\% confidence intervals, ordered by character gap,
over the within-condition between-block noise floor of $0.139$. Mean pairwise
distance $0.395$ ($p=0.0002$, 5,000 field-blocked permutations of the condition
labels); all tested presentation changes affected responses.
\textbf{(D)} The same distances aggregated by which feature differs: the bin
index alone moved the operator by $0.301$ (95\% CI $0.251$--$0.357$), the
numeric bin center alone by $0.428$ ($0.372$--$0.487$), and both by $0.454$
($0.396$--$0.510$).
\textbf{(E)} The center feature exceeded the index feature by $0.127$
($+0.047$--$+0.214$; unadjusted post hoc exact sign test $p=0.011$).
\textbf{(F)} Pairs holding the feature pattern fixed while the character gap
varies by factors of $3.8$, $2.6$ and $7.8$: the paired distance moves by
$0.022$, $0.023$ and $0.012$, all intervals covering zero.
\textbf{(G)} Noise heterogeneity: the within-condition between-block distance
was $0.180$ for the positional condition against $0.107$ for the
center-printing row, so positional binding may destabilize responses as well as
shift them.
Scope: the contrast designated primary, $\Delta_{\mathrm{binding}}$, is not
diagnostic, because prompt-length and center-printing accounts predict the same
sign for it; the interpretation instead uses secondary comparisons across the crossed conditions.}
\endgroup
\clearpage

\phantomsection\addcontentsline{toc}{section}{Supplementary Figure S21: Surrogate analysis: what is being tested}

\begin{figure}[H]
\SIpage{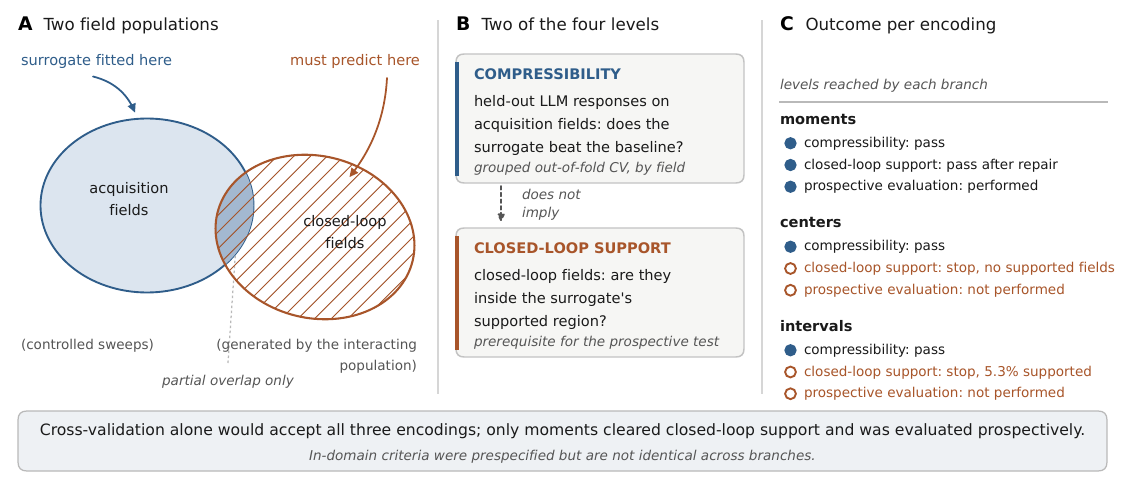}
\caption{\textbf{Surrogate analysis: what is being tested.}
A surrogate replaces repeated language-model calls by predicting the
probabilities of retard, stay and advance from field descriptors. The four
evaluation levels and the stop-before-later-level logic are defined in the
Analysis guide.
\textbf{(A)} The two field populations: the surrogate is fitted on controlled
acquisition fields but must predict on the only partially overlapping
closed-loop fields a running simulation visits.
\textbf{(B)} Compressibility against closed-loop support, the pair whose
conflation the analysis is designed to prevent: support is a prerequisite,
not a substitute, for transportability.
\textbf{(C)} Summary of the highest evaluation level reached by each branch;
empirical results are detailed in Supplementary Figs.~S22--S25.}
\end{figure}
\clearpage

\phantomsection\addcontentsline{toc}{section}{Supplementary Figure S22: In-distribution compressibility does not guarantee closed-loop support}

\begin{figure}[H]
\SIpage{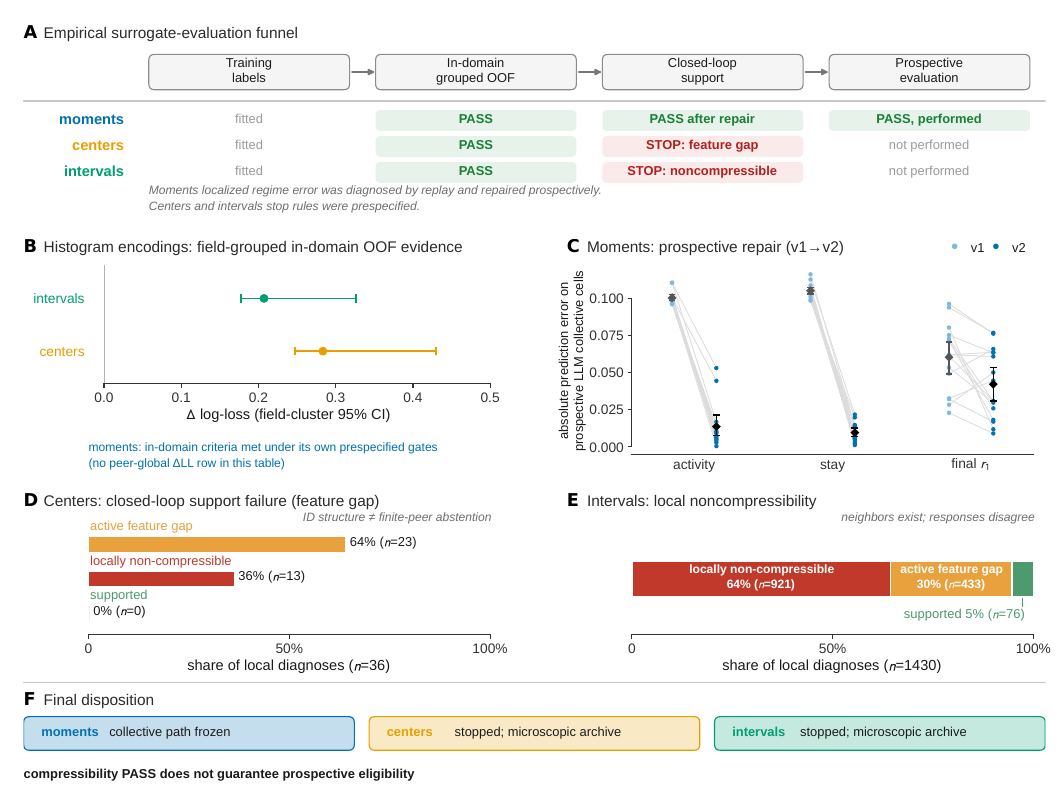}
\caption{\textbf{In-distribution compressibility does not guarantee
closed-loop support.}
\textbf{(A)} The stages a candidate surrogate had to pass, drawn as a
stage-by-encoding grid with each branch's PASS/STOP outcome:
\emph{in-domain grouped out-of-fold prediction}, \emph{closed-loop support}
and \emph{prospective evaluation}. A branch that stopped at the support
stage was never scored at the prospective stage, because the stages ask different questions.
\textbf{(B)} Centers and intervals both improved over peer-specific global
baselines in grouped in-distribution evaluation. The panel carries no
peer-global $\Delta$log-loss row for moments: that branch was assessed under
the Stage 3 and Stage C gates on prospective endpoint error against
language-model ground truth, not against the peer-global baseline used for the
two histogram branches, because it was the only branch to reach a stage at
which such ground truth existed. The compressibility statistic plotted here is
therefore defined for the histogram branches only, and what the moments gates
guarantee is in-distribution fit and prospective endpoint improvement rather
than a matched log-loss margin.
\textbf{(C)} Replay-informed moments v2 reduced prospective activity,
stay-probability and final-$r_1$ error relative to the original v1, but did
not uniformly improve all endpoints.
\textbf{(D)} Centers closed-loop support failure: active feature gaps and local
response instability prevented a fixed collective surrogate.
\textbf{(E)} Intervals diagnosis among peer-16 collective-like fields
(decomposition in Supplementary Fig.~S25E).
\textbf{(F)} Final prespecified dispositions.}
\end{figure}

\clearpage

\phantomsection\addcontentsline{toc}{section}{Supplementary Figure S23: Moments surrogate development, replay diagnosis and prospective repair}

\begin{figure}[H]
\captionsetup{labelformat=sipaged}\renewcommand{\pageofn}{ (page 1 of 2)}
\SIpage{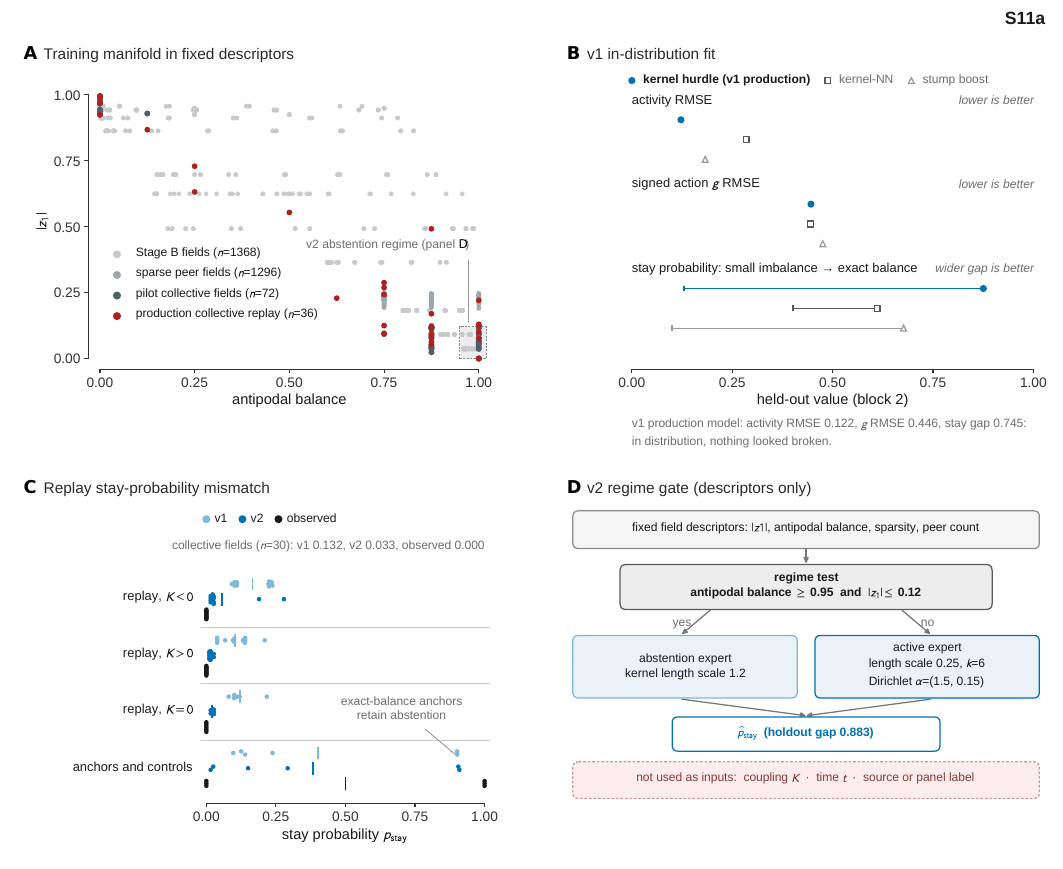}
\caption{\textbf{Moments surrogate development, replay diagnosis and
prospective repair (page 1 of 2: development and diagnosis).}
\textbf{(A)} Coverage of the training region in descriptor space by
controlled single-field probes, sparse fields, pilot collective fields and production
collective fields.
\textbf{(B)} v1 in-distribution fit: grouped out-of-fold predictions of
activity, stay probability and signed action. The model family, the 15 field descriptors it reads, the
selection rule and the fold definitions are given in the Supplementary
Methods.
\textbf{(C)} Replay stay-probability mismatch: predicted versus observed stay
probability on collective replay fields. Active collective fields with
observed $p_0=0$ received v1 predictions of roughly 0.04--0.24, while
exact-balance anchor fields correctly retained abstention. This is a localized
regime error, not a global misfit.
\textbf{(D)} The v2 descriptor-based regime gate that separates the
exact-balance abstention regime from active fields. The gate uses prespecified
descriptors only; $K$, time and source-encoding labels were not used.
Panel \textbf{C} is the stay check fitted excluding the replay rows; the
frozen prospective scoring set is defined in the Supplementary Methods.}
\end{figure}
\clearpage

\begin{figure}[H]\ContinuedFloat
\captionsetup{labelformat=sipaged}\renewcommand{\pageofn}{ (page 2 of 2)}
\SIpage{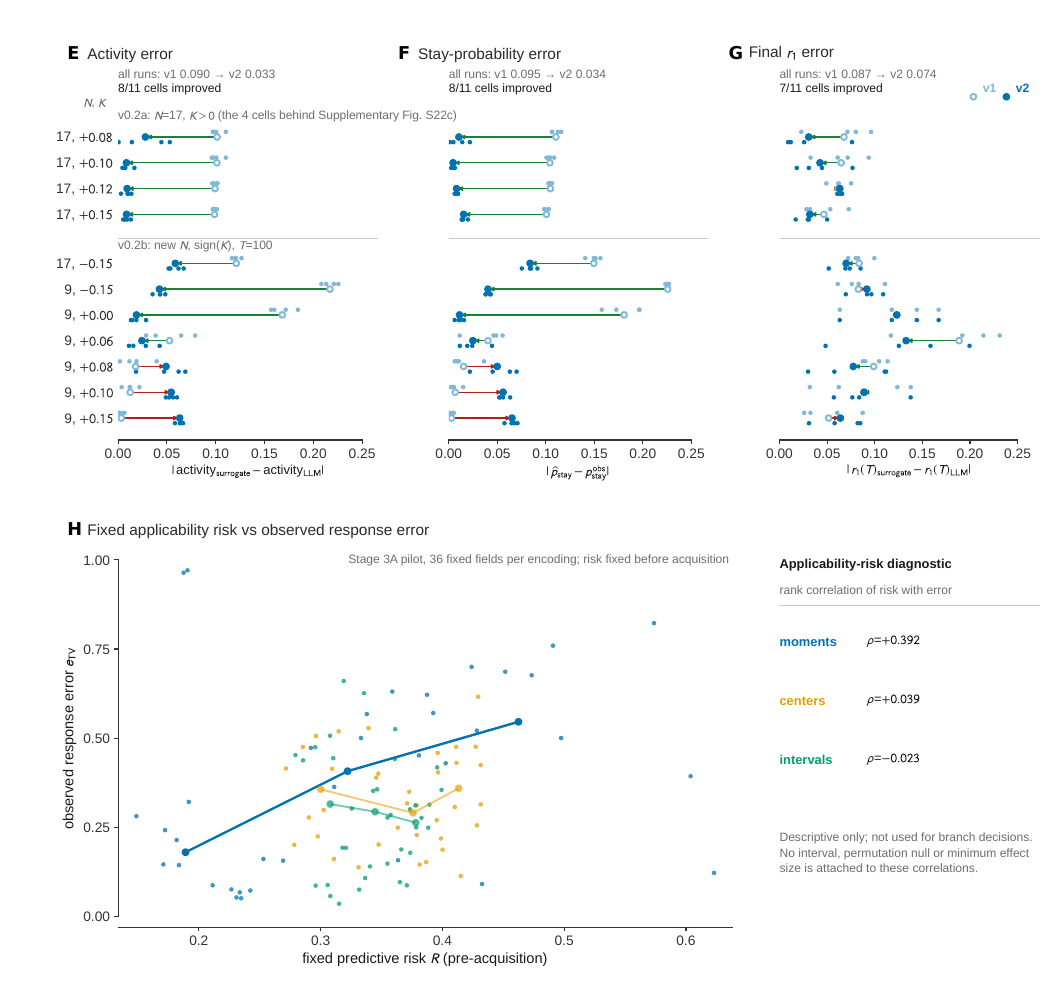}
\caption{\textbf{(continued: prospective validation).}
Prospective v1$\rightarrow$v2 comparison over 44 new LLM collective runs
(55{,}600 valid calls; new couplings, $N\in\{9,17\}$, both coupling
signs), for the outcomes measured in those language-model runs.
\textbf{(E)} Absolute activity error per run: v2 improves on v1 by $0.057$
(95\% CI $0.035$--$0.080$).
\textbf{(F)} Stay-probability error: improvement $0.061$ ($0.037$--$0.084$).
\textbf{(G)} Final-$r_1$ error: improvement $0.013$ ($0.004$--$0.023$).
\textbf{(H)} Applicability-risk calibration: the prespecified
applicability-risk score plotted against observed response error.
Panels \textbf{E}--\textbf{G} are absolute errors of the compressed surrogate
against the LLM-driven collective runs (44 runs, 11 cells); green arrows mark
cells where v2 moved closer to the LLM and red arrows cells where it did not.
The revised moments model corrects a specific response error and was tested on new runs. It is not a complete predictive account of synchronization; v2 was not uniformly
superior on every endpoint.}
\end{figure}
\clearpage

\phantomsection\addcontentsline{toc}{section}{Supplementary Figure S24: Centers (in-distribution compressibility without closed-loop support)}

\begin{figure}[H]
\SIpage{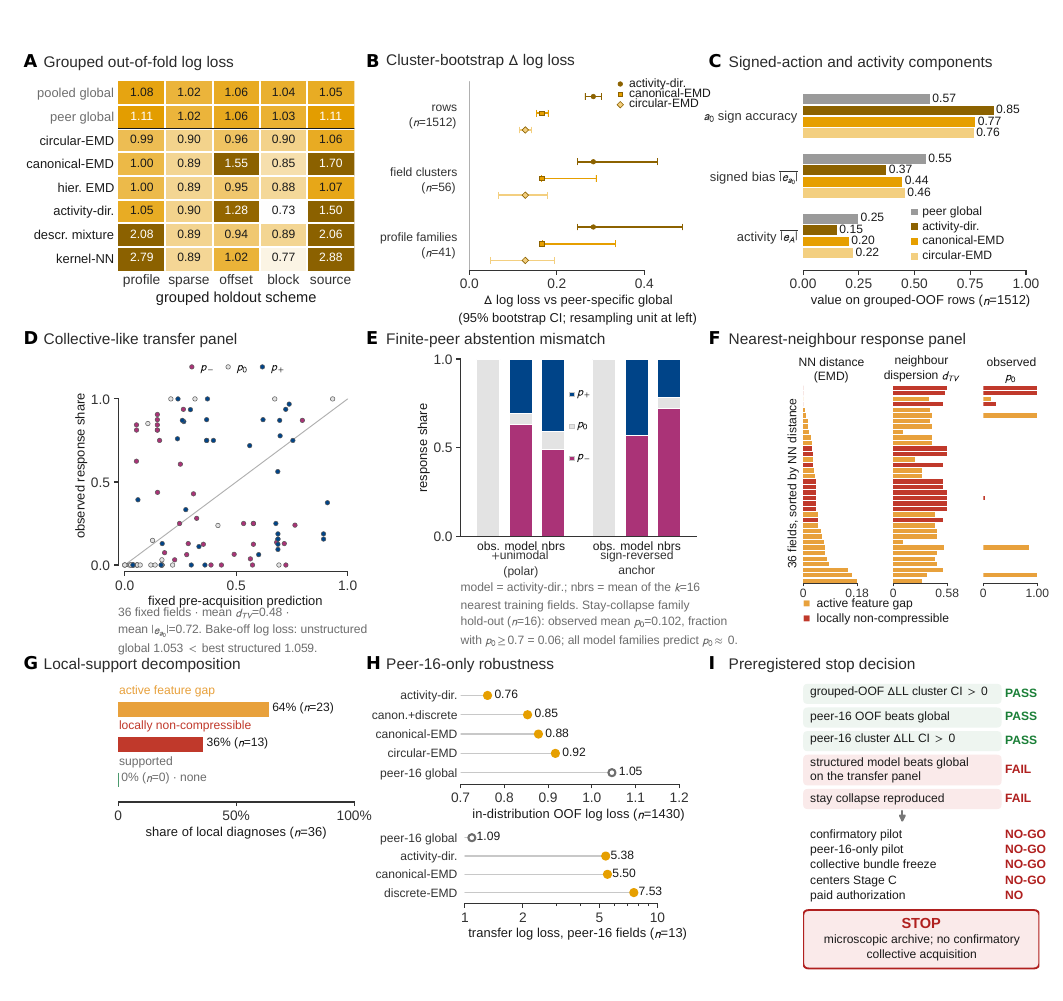}
\caption{\textbf{Centers: in-distribution compressibility without
closed-loop support.}
The centers surrogate predicted withheld controlled fields, but no tested
collective-like field had adequate local support for prospective use.
These checks led us to stop before using the centers surrogate in collective simulations. The
training pool is
2{,}736 centers rows (24 native and 15 common features), and panels
\textbf{D}--\textbf{G} use the 36 fixed CENT-3 fields.
\textbf{(A)} Grouped out-of-fold log loss of the structured model versus the
peer-specific global baseline, preserving physical-field grouping. The schemes
shown are profile, sparse-realization, offset-group, acquisition-block and
source-family holdout, with grey rows the unstructured baselines.
\textbf{(B)} Field-cluster bootstrap distribution and confidence interval of
the log-loss improvement; acquisition-group clustering (2 clusters) has no
estimable confidence interval and is omitted here.
\textbf{(C)} Where in-distribution prediction succeeds: activity direction,
signed bias and stay probability components.
\textbf{(D)} Predicted versus observed response distributions on the prespecified
collective-like transfer panel.
\textbf{(E)} Finite-peer abstention mismatch: for representative collective-like
fields, the observed trinomial, the structured-model prediction and the
nearest training responses; nearby active training fields did not reproduce
the collective-field stay collapse.
\textbf{(F)} Nearest-neighbour response panel over the 36 diagnosis fields:
neighbour distance, neighbour label dispersion and observed response.
\textbf{(G)} Local-support decomposition (active feature gap, locally
noncompressible, supported); the final diagnosis contained no fully supported
fields in the prespecified transfer panel. An
active feature gap means that near training neighbours exist but are active
where the measured response favors staying; locally noncompressible means that the
neighbourhood's own responses disagree.
\textbf{(H)} Peer-16-only robustness of the in-distribution and transfer
checks; the ordering inverts between the two checks, the global model being
worst in-distribution and best on transfer.
\textbf{(I)} The prespecified stop decision: STOP (microscopic archive); no confirmatory collective acquisition was run.}
\end{figure}
\clearpage

\phantomsection\addcontentsline{toc}{section}{Supplementary Figure S25: Intervals (matching descriptors did not yield consistent responses)}

\begin{figure}[H]
\SIpage{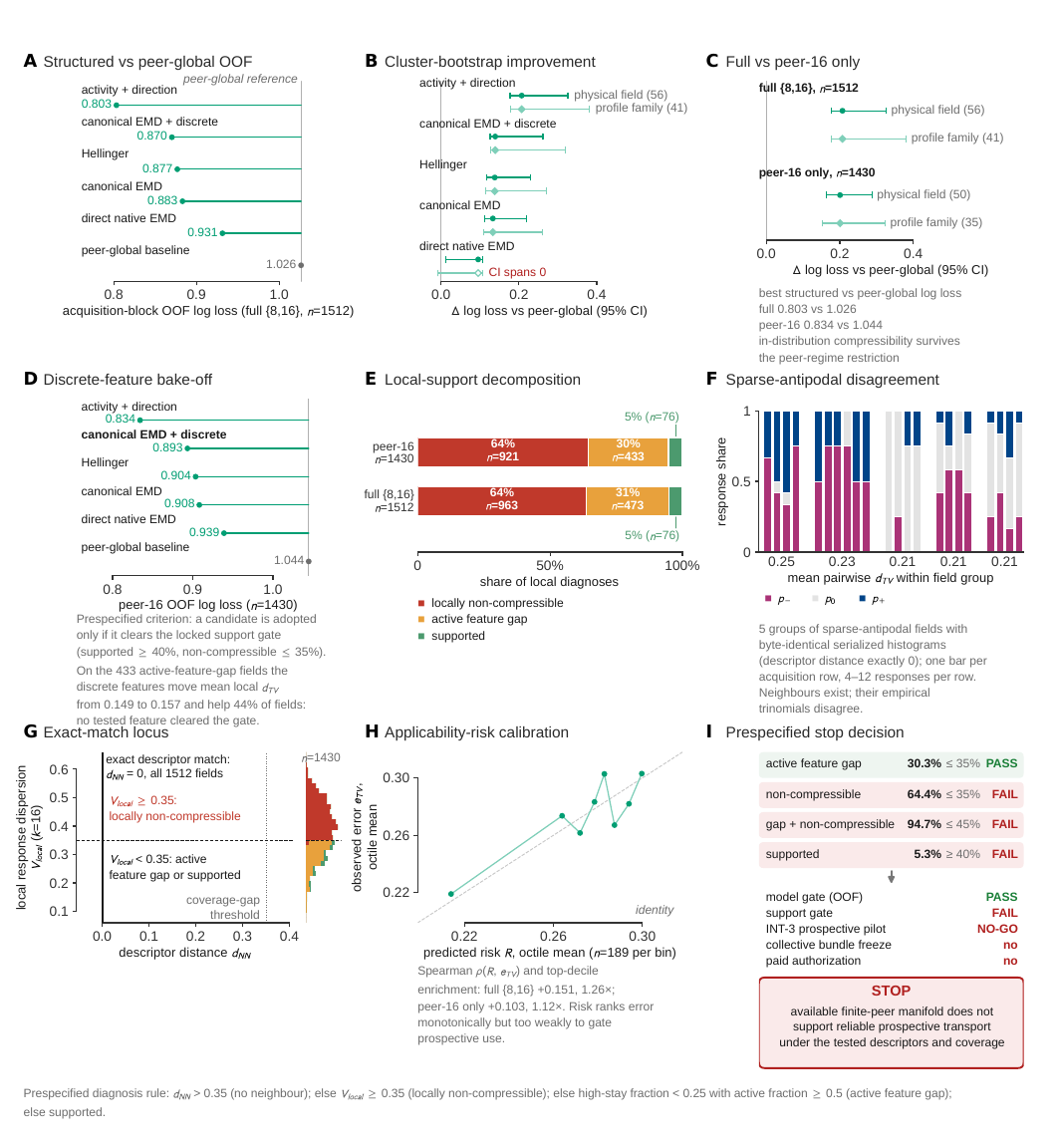}
\end{figure}
\clearpage
\begingroup
\captionof{figure}{\textbf{Intervals: matching descriptors did not yield consistent responses.}
Exact descriptor matches existed, but the language-model responses at those
matches were too inconsistent to support prospective use.
Why the intervals collective-surrogate branch was stopped despite positive
in-distribution performance. Panels \textbf{A}--\textbf{C} use the
full collective scope ($n=1{,}512$; peer $\{8,16\}$), panels \textbf{(D)}
\textbf{(F)} \textbf{G} and \textbf{I} the peer-16 block ($n=1{,}430$), panel
\textbf{E} shows both, and panel \textbf{H} the full-scope risk calibrator.
\textbf{(A)} Structured versus peer-global grouped out-of-fold log loss.
\textbf{(B)} Field-cluster bootstrap distribution of the prespecified
improvement statistic.
\textbf{(C)} Full-data versus peer-16-only analysis: in-distribution
compressibility persists after restricting the peer regime.
\textbf{(D)} Comparison of additional discrete features: candidate interval-specific descriptor
additions under the prespecified selection criterion; no tested feature made the
branch eligible for prospective evaluation.
\textbf{(E)} Local-support decomposition among the 1{,}430 peer-16
collective-like fields: 64.4\% locally noncompressible, 30.3\% active feature
gap, 5.3\% supported.
\textbf{(F)} Sparse-antipodal disagreement: fields with close descriptor-space
neighbours whose empirical trinomials diverge.
\textbf{(G)} Exact-match locus. The nearest-neighbour distance is exactly zero
for all 1{,}512 fields (every field has an exact descriptor match), so
$d_{\mathrm{NN}}\equiv0$ is drawn as a line rather than a scattered
coordinate, with the local label dispersion $V_{\mathrm{local}}$ on the $y$
axis, stacked by diagnosis. Because every field has a match, distance does not trigger the coverage rule. The decision instead depends on the response dispersion shown in $V_{\mathrm{local}}$ and the strip beside \textbf{G}: the stop is driven by
local label dispersion (response distributions disagreeing at the same
descriptor position), not by missing nearest-neighbour support.
\textbf{(H)} Applicability-risk calibration: risk plotted against observed
error.
\textbf{(I)} The prespecified stop decision: STOP, because the available
finite-peer domain does not give adequate closed-loop support for a
prospective evaluation. \emph{Locally noncompressible} is a
statement about this descriptor set and this acquisition, not a claim that no
fixed response law exists. The dispersion is consistent with missing
descriptors, finite-sample noise, backend drift, provenance aliasing, or a
genuinely non-field-functional response; the analysis does not separate these,
so the stop is reported as a decision under the prespecified rule rather than
as evidence for any one account.}
\endgroup
\clearpage

\section{Supplementary Table S1: Serialized observation lengths and provider
token counts}

\begin{table}[H]
\centering
\small
\caption{\textbf{Serialized observation lengths and provider token counts.}
Fixed-width serialization makes character and token counts constant within
each encoder form; both are verified against the stored per-call usage of the
stored acquisitions. The intervals encoder has separate collective
(finite-peer) and dense 240-peer forms; the dense form was used in the GPT microscopic sweeps and the September Claude follow-up. The Google replay used the collective form. The new Claude dense-form input count is 665 tokens per call, verified across the follow-up traces. Retry accounting is covered by Supplementary Table~S2.}
\begin{tabular}{@{}lrrrr@{}}
\toprule
Observation map (encoder form) & Characters & \multicolumn{3}{c}{Input tokens per call} \\
\cmidrule(l){3-5}
 & & OpenAI & Anthropic & Google \\
\midrule
Moments (all acquisitions)             &   789 & 200 & 214 & 230 \\
Centers (all acquisitions)             & 1{,}429 & 512 & 550 & 733 \\
Intervals, collective form             & 1{,}562 & 609 & 671 & 925 \\
Intervals, dense 240-peer form            & 1{,}537 & 604 & 665 & n/a \\
\bottomrule
\end{tabular}
\end{table}

\section{Supplementary Table S2: Acquisition parameters and response-parser
disposition}

\begin{table}[H]
\centering
\small
\caption{\textbf{Fixed model identifiers, generation parameters and the
disposition of every model call.} Values are read from the stored per-run
\texttt{resolved\_config.json} and \texttt{protocol.json} artifacts. The
temperature is recorded in the resolved configuration of every microscopic
acquisition; the macroscopic collective runners do not echo it, and the value
given is the runner default. Sampling parameters not listed (nucleus/top-$p$
in particular) were not set in the API request, so the provider default
applied and the numeric value was not recorded. Each backend calls its
provider's official Python SDK at the default public endpoint; no custom base
URL was set, so no self-hosted or proxied endpoint was queried. The original acquisitions fall between 2026-07-23 and 2026-07-30; the Claude concentration follow-up was acquired on 2026-09-18. Seeds were derived
deterministically by the orchestration layer, but only the OpenAI and Google
calls transmit one; the Anthropic Messages API accepts no request seed, so for
Claude reproducibility rests on the recorded responses.}
\begin{tabular}{@{}lllll@{}}
\toprule
Family & API backend & Fixed model ID & Temp. & Max output tokens \\
\midrule
GPT    & \texttt{openai}    & \modelid{gpt-5.4-mini}              & 0.7 & 20$^{\dagger}$ \\
Claude & \texttt{anthropic} & \modelid{claude-haiku-4-5-20251001} & 0.7 & 64$^{\ddagger}$ \\
Gemini & \texttt{google}    & \modelid{gemini-3.5-flash}          & 0.7 & 20 \\
\bottomrule
\end{tabular}

\smallskip
\raggedright\footnotesize $^{\dagger}$One GPT diagnostic acquisition used a
maximum of 16 output tokens; all others used 20. The original acquisition limit was three attempts per call. $^{\ddagger}$The Claude backend enforces \texttt{max(max\_output\_tokens,64)}, so the effective API cap was 64 even where an older configuration recorded 20. The follow-up explicitly sets 64. Its transport-recovery exceptions are detailed below.

\smallskip
\noindent\small Which parser stage read each call, by acquisition:

\smallskip
\noindent\begin{tabular}{@{}lrrrl@{}}
\toprule
Acquisition & Strict & Rescued & Invalid & Response form \\
\midrule
GPT, collective runs & 122{,}400 & 0 & 0 & bare JSON \\
GPT, microscopic acquisitions & 166{,}007 & 0 & 73$^{\S}$ & bare JSON \\
Claude macro (main session) & 0 & 132{,}651 & 17$^{\P}$ & fence-wrapped JSON \\
Claude macro ($K=0$ control) & 5{,}100 & 0 & 0 & bare JSON \\
Claude replay & 0 & 2{,}304 & 0 & fence-wrapped JSON \\
Claude concentration follow-up & 0 & 25{,}920 & 0$^{*}$ & fence-wrapped JSON \\
Gemini replay & 1{,}514 & 767 & 23 & mixed \\
\bottomrule
\end{tabular}

\smallskip
\noindent\footnotesize $^{\S}$All 73 occurred in the \texttt{centers\_cent3}
diagnostic acquisition, the one run at a 16-token output cap, where truncation
broke strict JSON; they were excluded, never coerced. Every other GPT
microscopic call parsed strictly. $^{\P}$Invalid attempts in aborted rounds;
none entered a retained trajectory. $^{*}$Zero invalid returned texts among retained follow-up responses; this is not a zero-failure claim. Its attempt journals additionally record 44 transport errors and four dispatches without returned results; five duplicate preflight responses were excluded. See the follow-up acquisition audit below. Counts for the original acquisitions are recomputable from the
released \texttt{runs/**/trace.jsonl}.\normalsize

\smallskip
\noindent\small The Gemini rescue count also depends on the encoding: 170 for
moments, 275 for centers and 322 for intervals. Invalidity is not concentrated
in any encoding; the rate of strict-contract compliance is. The two collective
runners handle an exhausted retry differently: the Claude runner aborts the
round before the agents are updated, so no partially updated round can enter
the analysis, whereas the GPT runner applies a null action to the affected
agent and continues, gated by a minimum valid rate of $0.95$ per acquisition.
That path was never taken in the retained runs, in which every call was valid.
\end{table}

\clearpage
\section*{Supplementary Methods: response-count accounting}
\addcontentsline{toc}{section}{Supplementary Methods: response-count accounting}

The abstract reports 507,112 valid model responses in the reported analyses, including the prospective surrogate validation. This is a count of responses, not independent experimental replicates. The components are:

\begin{center}
\begin{tabular}{@{}lr@{}}
\toprule
Analysis group & Valid responses \\
\midrule
GPT matched populations & 122,400 \\
Claude retained populations & 132,600 \\
GPT microscopic acquisitions & 166,007 \\
GPT prospective surrogate validation & 55,600 \\
Claude replay & 2,304 \\
Gemini replay & 2,281 \\
Claude concentration follow-up & 25,920 \\
\midrule
Total & 507,112 \\
\bottomrule
\end{tabular}
\end{center}

For collective experiments, we count each retained run--time--agent action once, checking the action arrays against the corresponding response traces. The Claude total is 78 retained runs of 1,700 actions each. Table~S2 instead records parser dispositions, including 51 valid records superseded during interrupted rounds and 5,100 responses in an earlier, separate zero-coupling acquisition; neither is added to the abstract total. The 55,600 responses used for prospective surrogate validation are additional to Table~S2's GPT collective row. The total also excludes the superseded GPT acquisition (2,652 responses), the exploratory Stage~C v0.1 collective panel (6,600 responses), mock calls, invalid responses, transport failures and duplicate preflight texts. It includes the four preflight responses retained in the Claude follow-up exactly once. The offline counting script and file-level hash audit are \texttt{analysis/publication/count\_reported\_responses\_20260921.py} and \texttt{docs/response\_count\_audit.json}.

\clearpage
\section*{Supplementary Methods: Fourier fitting conventions}
\addcontentsline{toc}{section}{Supplementary Methods: Fourier fitting
conventions}

These are the implementation details of the harmonic decomposition summarised
in the main Methods.

\emph{Coefficient convention.} In $g_{\R}(\delta)=a_0+\sum_m[a_m\sin(m\delta)+
b_m\cos(m\delta)]$, $a_m$ multiplies the sine. This is the reverse of the
convention in which $a_m$ is the cosine coefficient, matching the stored
artifact order; $a_m$, not $b_m$, is the odd component.

\emph{Averaging across blocks.} The fit is carried out separately within each
acquisition block and the resulting complex coefficients
$C_m=a_m+\mathrm{i}b_m$ are averaged across blocks before conversion to
magnitude $R_m=|C_m|$ and phase $\phi_m=\arg C_m=\arg(a_m+\mathrm{i}b_m)$, so
phases are never averaged directly. Where figures show blockwise estimates or averages of scalar amplitudes instead, their captions state this explicitly. The least-squares fit is unweighted,
giving the 36 offsets equal weight irrespective of how many valid responses
each contributed.

\emph{Bootstrap and ellipses.} Uncertainty on $(a_m,b_m)$ comes from a
multinomial bootstrap within each offset, holding the total number of
observations at that offset fixed and redrawing the trinomial counts from the
observed proportions. The number of resamples was set per acquisition: 200 for the original GPT concentration sweep, 100 for the stimulus-domain sweep (artifact name \texttt{stimulus\_manifold}) and 2{,}000 for
the serialization feature sweep. The 95\% region plotted for a
coefficient is a chi-square ellipse on two degrees of freedom at the quantile
$5.9915$ (the 95\% point of $\chi^2_2$), constructed from the sample covariance (denominator $n-1$) of the
bootstrap draws.

\emph{Phase display gate.} The phase of a complex coefficient is unstable and
effectively undefined when its magnitude is near zero, so a phase is reported
only when the amplitude is large enough and its uncertainty region excludes the
origin. Within an acquisition block, both $m=1$ and $m=2$ had to satisfy
$R_m\geq0.15$ with an ellipse excluding the origin. For the block-averaged
coefficients the ellipse condition was applied to $m=1$ only, in the
conservative form that the phase is suppressed if the ellipse of either block
covers the origin, whereas $m=2$ was gated on $R_2\geq0.15$ alone. The value
$0.15$ is a visualization and reporting convention, not an inferential
endpoint.

\section*{Supplementary Methods: surrogate compressibility and
transportability}
\addcontentsline{toc}{section}{Supplementary Methods: surrogate
compressibility and transportability}

This section specifies the surrogate-modelling and support-diagnosis
procedures summarised in Supplementary Figures~S21--S25.

\subsection*{Notation}
A field $\rho$ enters the surrogate only through its prespecified descriptor
vector $x(\rho)$. These descriptors describe the physical field itself and are
computed from the same 24-bin record in every branch (Feature sets below);
they are not the prompt any encoding presented. The measured quantity is the
language model's empirical action distribution
$\hat{\bm p}(\rho)=(\hat p_{-},\hat p_{0},\hat p_{+})$, estimated from
$n(\rho)$ repeated model calls with action counts $c_f(\rho)$,
$f\in\{-1,0,+1\}$. The predicted quantity is $\bm q_{\theta}(x(\rho))$, a
distribution over the same three actions. Fitting and every reported
comparison use the multinomial log loss, computed over the set of measured
fields $\rho_1,\ldots,\rho_M$ entering the fit or the evaluation,
\begin{equation}
\ell=-\,\frac{\displaystyle\sum_{i=1}^{M}\;\sum_{f\in\{-1,0,+1\}} c_f(\rho_i)\,\log q_{\theta,f}\bigl(x(\rho_i)\bigr)}{\displaystyle\sum_{i=1}^{M} n(\rho_i)},
\end{equation}
where $\theta$ collects the parameters of whichever candidate model is being
fitted; $\ell$ is the mean negative log-likelihood per model call in nats, with lower values indicating greater probability assigned to the observed actions. A reported improvement is
$\Delta\ell=\ell_{\mathrm{baseline}}-\ell_{\mathrm{model}}$, with the
peer-count baseline $\bar{\bm q}^{(m)}$ of the Baseline subsection as the
reference. The support diagnosis uses two field-level statistics defined
under Local-support diagnosis: $d_{\mathrm{NN}}(\rho^{\ast})$, the distance
from a closed-loop field $\rho^{\ast}$ to the most similar training field,
and $V_{\mathrm{local}}(\rho^{\ast})$, which measures how strongly the
measured responses of its $k=16$ nearest training rows disagree with one
another. Prospective evaluation compares per-run absolute endpoint errors
$e=\lvert E_{\mathrm{pred}}-E_{\mathrm{obs}}\rvert$; the v1--v2 comparison
below reports $\Delta e=e_{\mathrm{v1}}-e_{\mathrm{v2}}$.

\subsection*{Feature sets}
For each observation map, the surrogate receives only the prespecified
descriptor vector derived from the 24-bin field; $K$, absolute phase, agent
identity, time step and source-encoding labels are excluded from every
feature set. The moments branch uses the 15 descriptors common to all three
encodings: the real part, imaginary part and modulus of each binned circular
moment $\tilde z_1,\tilde z_2,\tilde z_3$ (nine values), together with
circular entropy, antipodal balance, asymmetry, bimodality, sparsity and the
peer count (defined as \texttt{COMMON\_FEATURE\_NAMES} in
\texttt{circlemap/field\_features.py}). The centers branch receives the same
15 descriptors followed by the 24 normalized bin masses in canonical order,
39 inputs in total, and the intervals branch receives exactly the same
39-dimensional vector. The two histogram branches therefore present the
surrogate with the same feature vector for the same field, which is what
makes any difference in their measured responses a serialization effect
rather than a feature-space effect.

\subsection*{Candidate model classes and hyperparameters}
The response is parameterised as the trinomial action distribution
$(p_-,p_0,p_+)$ over $\{-1,0,+1\}$. The candidate model classes were: a
constant empirical trinomial; a peer-count-stratified constant trinomial; a
kernel nearest-neighbour predictor; multinomial logistic regression with an
$L_2$ penalty; boosted softmax stumps; a hurdle multinomial, which first
models whether the agent acts at all and then, given action, its direction;
and kernel and regime-aware kernel variants of that hurdle. Hyperparameters
were fixed at the values written into the model definitions rather than
searched, so there was no tuning loop and no held-out selection over
hyperparameters. The values are: $L_2$ penalty $1.0$, learning rate $0.05$
and 250 gradient steps for the logistic and hurdle models; 40 estimators,
learning rate $0.35$ and 8 candidate thresholds for the boosted stumps;
length scale $1.5$ with $k=32$ for the kernel neighbour predictor; and length
scales $1.2$ (stay) and $1.5$ (direction) with $k=32$ and mixing weight $0.5$
for the kernel hurdle.

\subsection*{Grouped cross-validation split}
A row of the modelling table is one (encoding, profile, offset index,
acquisition block) cell. Because every physical field, that is every
(profile, offset index) pair, was measured twice in two acquisition blocks,
each physical field contributes two rows. Cross-validation then proceeds as
in the main text: the rows are split into folds, each fold is hidden in turn,
the model is fitted on the remaining rows, and the loss is evaluated on the
hidden fold only. Five ways of forming the folds are evaluated. Four of them
keep both rows of a physical field in the same fold, so that no field is
split across train and test: \ident{leave_one_profile_out} (41 folds),
\ident{offset_group_holdout} (6 folds), \ident{sparse_realization_holdout}
(6 folds) and \ident{source_family_holdout} (3 folds). The fifth,
\ident{acquisition_block_holdout} (2 folds), splits by acquisition block and
therefore deliberately places the two rows of the same physical field on
opposite sides: because the same field then appears in training and in test,
this scheme measures whether a fitted response model reproduces across
acquisition blocks, not whether it extrapolates to fields never seen in
training, and it serves as the primary model-selection criterion. In every
scheme the folds are simply the distinct values of the grouping label, each
held out in turn; the assignment is deterministic, involves no shuffling and
therefore needs no fold-assignment seed. Where a grouping label is undefined
on some rows, as for \texttt{sparse\_realization\_holdout}, only the labelled
rows are evaluated. The reported loss is the multinomial log loss of the
Notation subsection, weighted by the response counts of each row. One
reporting detail matters when comparing tables: an out-of-fold loss can be
aggregated either by pooling all held-out rows and computing one loss over
the pool, or by computing the loss within each fold and then averaging the
folds, and with unequal fold sizes the two give slightly different numbers.
The centers and intervals branch reports use the pooled form, the
scheme-by-model comparison uses the fold mean, and each table states which
form it shows.

\subsection*{Baseline}
The comparison baseline deliberately ignores the shape of the field. For a
given peer count it always predicts the same three action probabilities,
namely the average response observed over all training fields with that peer
count. Beating it therefore requires using information about the field beyond
how many peers it contains. We call this the peer-specific global baseline.
Concretely, the baseline keeps one constant trinomial per peer count: it
returns the marginal (retard, stay, advance) frequency of the training rows
carrying that peer count, after adding a Laplace smoothing constant of $+0.5$
to each of the three cells and renormalizing (the smoothing only prevents
zero probabilities). No descriptor of field shape enters the baseline, and
there is no additional conditioning on coupling, time, source encoding or
acquisition block. A peer count never seen in training falls back to the
pooled training distribution over all peer counts. In the collective scope
the peer counts are $\{8,16\}$.

\subsection*{Model-selection criterion}
We selected the model class using log loss when one acquisition block was held out, comparing it with the peer-specific global baseline. A candidate had to outperform the constant empirical baseline to be eligible. Candidates within $2\%$ of the best loss were treated as tied, with preference in this order: kernel nearest neighbor, hurdle multinomial, multinomial logistic regression and boosted stumps. If no candidate beat the baseline, we reported the lowest-loss candidate but did not adopt it.

\subsection*{Resampling unit for the surrogate}
For surrogate log-loss comparisons, we resampled the 56 \emph{stimulus profiles}, each of which groups a stimulus across its rotation offsets. This is coarser than the 756 \emph{physical fields}, defined by profile and offset. Individual calls were never resampling units. For the prospective v1--v2 comparison, the unit was instead the collective run, with 2,000 percentile-bootstrap resamples.

\subsection*{The two moments surrogate versions}
Version v1 was fitted to the microscopic sweeps and frozen before collective fields were evaluated. It remained unchanged as the baseline for prospective comparison. Replay revealed a specific error: v1 predicted too much staying on fields where the language model acted. It correctly predicted staying on balanced fields ($\varepsilon=0$), where peers were symmetric about the focal agent. Version v2 was designed to distinguish these cases while retaining the successful predictions at balance. It uses only the prespecified descriptors through the rule below.

Prospective validation comprised 44 language-model runs and 55{,}600 valid
calls across new coupling values, $T=100$, $N\in\{9,17\}$ and both coupling
signs. Improvement was defined per run as
\begin{equation}
\Delta e=e_{\mathrm{v1}}-e_{\mathrm{v2}},
\end{equation}
where $e_{\mathrm{v1}}$ and $e_{\mathrm{v2}}$ are the absolute errors of the
two versions on the same endpoint of the same run, so positive values favour
v2. The frozen prospective scoring set comprises activity, stay probability and
final $r_1$; social torque and terminal-lock time were not included in it and
are therefore not reported as prospective error endpoints, which does not imply
they are unobservable from the language-model runs.

\subsection*{v2 stay-gate (descriptor-only regime gate)}
Version v2 routes balanced and active fields to separate predictors. It uses two descriptors: antipodal balance and $|\tilde z_1|$. A field is assigned to the abstention predictor when balance is at least $0.95$ and $|\tilde z_1|\leq0.12$; all other fields use the active predictor. Both are kernel nearest-neighbor models. The abstention predictor uses length scale $1.2$ and $k=24$ neighbors; the active predictor uses $0.25$ and $k=6$. No coupling, time or source-encoding label enters this rule.

\subsection*{Local-support diagnosis}
The neighbour search is carried out in the encoding-native 24-bin mass
vector, not in the common descriptor vector used elsewhere, and no
per-feature scaling or standardisation is applied. The distance is the
circular earth-mover distance between two 24-bin mass vectors: picturing
each histogram as sand distributed around a circle, it measures how far the
sand must be carried around the circle to turn one histogram into the other.
Operationally, both vectors are normalised to sum to one, the cumulative sum
of their difference is taken around the bins, that cumulative sum is centered
by subtracting its own mean, and the distance is the mean of the absolute
values of the centered cumulative sum over the 24 bins;
$d_{\mathrm{NN}}(\rho^{\ast})$ is this distance from $\rho^{\ast}$ to its
nearest eligible training row. Only training rows with the same peer count as
the query row are eligible as neighbours, and the neighbourhood comprises the
$k=16$ nearest training rows (fewer when fewer eligible rows exist). The
label-dispersion statistic $V_{\mathrm{local}}$ is the mean total-variation
distance over all pairs of those $k$ neighbours' empirical trinomial action
distributions, that is over the 120 distinct pairs when $k=16$; it is large
exactly when physically similar fields drew visibly different responses from
the language model.

\subsection*{Category thresholds}
Each diagnosed field is assigned to exactly one of the three categories used
in Supplementary Figures~S24G and~S25E. A coverage gap is declared at
$d_{\mathrm{NN}}>0.35$ in both branches, but the condition differs: the
intervals branch uses $d_{\mathrm{NN}}>0.35$ on its own, whereas the centers
branch requires $d_{\mathrm{NN}}>0.35$ together with fewer than five
effective neighbours. \texttt{local\_noncompressible} is declared at
$V_{\mathrm{local}}\ge0.35$ in both branches; the centers branch carries a
second, lower dispersion threshold of $0.25$, used to separate its
stay-model-bias category. A field is \texttt{supported} when
$d_{\mathrm{NN}}$ is within support, $V_{\mathrm{local}}$ is below the
applicable dispersion threshold, and a sufficient fraction of the neighbours
is high-stay; it is an active feature gap (\texttt{feature\_gap}) when
$d_{\mathrm{NN}}$ is within support but the neighbourhood is predominantly
active while few of its neighbours are high-stay. The neighbour-level stay
and activity cutoffs and the required neighbour fractions entering these two
composite categories are read from the locked support-gate constants in the
released manifest.

\noindent\emph{Decision order.} The categories are mutually exclusive because
they are evaluated as an ordered cascade: the first condition that fires
assigns the label and the remaining conditions are not consulted. The code applies these tests in the following branch-specific order:

\smallskip
\noindent\begin{tabular}{@{}llll@{}}
\toprule
Branch & First test & Second test & Third test \\
\midrule
intervals & coverage gap & local dispersion & active feature gap \\
centers & coverage gap & active feature gap & local dispersion \\
\bottomrule
\end{tabular}

\smallskip
\noindent
A field reaching the end of the cascade without firing any condition is
labelled \texttt{supported}; the branch-specific conditions are those fixed in
Category thresholds above.

The order matters in principle, because a field could satisfy both the
dispersion and the active-feature-gap conditions and would then receive
different labels under the two orders. It does not change either stop decision
here: for centers the diagnosis contained no supported fields under either
order, and for intervals the coverage condition cannot fire at all, because
$d_{\mathrm{NN}}$ is exactly zero for every field in both stored support tables
(Supplementary Fig.~S25G).

\bigskip

\section*{Reproducibility and decision audit}
\addcontentsline{toc}{section}{Reproducibility and decision audit}

This section records when protocols and decisions were fixed and which software produced the results. It allows the reported analyses to be checked against the repository records.

\subsection*{Prespecification and acquisition integrity}

Throughout, an \emph{artifact} is a single file written by the pipeline and
never edited afterwards: a protocol or configuration file, a serialized prompt
set, a raw run directory, or a table of computed statistics. Protocols,
observation maps, prompt text, model IDs, physical seeds, analysis endpoints and
stopping rules were versioned and SHA-256 locked before the corresponding
acquisitions, and the locks were verified by the runners at acquisition time. We use \emph{prespecified} throughout in this
internal sense: these records were written, versioned and hash-locked in the
project repository before the acquisition or analysis they govern, but they
were not deposited with a public registry, so the ordering is documented by our
own records rather than externally attested. For the identical-field replay the
guarantee is also structural (Supplementary Fig.~S11).

Raw responses, parser status, retry history, serialized observations, actions
and phase trajectories were retained. No failed response was silently converted
into a valid action.

The repository carries the operational record that this section summarizes: the
machine-readable hash manifest (individual hash values and artifact paths), the individual request URLs, the SDK patch levels where they were recorded, and
the complete per-call retry history behind Supplementary Table~S2. It also carries the
acquisition traces themselves, one record per model call across all 15
acquisition groups, so every count reported here can be recomputed from the
raw responses rather than taken on trust.

\subsection*{Stop-or-repair chronology}
The three branch outcomes (moments prospective repair, centers STOP with
microscopic archive, and intervals STOP) followed a prespecified
chronology:
descriptor-space coverage check, in-distribution grouped-CV compressibility
test, collective-like transfer test, local-support diagnosis, and the prespecified
stop-or-repair decision. The table below gives the time at which each step's
artifact was written, together with the artifact that carries it, so that the
ordering can be checked directly against the released tree. Times are UTC. The
two STOP rows are the closure times recorded inside the manifests themselves
(\texttt{closed\_at\_utc}); the remaining rows are the write times of the named
artifacts.

\smallskip
\noindent\begin{tabular}{@{}llll@{}}
\toprule
Step & Moments (repair) & Centers (STOP) & Intervals (STOP) \\
\midrule
Descriptor-space coverage &
2026-07-23 15:33 &
2026-07-24 04:21 &
2026-07-24 05:58 \\
In-distribution grouped CV &
2026-07-23 15:53 &
2026-07-24 04:22 &
2026-07-24 05:59 \\
Collective-like transfer &
2026-07-23 23:45 &
2026-07-24 04:47 &
not acquired$^{\ddagger}$ \\
Local-support diagnosis &
2026-07-23 23:49 &
2026-07-24 05:41 &
2026-07-24 06:07 \\
Stop-or-repair freeze &
2026-07-24 01:05 &
2026-07-24 05:58:41 &
2026-07-24 06:07:23 \\
\bottomrule
\end{tabular}

\smallskip
\noindent\footnotesize The artifact paths for each row are listed in the
repository manifest. $^{\ddagger}$The intervals branch stopped before any
prospective transfer acquisition was authorized, so no paid transfer
acquisition exists for that branch.
\normalsize

\subsection*{Software and library versions}
\begin{itemize}\setlength{\itemsep}{1pt}
\item Language and runtime: the original analyses used CPython 3.11 on Windows. The September Claude follow-up and integrated figures used CPython 3.13.14, numpy 2.5.1, pandas 3.0.3 and matplotlib 3.11.0. The following library versions describe the original analyses. All local random draws in
the simulation and statistical analyses use explicitly seeded \texttt{numpy}
generators; provider-side language-model sampling is governed separately by the
API settings reported in Supplementary Table~S2.
\item Numerical and statistical libraries: the analyses import
\texttt{numpy}, \texttt{pandas} and \texttt{matplotlib} and nothing else.
Every permutation test, bootstrap, exact sign test and rank correlation is
implemented directly against \texttt{numpy}, so no result depends on the
defaults of a statistics package; neither \texttt{scipy} nor
\texttt{scikit-learn} is a dependency. The project environment carries
\texttt{numpy}~2.4.6 and \texttt{pandas}~3.0.3; the figures were rendered with
\texttt{matplotlib}~3.10.9.
\end{itemize}

\clearpage
\section*{Persistence of polar order (post hoc descriptive check)}
\addcontentsline{toc}{section}{Persistence of polar order (post hoc descriptive check)}
This post hoc descriptive analysis uses the existing positive-coupling trajectories, not additional model queries. For each of the six core physical seeds per coupling and encoding in each model, we count consecutive terminal state observations satisfying $r_1>0.9$ and summarize the last 20 observations, $t=81,\ldots,100$. A count of $w$ observations spans $w-1$ update intervals. We report counts and continuous order without a new significance test. The original endpoint and terminal-lock definitions, and all confirmatory $p$-values, remain unchanged. The reproducible script is \ident{analysis/publication/build_figures.py}; run-level results and summaries are in \ident{figures/figure_work/persistence_by_run.csv} and \ident{figures/figure_work/persistence_summary.json}. This check measures persistence within the observed horizon, not asymptotic frequency entrainment.

\begin{center}
\small
\begin{tabular}{@{}llrrrrr@{}}
\toprule
Model & Encoding & $K$ & Endpoint & Last 20 & Last 50 & Mean $r_1$, last 20 \\
\midrule
GPT & moments & 0.08 & 6/6 & 6/6 & 6/6 & 0.999 \\
GPT & moments & 0.15 & 6/6 & 6/6 & 6/6 & 0.996 \\
GPT & centers & 0.08 & 0/6 & 0/6 & 0/6 & 0.727 \\
GPT & centers & 0.15 & 0/6 & 0/6 & 0/6 & 0.764 \\
GPT & intervals & 0.08 & 0/6 & 0/6 & 0/6 & 0.378 \\
GPT & intervals & 0.15 & 0/6 & 0/6 & 0/6 & 0.502 \\
Claude & moments & 0.08 & 0/6 & 0/6 & 0/6 & 0.595 \\
Claude & moments & 0.15 & 0/6 & 0/6 & 0/6 & 0.627 \\
Claude & centers & 0.08 & 2/6 & 2/6 & 2/6 & 0.863 \\
Claude & centers & 0.15 & 6/6 & 6/6 & 6/6 & 0.986 \\
Claude & intervals & 0.08 & 5/6 & 4/6 & 4/6 & 0.945 \\
Claude & intervals & 0.15 & 6/6 & 6/6 & 5/6 & 0.993 \\
\bottomrule
\end{tabular}
\end{center}
Endpoint uses $r_1(T)\geq0.9$; Last 20 and Last 50 count runs with $r_1>0.9$ at every observation in the respective trailing window. These are 20 and 50 observations, spanning 19 and 49 update intervals. The final column averages within the last 20 observations and then across the six physical seeds. At $K=0.15$, the GPT--Claude reversal persists in all six seeds over the last 20 observations. At $K=0.08$, one Claude intervals run counted at the endpoint does not meet this persistence criterion. The all-run minimum for GPT moments is 68 consecutive terminal observations, consistent with the window-count convention used in Supplementary Fig.~S4.

\clearpage
\section*{Claude controlled-field follow-up: acquisition and interpretation}
\addcontentsline{toc}{section}{Claude controlled-field follow-up: acquisition and interpretation}
This descriptive follow-up was designed after the GPT--Claude collective reversal was known. It tests whether the response patterns identified in GPT transfer to Claude, without adopting the GPT directional phenotype rules as confirmatory Claude hypotheses. The protocol is \ident{experiments/stage_b_response_law/protocol_claude_kappa_v2.json}; raw and accepted records and analyses are under \ident{runs/claude_kappa_v2/anthropic/}. The exact GPT prompts, hashes, five concentrations and 36 mirror-paired jittered offsets were reused, with 24 responses per field--encoding cell in each of two sequential acquisition blocks. Each block's frozen shuffled task order was retained, with concurrent requests. The Claude model ID was \ident{claude-haiku-4-5-20251001}, temperature 0.7 and effective output cap 64. Anthropic received no sample seed. The model returned the specified ID for every accepted response. GPT's sweep and Claude's collective experiments were acquired in July 2026; this sweep was acquired on 18 September 2026. Model family and acquisition period are therefore not independently randomized in the cross-model comparison.

\subsection*{Syntax amendment and complete attempt accounting}
The initial parser required bare JSON. All nine returned preflight texts instead contained an outer Markdown JSON fence, across four distinct tasks. Before full acquisition, the syntax rule was amended to remove exactly one complete outer fence and then require the original single-field \ident{social_action} JSON schema and valid action label. This is a disclosed amendment after inspecting response syntax, not an externally preregistered rule. The first received response for each preflight task was imported (four responses); the five later duplicate retries remain in the immutable v1 journal and were excluded. No action label was used to choose which preflight response to retain. All 25,920 retained responses required fence removal; none was bare-JSON compliant. No prose extraction, extra-key acceptance or fallback-to-stay rule was used.

\begin{center}\small
\begin{tabular}{@{}lrr@{}}\toprule
Quantity & Block 1 & Block 2 \\\midrule
Accepted responses & 12,960 & 12,960 \\
Complete field--encoding cells (24 each) & 540 & 540 \\
Dispatched attempts in v2 journal & 12,984 & 12,984 \\
Returned result records & 12,980 & 12,984 \\
Transport-error result records & 20 & 24 \\
Dispatches without returned result & 4 & 0 \\
First-attempt valid rate, amended parser & 99.907\% & 99.938\% \\
Imported first preflight responses & 4 & 0 \\
\bottomrule\end{tabular}\end{center}

The two v2 journals contain 25,968 dispatches, including the four imported first attempts. Adding the five excluded preflight duplicates gives 25,973 distinct recorded dispatches across v1 and v2. These comprise 25,920 accepted responses, five excluded duplicate texts, 44 transport-error results and four dispatches with no recorded result. The last four occurred when acquisition was stopped to increase concurrency from four to eight workers, after 12,254 accepted block-1 samples. Their action outcomes are unknown; the same tasks were subsequently retried. For four block-1 tasks and eight block-2 tasks that exhausted the original three-attempt cap without a valid response, a logged transport-recovery amendment raised the cap to six; all succeeded on attempt four. Thus the original attempt cap was exceeded, but cell sizes and prompts were unchanged, and no accepted response was replaced or selected according to its action. First-attempt validity exceeded the frozen 99\% quality threshold in both blocks; every cell reached exactly 24 valid responses. Token totals in the logs omit errors or unknown dispatches without usage records and must not be treated as complete billing totals.

\clearpage
\subsection*{Analysis and scope}
The unconstrained least-squares Fourier fit used $M=6$, with $M=2,4,8,12$ sensitivity fits, separately for each block. Existing within-offset multinomial bootstrap ellipses used 2,000 replicates. They describe sampling uncertainty conditional on the measured fields and observed multinomial probabilities, not uncertainty from model updates or unseen field shapes; empirical zero-count actions may be underestimated. Fourier coefficients can exceed the response range even though the mean action lies in $[-1,1]$. Display clipping affects only drawn curves, never fitted coefficients. The new figures show blocks separately and their connecting lines are not uncertainty intervals. No new confirmatory $p$-values are reported.

Across all five concentrations and both blocks, Claude histogram $a_1$ values ranged from 1.226 to 1.276, whereas moments ranged from 0.944 to 1.028. At $\kappa=12$, the block-specific moments $(a_1,b_1)$ values were $(0.997,0.775)$ and $(1.003,0.766)$, compared with centers $(1.275,0.063)$ and $(1.276,0.057)$ and intervals $(1.226,0.296)$ and $(1.227,0.294)$. The moments first-harmonic angle $\operatorname{atan2}(b_1,a_1)$ was $37.9^\circ$ and $37.4^\circ$; across all five fitted orders and both blocks its range was $37.3^\circ$--$38.5^\circ$. Centers ranged from $2.6^\circ$ to $3.3^\circ$ and intervals from $13.4^\circ$ to $15.3^\circ$ across these orders. These ranges are sensitivity ranges, not confidence intervals or agent-to-agent phase lags. In moments, the fitted local slope at zero changes sign across orders; it remains a diagnostic, not a stability criterion.

The strong positive histogram response is qualitatively consistent with the collective reversal, but moments also has positive $a_1$ and near-unit activity without reaching high polar order. The even component and phase displacement therefore supply an additional candidate explanation, not a validated mechanism. The curves describe rotated unimodal fields; collective trajectories can visit other field shapes and accumulate stochastic feedback. A causal or quantitative account requires out-of-sample closed-loop predictions or an intervention that tests this candidate while controlling the rest of the response law.

The reproducible offline audit and figure script is \ident{analysis/publication/integrate_claude_figures.py}. It verifies complete cells, unique retained tasks, the first-response preflight rule, model identifiers and reconstruction of the frozen Fourier coefficients for both families. Outputs under \ident{figures/figure_work/} are:
\begin{itemize}
\item Acquisition audit: \ident{claude_integration_audit.json}.
\item Endpoint table: \ident{claude_integration_endpoints.csv}.
\item Sensitivity table: \ident{claude_kappa12_order_sensitivity.csv}.
\end{itemize}
These data and scripts are included in the public \texttt{v1.2.0} release, archived at \url{https://doi.org/10.5281/zenodo.22899385}.

\begin{figure}[H]
\centering\includegraphics[width=180mm]{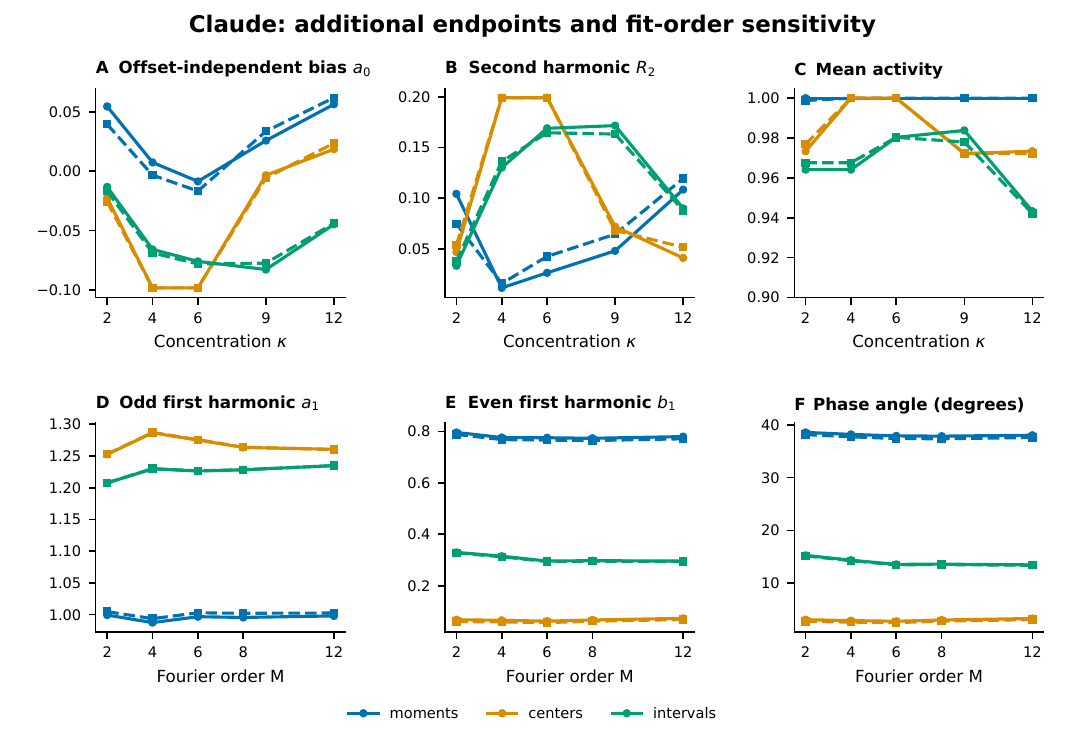}
\caption{\textbf{Claude concentration dependence and Fourier-order sensitivity.} \textbf{(A--C)} Offset-independent bias $a_0$, second-harmonic amplitude $R_2$ and mean activity across the five concentrations. \textbf{(D--F)} First-harmonic sine and cosine coefficients and their angle at $\kappa=12$ across five fitted orders. Colors denote moments, centers and intervals; solid/circles and dashed/squares denote acquisition blocks 1 and 2. Lines connect point estimates and are not uncertainty intervals. The phase displacement of moments persists across orders; these angles describe first harmonics, not equilibrium phase lags. The local derivative at zero is order-sensitive and is not used to classify stability. Blockwise conditional bootstrap ellipses and the complete probability tables are supplied in the accompanying analysis files.}
\label{fig:claudekappasensitivity}
\end{figure}

\clearpage
\begin{figure}[H]
\centering\includegraphics[width=180mm]{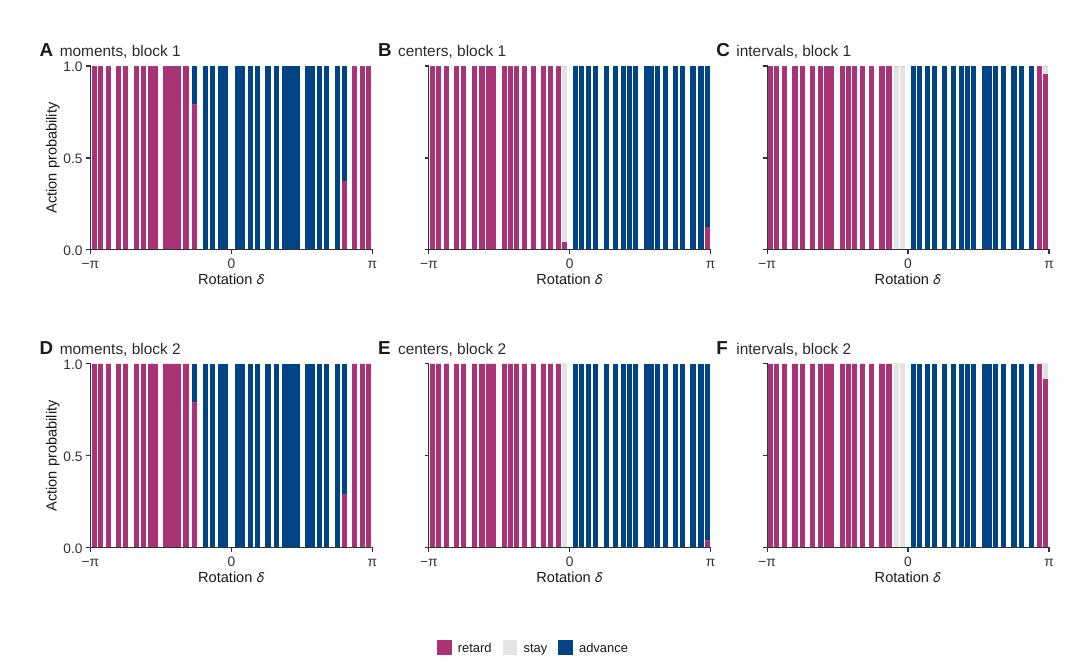}
\caption{\textbf{Full Claude action distributions expose the shifted moments response.} Each bar shows the empirical probabilities of retard, stay and advance at one recorded rotation offset for $\kappa=12$ (24 responses per bar). Columns correspond to moments, centers and intervals; rows show the two acquisition blocks. The horizontal axis is the same physical rotation for all encodings and both models. Moments changes sign at different rotations from the histograms, with almost no abstention. Complete probability tables for all five concentrations accompany the acquisition records; no pooling of acquisition blocks is used. These distributions describe the queried unimodal fields and do not by themselves predict collective stability.}
\label{fig:claudekappaactions}
\end{figure}

\end{document}